\documentclass{aa}  

\usepackage{graphicx}
\usepackage{pdflscape}
\usepackage{txfonts}
\begin{document} 

\title{CO Depletion in Infrared Dark Clouds}

   \author{G. Cosentino\inst{1,2}\thanks{E-mail:cosentino@iram.fr},
          J. C. Tan\inst{3,4},
          C. Gainey\inst{5},
          C. Y. Law\inst{6},
          C.-J. Hsu\inst{3},
          D. Xu\inst{4},
          W. Lim\inst{7},
          I. Jim\'enez-Serra\inst{8},\\
          A. T. Barnes\inst{2},
          F. Fontani\inst{6,9,10},
          J. D. Henshaw\inst{11},
          P. Caselli\inst{10},
          S. Viti\inst{12}}
    \authorrunning{Cosentino et al.}
    \titlerunning{CO depletion in IRDCs}
   \institute{Institut de Radioastronomie Millimétrique, 300 Rue de la Piscine, 38400 Saint-Martin-d’Hères, France
   \and
   European Southern Observatory, Karl-Schwarzschild-Strasse 2, D-85748 Garching, Germany
    \and 
   Department of Space, Earth and Environment, Chalmers University of Technology, SE-412 96 Gothenburg, Sweden
    \and 
    Department of Astronomy, University of Virginia, 530 McCormick Road Charlottesville, 22904-4325 USA
    \and
    Department of Astronomy, Yale University, New Haven, CT 06511, USA
    \and
    INAF  Osservatorio Astronomico di Arcetri, Largo E. Fermi 5, 50125 Florence, Italy
    \and
    California Institute of Technology, Pasadena, CA 91125, USA
    \and
    Centro de Astrobiolog\'{i}a (CSIC/INTA), Ctra. de Torrej\'on a Ajalvir km 4, Madrid, Spain
    \and
    Laboratory for the study of the Universe and eXtreme phenomena (LUX), Observatoire de Paris, 5, place Jules Janssen, 92195 Meudon, France
    \and
    Max Planck Institute for Extraterrestrial Physics, Giessenbachstrasse 1, 85748 Garching bei M\"{u}nchen, Germany
    \and
    Astrophysics Research Institute, Liverpool John Moores University, 146 Brownlow Hill, Liverpool L3 5RF, UK
    \and
    Leiden Observatory, Leiden University, PO Box 9513, 2300 RA Leiden, The Netherlands
    }
 
   \date{Received September 15, 1996; accepted March 16, 1997}

 
\abstract
{Infrared Dark Clouds (IRDCs) are cold, dense structures likely representative of the initial conditions of star formation. Many studies of IRDCs employ CO to investigate cloud dynamics. However, CO can be highly depleted from the gas phase in IRDCs, impacting its fidelity as tracer. The CO depletion process is also of great interest in astrochemistry, since CO ice in dust grain mantles provides the raw material for forming complex organic molecules.}
{We study CO depletion toward four IRDCs to investigate how it correlates with volume density and dust temperature, calculated from {\it Herschel} far-infrared images.}
{We use $^{13}{\rm CO}\: J=1\rightarrow0$ and $2\rightarrow1$ maps to measure CO depletion factor, $f_D$, across IRDCs G23.46-00.53, G24.49-00.70, G24.94-00.15, and G25.16-00.28. We also consider a normalized CO depletion factor, $f_D^\prime$, which takes a value of unity, i.e., no depletion, in the outer, lower density, warmer regions of the clouds. We then investigate the dependence of $f_D$ and $f_D^\prime$ on gas density, $n_{\rm H}$ and dust temperature, $T_{\rm dust}$.}
{We find CO depletion rises as density increases, reaching maximum values of $f_D^\prime\sim10$ in some regions with $n_{\rm H}\gtrsim3\times10^5\:{\rm cm}^{-3}$, although with significant scatter at a given density. We find a tighter, less scattered relation of $f_D^\prime$ with temperature, rising rapidly for temperatures $\lesssim18\:$K. We propose a functional form $f_D^\prime = \:{\rm exp}(T_0/[T_{\rm dust}-T_1])$ with $T_0\simeq4\:$K and $T_1\simeq12\:$K to reproduce this behaviour.
}
{We conclude that CO is heavily depleted from the gas phase in cold, dense regions of IRDCs. Thus, if not accounted for, CO depletion can lead to underestimation of total cloud masses based on CO line fluxes by factors up to $\sim5$. These results indicate a dominant role for thermal desorption in setting near equilibrium abundances of gas phase CO in IRDCs, providing important constraints for both astrochemical models and the chemodynamical history of gas during the early stages of star formation.}

   \maketitle
%
\section{Introduction}\label{intro}

Infrared Dark Clouds (IRDCs) are cold \cite[$T\lesssim$20 K;][]{pillai2006}, dense \cite[$n_{\rm H}\gtrsim10^4$ cm$^{-3}$;][]{butlerTan2012} and highly extincted ($A_{V}\sim10-100\:$mag) regions of the Interstellar Medium (ISM), known to harbour the conditions of star and star cluster formation. First detected as dark features against the mid-IR Galactic background \citep{perault1996,egan1998}, IRDCs show low levels of star formation activity and have mass surface densities similar to those of massive star forming regions \citep{tan2014}. Furthermore, IRDCs host cold, dense, deuterated pre-stellar cores, i.e., the earliest phase of massive star formation \citep{tan2013,2017ApJ...834..193K}. For all these reasons, IRDCs have long been regarded as the birth places of massive stars and star clusters \citep{rathborne2006,foster2014, pillai2019,moser2020,yu2020}.

Despite this importance, the mechanisms that trigger star formation in these objects are still unclear \citep[e.g.,][]{tan2000,tan2014,hernandez2015,peretto2016,RetesRomero2020,morii2021}. 
Theoretical models and simulations have suggested that star formation can be efficiently ignited within IRDCs as a consequence of dynamical compression and gravitational instability of the gas. Among current theories, IRDCs have been proposed to form as the shock-compressed layer in the collision between pre-existing Giant Molecular Clouds (GMCs) \citep[e.g.,][]{tan2000,tasker2009,2010ApJ...710L..88T,2014ApJ...787...68S,wu2015,wu2017,2018PASJ...70S..56L,2024ApJ...977L...6F} with the collision being a consequence of GMCs orbital motion in a shearing Galactic disk. Other models also involve formation from compressive collisions, but driven by momentum from stellar feedback \citep[e.g.,][]{inutsuka2015}. Still, other models invoke IRDCs/dense clump formation as part of the same processes of hierarchical gravitational collapse that forms the surrounding GMCs \citep[e.g.,][]{vazquez2019}. All these mechanisms are expected to leave different imprints on the gas dynamical properties in the formed IRDCs.

Since molecular hydrogen ${\rm{H_2}}$, the primary constituent of molecular clouds, is not excited at the low temperatures of these objects, indirect methods of tracing cloud mass have been developed. Among these, the rotational transitions of CO, which is the next most abundant molecular, and its isotopologues have been used to estimate clouds structural and kinematic properties, assuming a certain CO to ${\rm{H_2}}$ abundance ratio \citep[e.g.][]{caselli1999,crapsi2005,fontani2006,hernandez2011,jimenezserra2014,sabatini2019}. 
However, at the low-temperatures and high-densities typically found in IRDCs, CO can be heavily depleted from the gas phase due to freeze-out onto dust grains. This may cause important quantities, such as cloud mass, to be significantly underestimated.. 

CO depletion also has major implications for the chemistry of star-forming regions. For example, if CO is highly depleted from the gas phase, then the abundance of $\rm H_2D^+$ can rise, leading to high levels of deuteration of remaining gas-phase species, such as $\rm N_2H^+$ and $\rm NH_3$ \citep[e.g.,][]{dalgarno1984,caselli2002,fontani2006,kong2015}. In addition, the formation of many complex organic molecules (COMs) is expected to occur within CO ice mantles of dust grains \citep[e.g.][]{HerbstDishoeck2009}.

Gas phase depletion of CO is typically quantified using the so-called CO depletion factor, $f_D$, i.e., the ratio between the \textit{expected} CO column density given a CO-independent measure of the column density of H nuclei, $N_{\rm H}$, and assuming standard gas phase CO abundance along a line of sight to the \textit{observed} CO column density.
Estimates of $N_{\rm H}$ in molecular clouds typically are made
via dust continuum emission at sub-mm and mm wavelengths \citep[e.g.,][]{lim2016} or via mid- and near-infrared dust extinction measurements \citep[e.g.,][]{butlerTan2012,kainulainen2013}. 

On the smaller, $\lesssim 0.1\:$pc scales of cores, CO depletion has been investigated toward both low-mass \citep[e.g.,][]{caselli1999,kramer1999,whittet2010,fordShirley2011,christie2012} and high-mass \citep[e.g.,][]{fontani2006,zhang2009,fontani2012,sabatini2019} examples. In both cases large values of CO depletion factors have been reported. In particular, in high-mass pre-stellar and early-stage cores \citet{fontani2012} have estimated values up to $f_D>80$, while \citet{zhang2009} found $f_D>100$.

On the larger scales of IRDCs, \citet{hernandez2011} and \citet{hernandez2012} used multi-transition single-dish observations of the C$^{18}$O isotope to obtain a parsec-scale CO depletion map of IRDC G35.39-00.33 and, under the assumption of local thermodynamic equilibrium (LTE) conditions, reported values of $f_D$ of up to $\sim3$. Toward the same cloud, \cite{jimenezserra2014} used the LVG approximation and reported CO depletion factors of $\sim5, 8$ and 12 in three selected positions. Similar results have been reported toward the IRDC G351.77-0.51 by \cite{sabatini2019}. Toward the IRDC G28.37+00.07, \cite{entekhabi2022} reported values of $f_D$ up to $\sim10$. \cite{feng2020} investigated CO depletion toward a sample of four IRDCs, presenting CO depletion maps that show $f_D$ values of up to 15. All these studies have confirmed that depletion of CO from the gas phase is significant not just toward the dense cores and clumps, but also in the inter-clumps regions of IRDCs. 

While it is generally known that CO depletion is affected by density and temperature variations, only a few previous studies have been dedicated to specific investigation of the dependence of $f_D$ on these properties. In particular, \cite{kramer1999} investigated the dependency of CO depletion on dust temperature toward the dense core IC~5146. They found that their data could be well fit with the following function:
\begin{equation}
f_{D} = A \: {\rm exp} (T_0/T)\label{eq:kramer}
\end{equation}
with $A=0.41^{0.67}_{0.25}$ and $T_0=14.1\pm0.6\:$K. However, we note that their data covered a relatively modest range of depletion factors, i.e., up to $f_D\sim 2.5$. Furthermore, their dust temperatures were estimated based on the ratio of 1.2~mm flux to near infrared dust extinction (with equivalent $A_V$ ranging up to about 30~mag, i.e., up to $\Sigma\simeq0.13\:{\rm g\:cm}^{-2}$), which is relatively sensitive to the choice of dust opacity model.
 
In this work we explore CO depletion in a sample of four IRDCs, namely G24.94-00.15, G23.46-00.53, G24.49-00.70, G25.16-00.28 (hereafter Clouds O, V, X and Y), which are part of a larger sample of 26 clouds (A-Z) \citep[][Cosentino et al., in prep.]{2009ApJ...696..484B,butlerTan2012}.
This paper is organised as follows. In Section 2, we present our sample, the data in hand and the technical details of our observations. In Section 3, we describe the method adopted and the analysis performed. In Sections 4 and 5, we present and discuss our results. In Section 6, we draw our conclusions. 

\section{The IRDC Sample}\label{target}

The four IRDCs studied in this work are selected from a larger sample of 16 clouds (K to Z) that will be presented in a forthcoming paper (Cosentino et al. in prep.) and which represent an extension to the sample of ten IRDCs of \citet{2009ApJ...696..484B,butlerTan2012} (named A to J). These 26 sources have been identified as dark features against the diffuse MIR Galactic background, and have been selected for being located relatively nearby (kinematic distance $\leq$ 7 kpc) and for showing the highest levels of contrast against the diffuse Galactic background emission as observed with Spitzer-IRAC at 8 $\mu$m \citep{churchwell2009}. 

\begin{figure*}
    \centering    
    \includegraphics[width=0.33\textwidth]{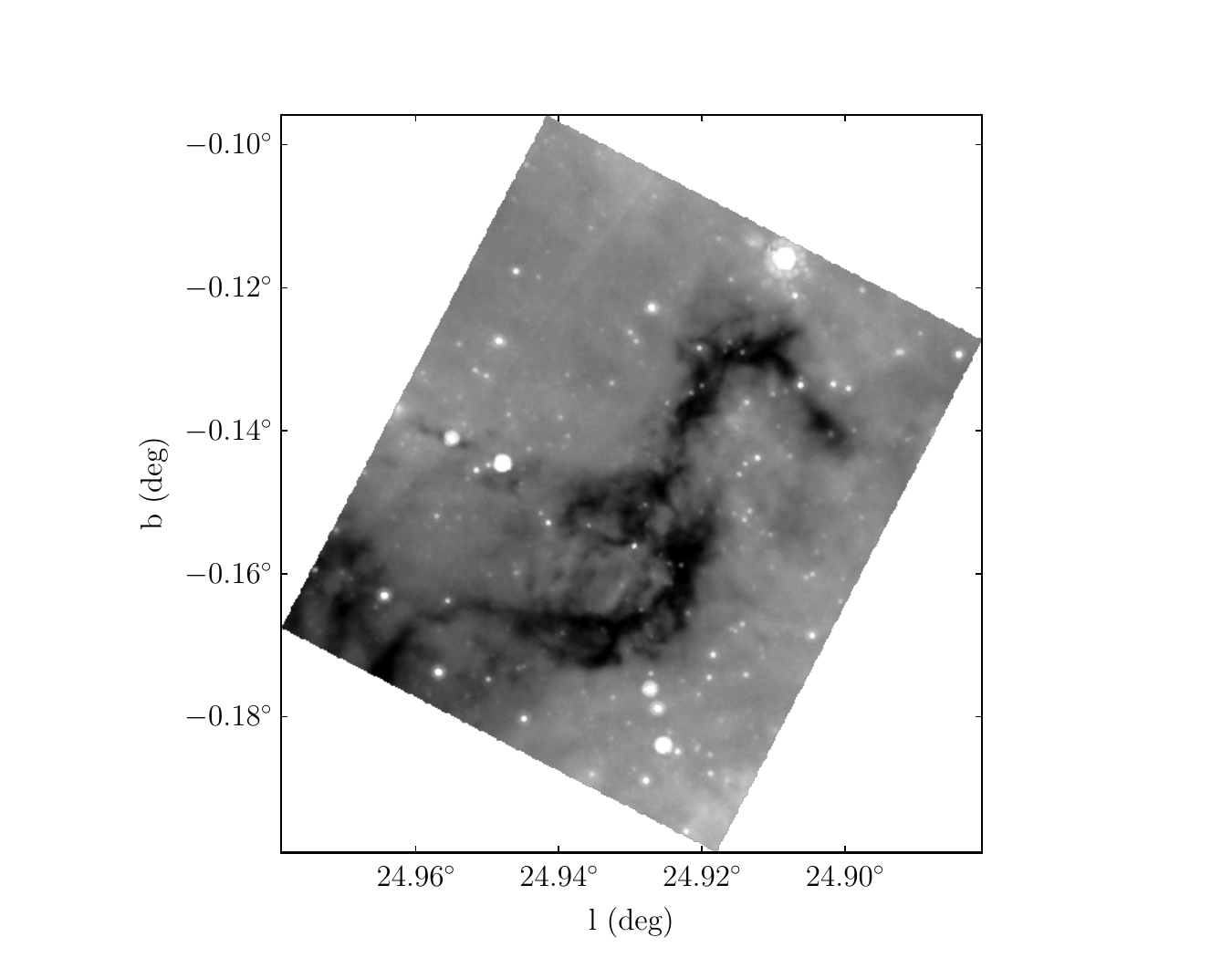}\includegraphics[width=0.33\textwidth]{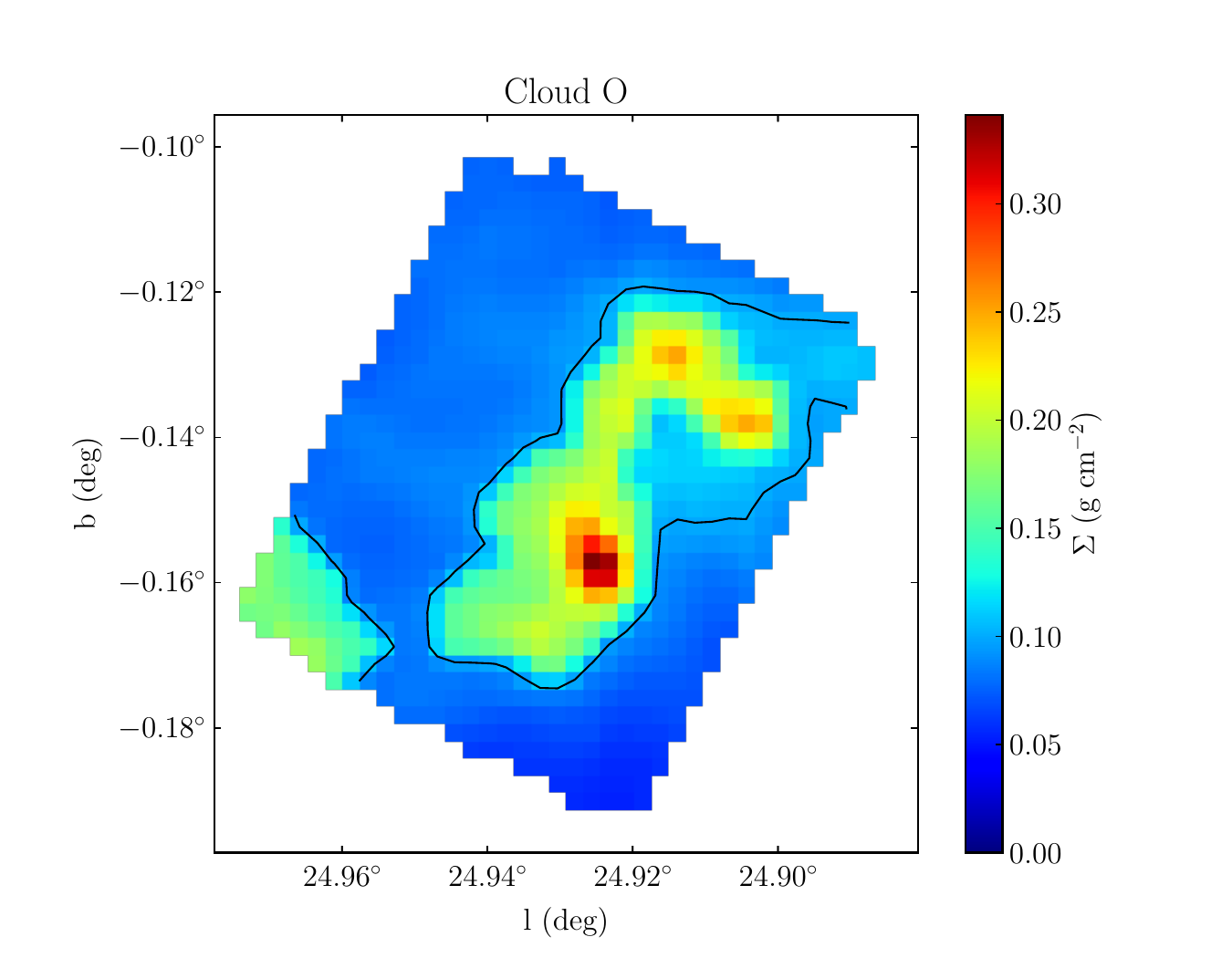}\includegraphics[width=0.33\textwidth]{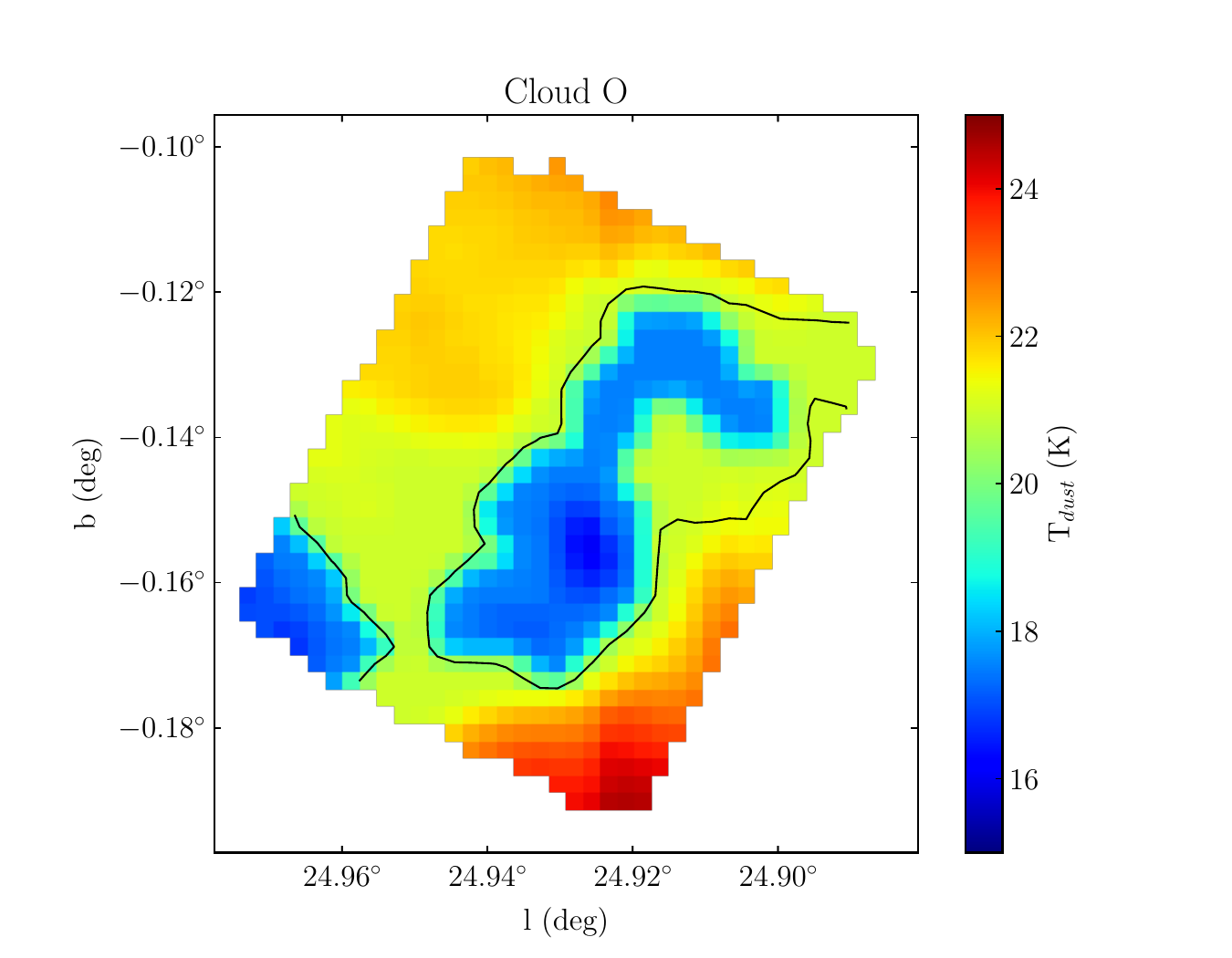}\\
    
    \includegraphics[width=0.33\textwidth]{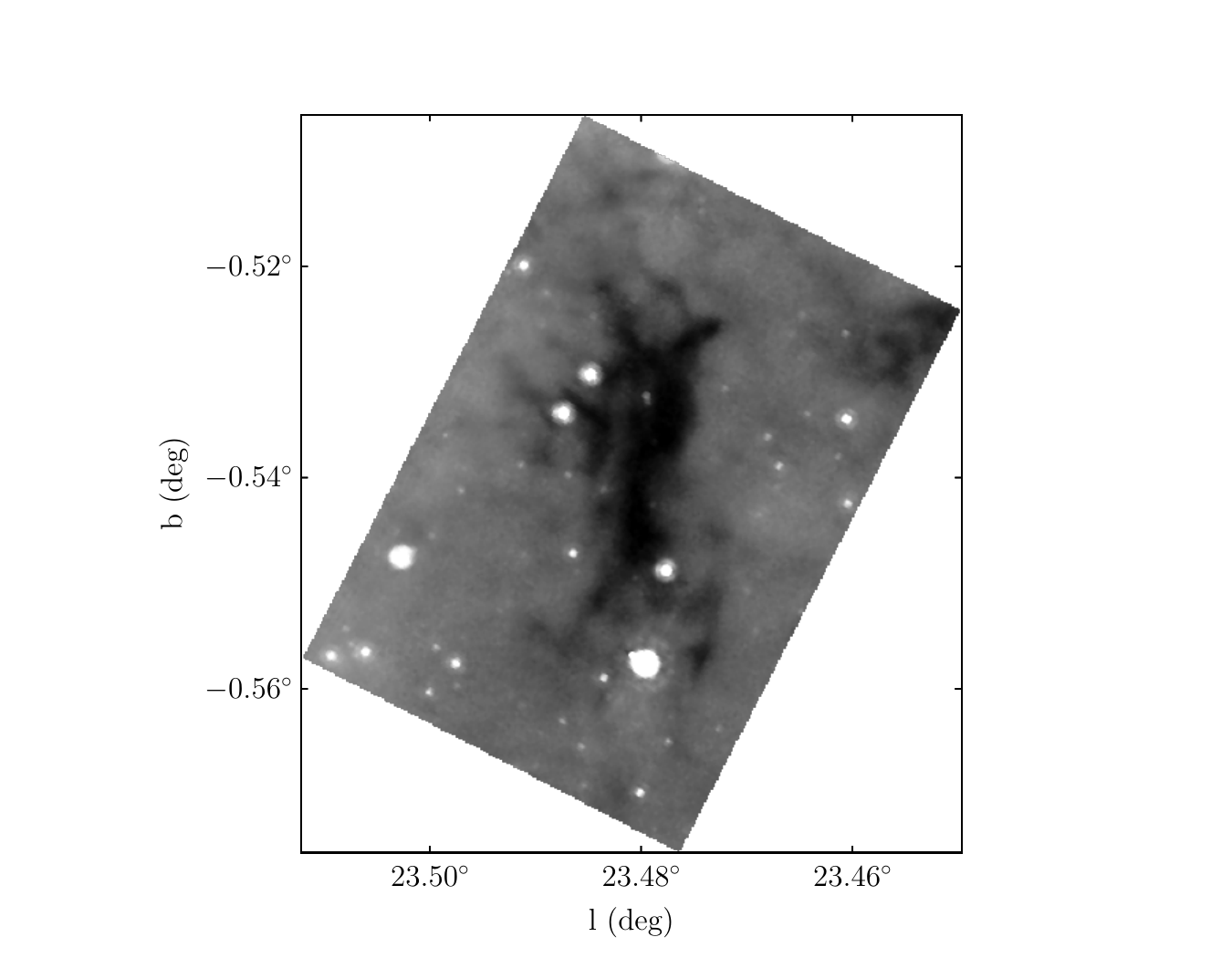}\includegraphics[width=0.33\textwidth]{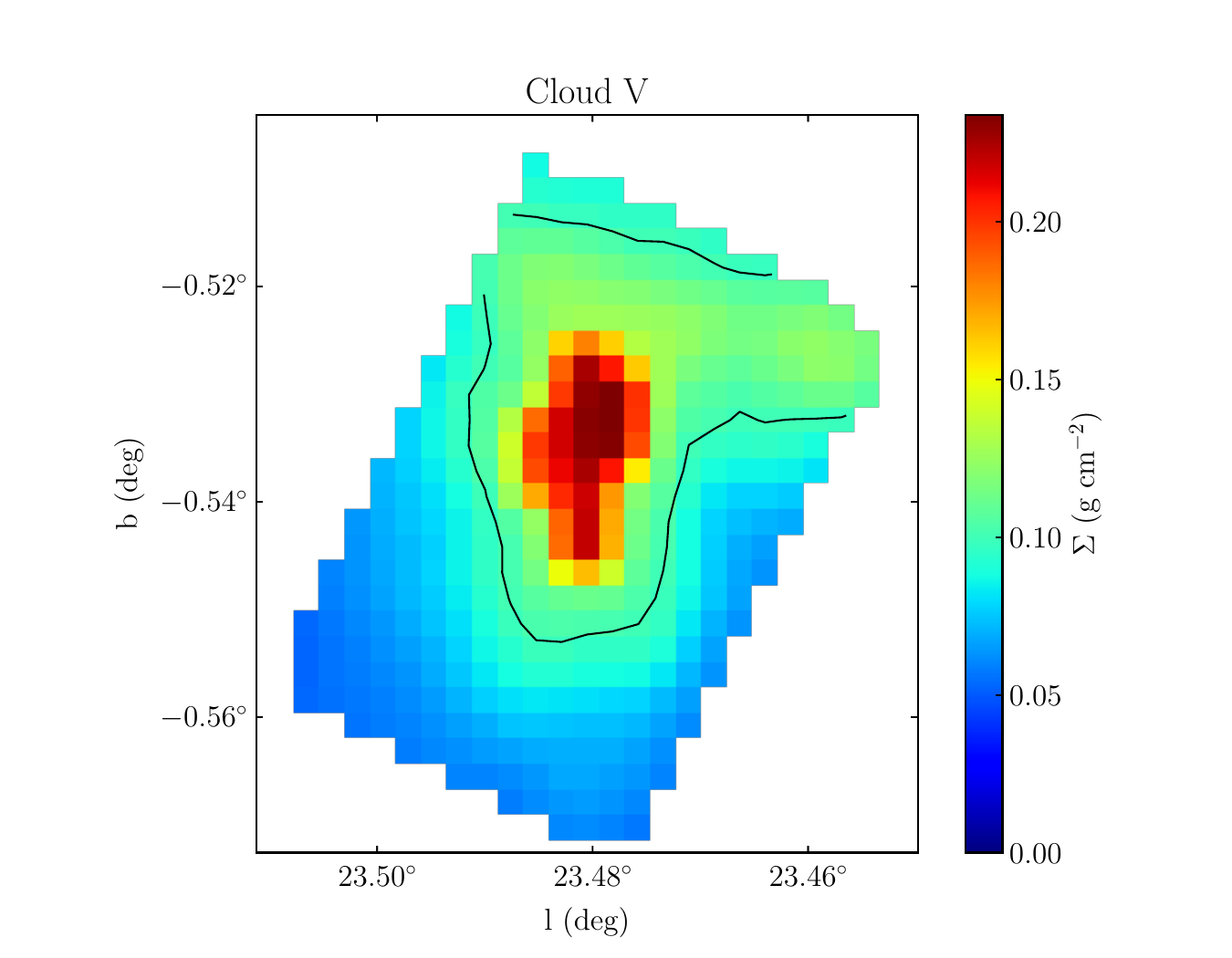}\includegraphics[width=0.33\textwidth]{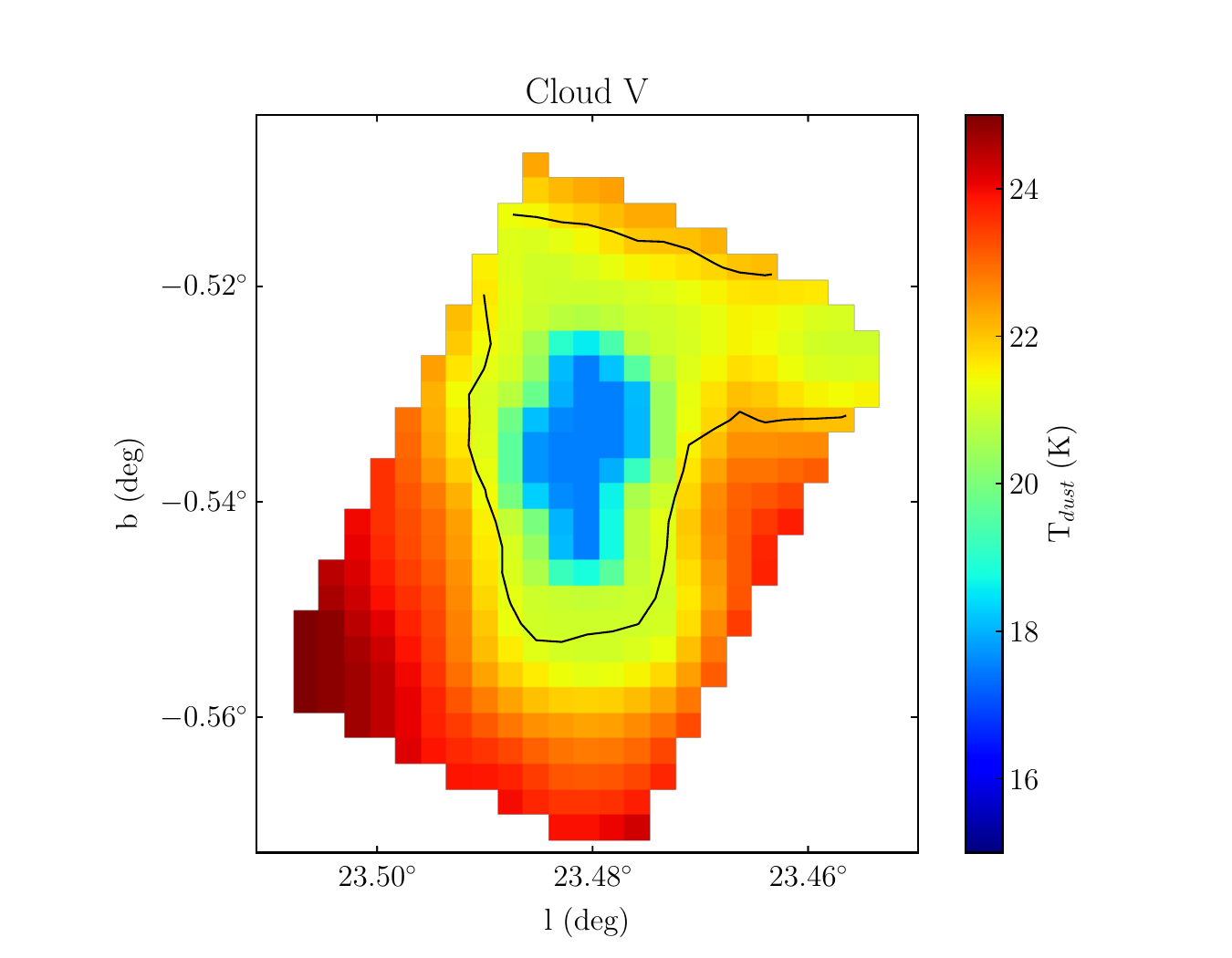}\\

    \includegraphics[width=0.33\textwidth]{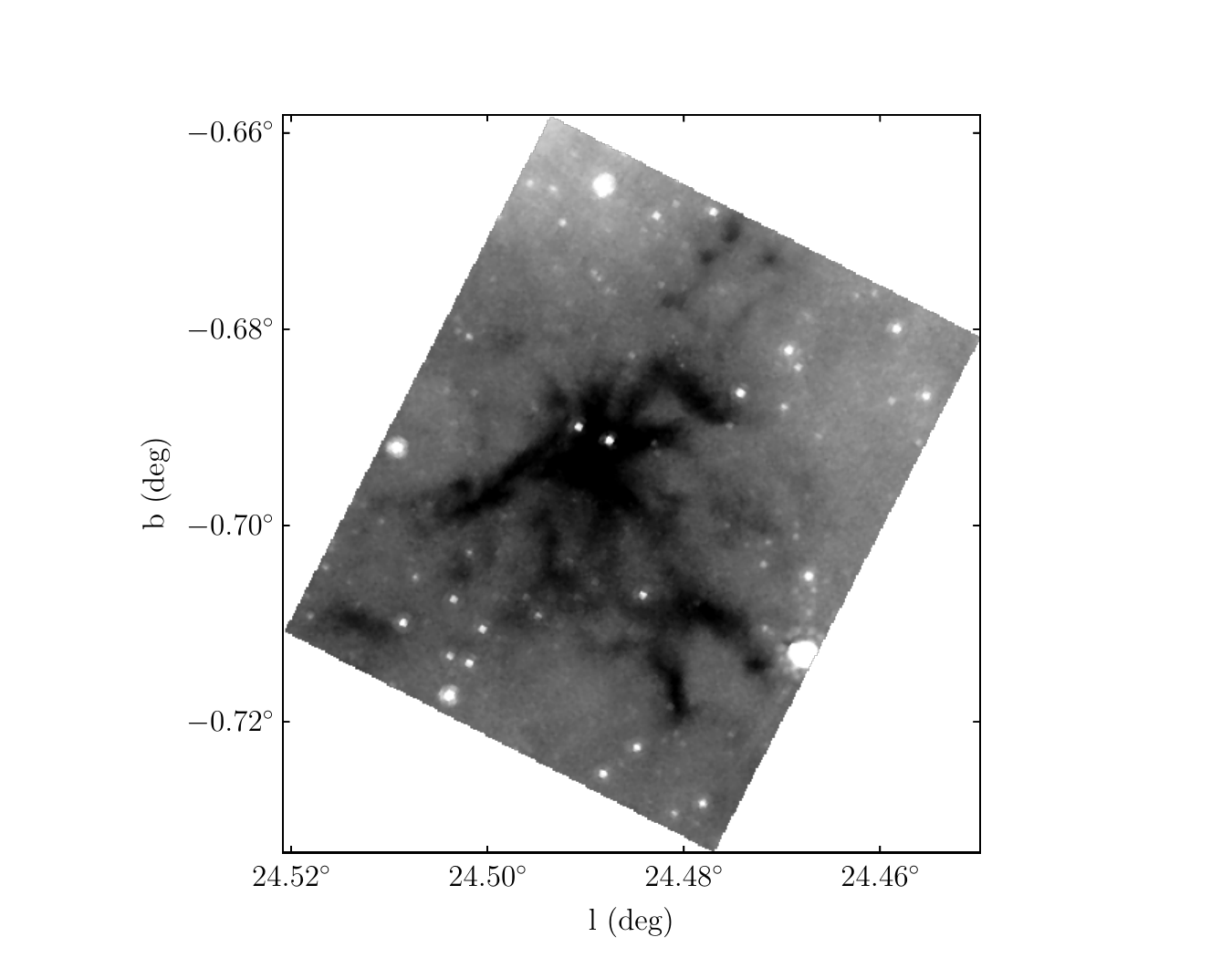}\includegraphics[width=0.33\textwidth]{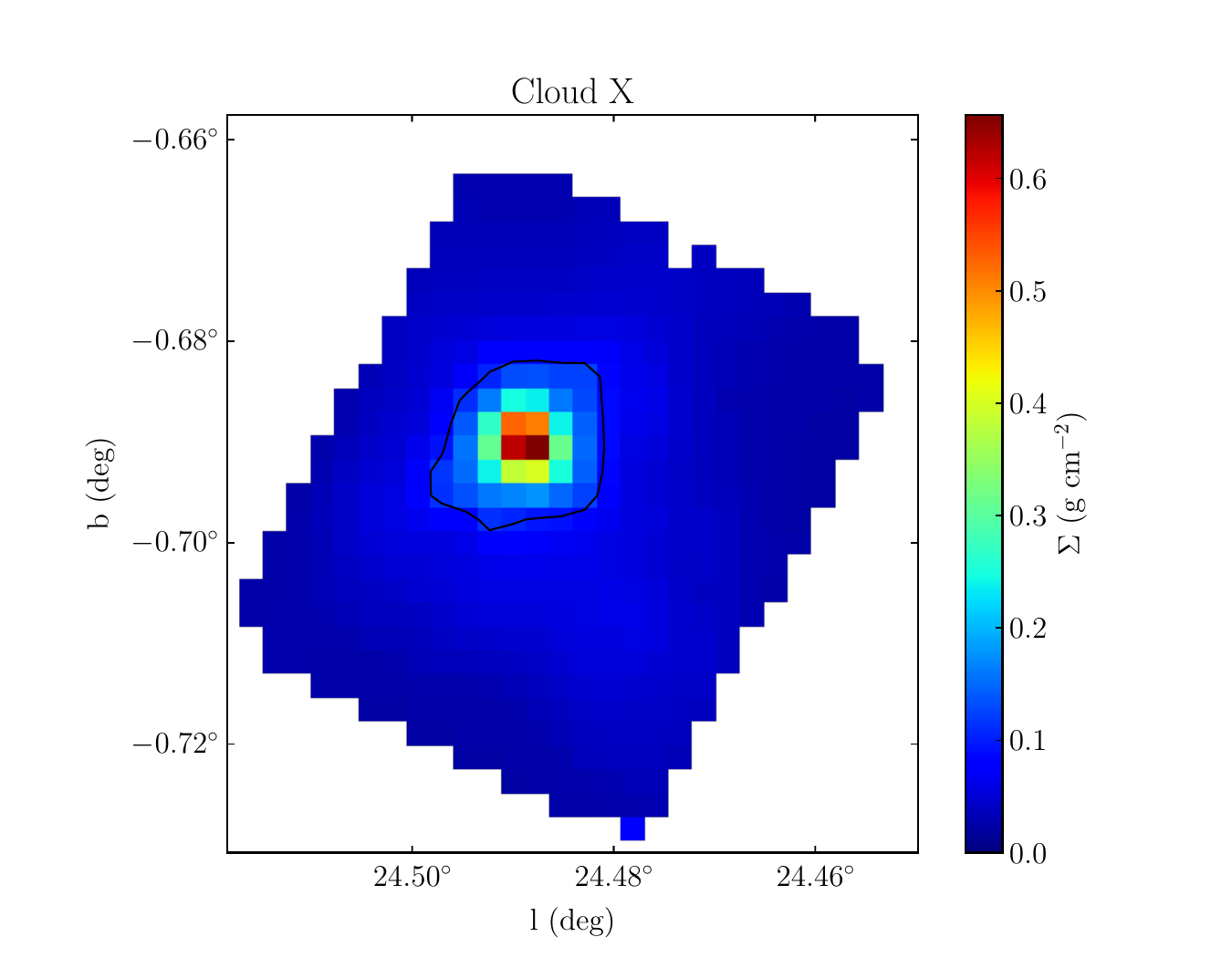}\includegraphics[width=0.33\textwidth]{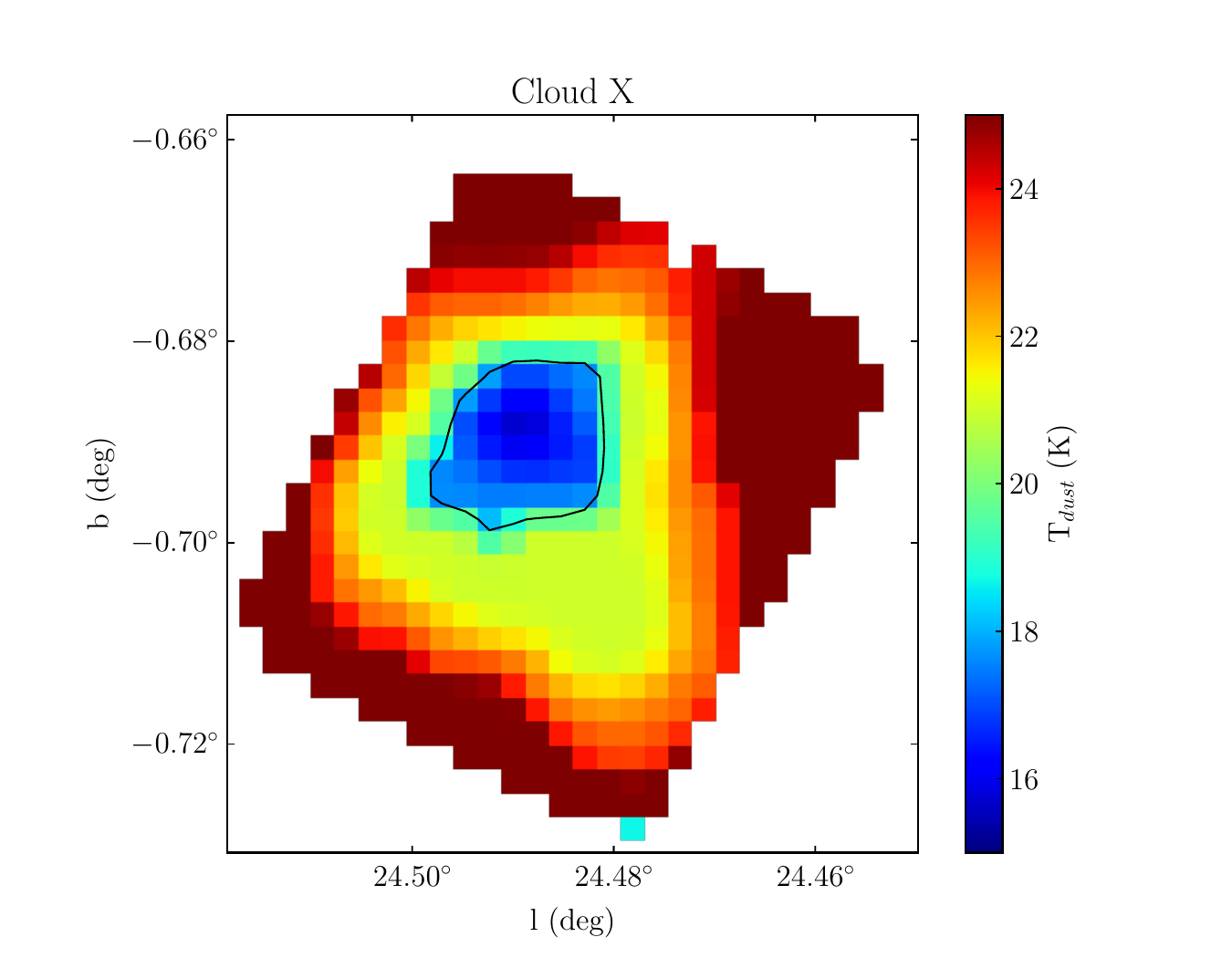}\\

    \includegraphics[width=0.33\textwidth]{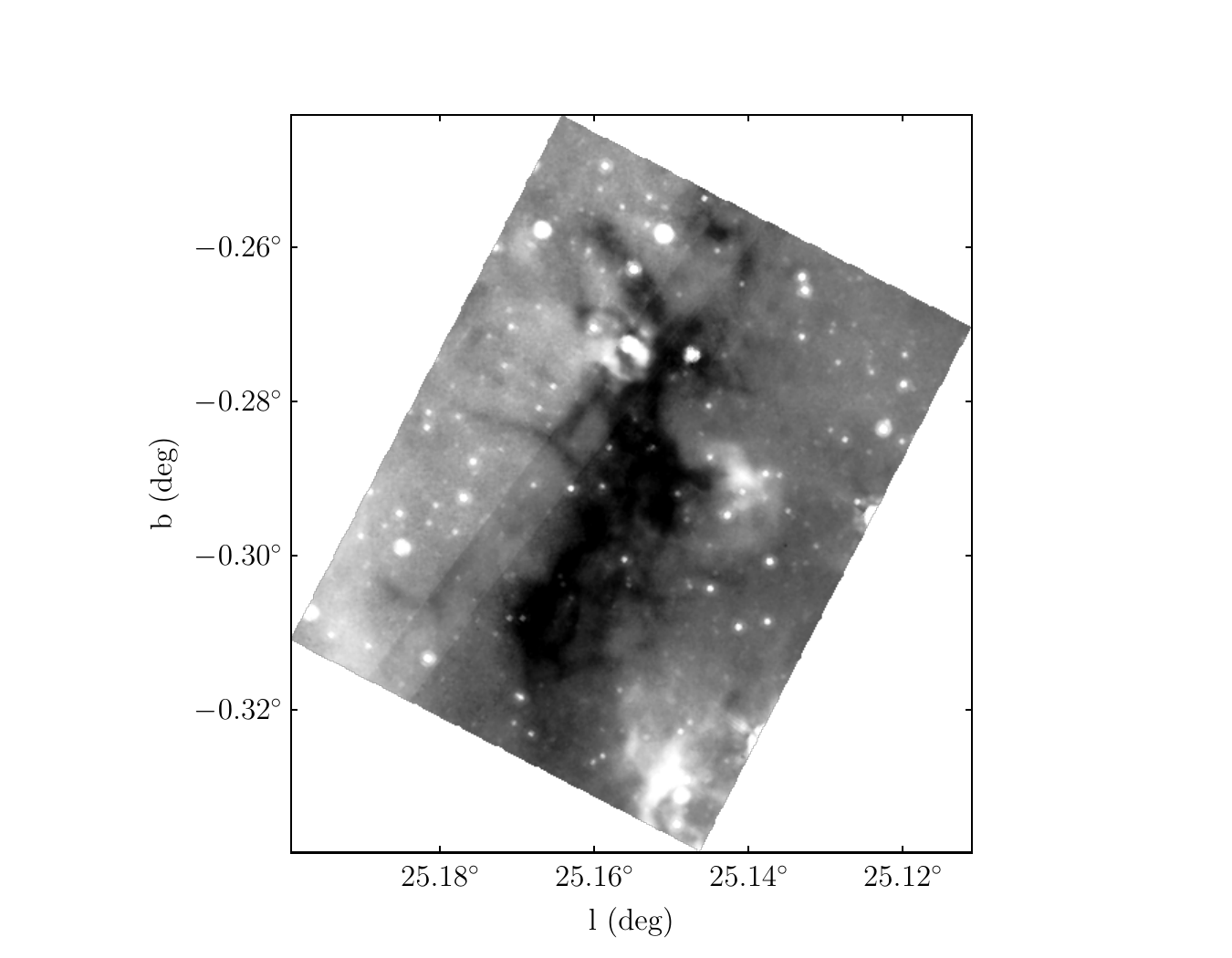}\includegraphics[width=0.33\textwidth]{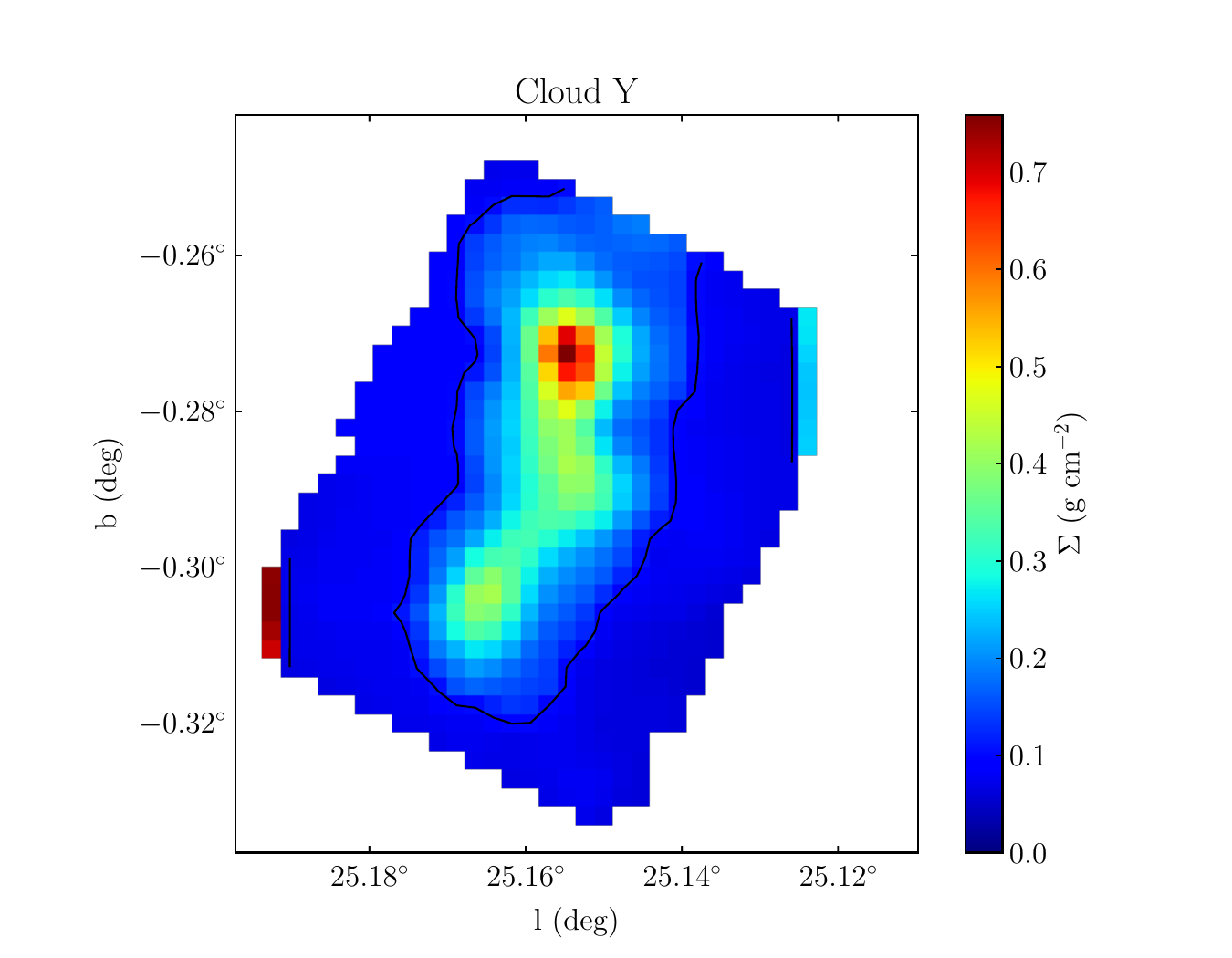}\includegraphics[width=0.33\textwidth]{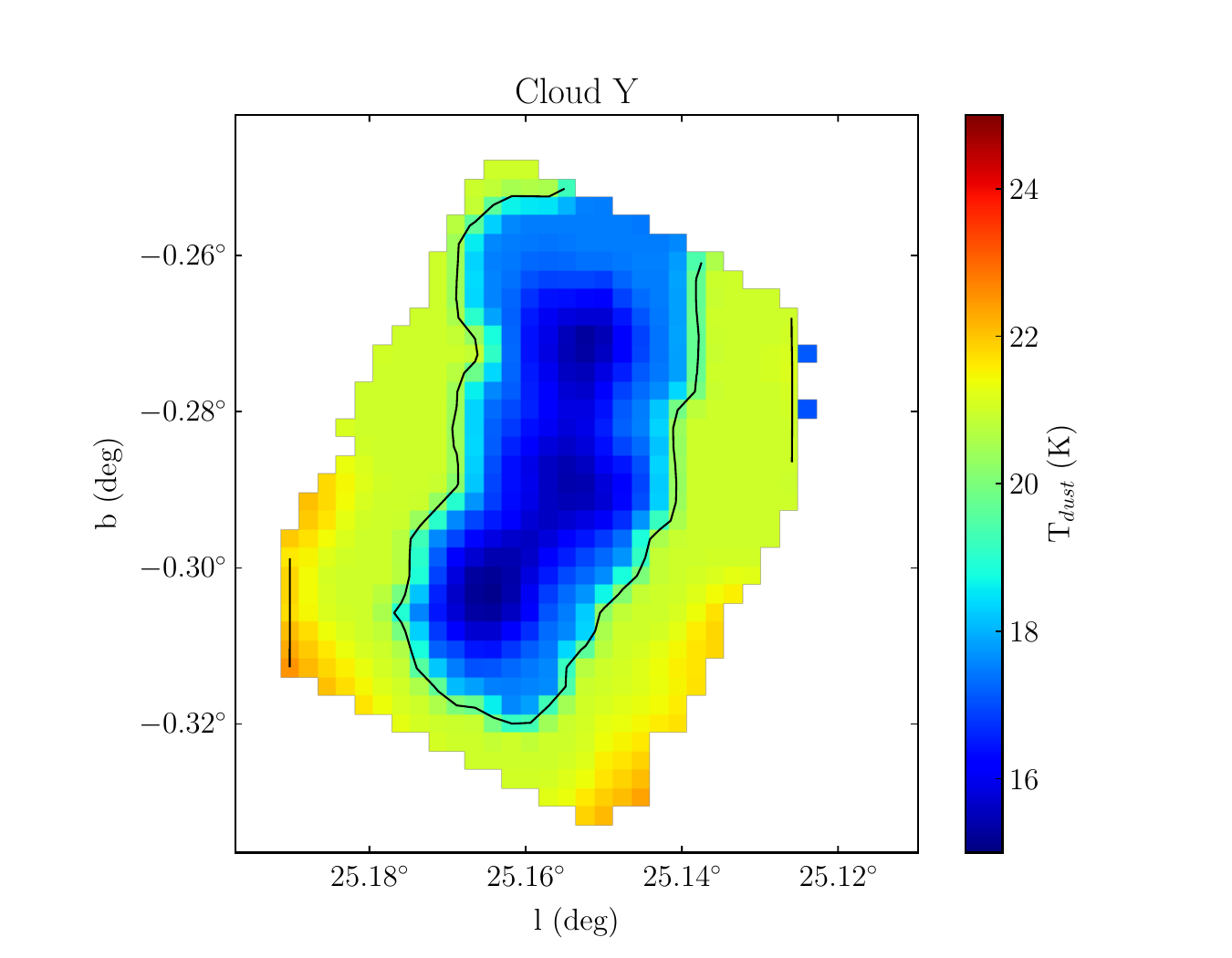}

    \caption{{\it Spitzer} 8 $\mu$m images (left column), {\it Herschel}-derived mass surface density maps ($\Sigma$; middle column) and dust temperature ($T_{\rm{dust}}$; right column) for the four IRDCs O, V, X, and Y (top to bottom). In the middle and right panels, the $\Sigma$=0.1 g cm$^{-2}$ black contours highlight the shape of each cloud.}
    \label{fig:fig1}
\end{figure*}

The selection of our four IRDCs, namely G24.94-00.15 (cloud O), G23.46-00.53 (cloud V), G24.49-0.70 (cloud X) and G25.16-0.28 (cloud Y) has been made to probe different regimes of mass surface density, $\Sigma$, and dust temperature, $T_{\rm{dust}}$. In addition, these IRDCs show relatively simple kinematics (Cosentino et al. in prep.). For the four clouds, $\Sigma$ and $T_{\rm{dust}}$ images were obtained from multi-wavelength {\it Herschel} images using the method described in \cite{lim2016}. Briefly, the images were first regridded to match the poorest angular resolution achieved by Herschel (36$^{\prime\prime}$). Next, the Galactic Gaussian method of background subtraction was used \citep[see][]{lim2016}. After this, the multi-wavelength emission, regridded to an 18$^{\prime\prime}$ pixel scale \citep[hires method of][]{lim2016} was fitted using a grey-body function and $\Sigma$ and $T_{\rm{dust}}$ were estimated. 
 The $\Sigma$ and $T_{\rm{dust}}$ Herschel-derived images of the four clouds are shown in Figure~\ref{fig:fig1}. To the best of our knowledge, no previous studies have been dedicated to these clouds.

\section{Observations and Data}

In May 2021 we used the 30m single dish antenna at Instituto de Radioastronomia Millimetrica (IRAM 30m, Pico Veleta, Spain) to map the $J=2\rightarrow$1 rotational transition of $^{13}$CO ($\nu$= 220.38 GHz) toward 16 IRDCs, as an extension to the \cite{butlerTan2012} A-J sample. Observations were performed in On-The-Fly observing mode with angular separation in the direction perpendicular to the scanning direction of 6$^{\prime\prime}$. For each cloud, the map central coordinates, map size, and utilized off positions are listed in Table~\ref{tab:tab1}. For the observations, we used the Fast Fourier Transform Spectrometer (FTS) set to provide a spectral resolution of 50 kHz, corresponding to a velocity resolution of 0.13 km s$^{-1}$ at the $^{13}$CO(2-1) rest frequency. Intensities were measured in units of antenna temperature, $T^*_A$, and converted into main-beam brightness temperature, $T_{\rm mb}= T^*_A (B_{\rm eff}/F_{\rm eff}$), using forward and beam efficiencies of $F_{\rm eff}$ and $B_{\rm eff}$ of 0.94 and 0.61, respectively. The final data cubes were created using the CLASS software within the GILDAS$\footnote{See http://www.iram.fr/IRAMFR/GILDAS.}$ package and have a spatial resolution of 11$^{\prime\prime}$ and a pixel size of 5.5$^{\prime\prime}\times$5.5$^{\prime\prime}$. The achieved rms per channel per pixel are reported in Table~\ref{tab:tab1}.

\begin{table*}
    \centering
    \begin{tabular}{ccccccccccccc}
    \hline\hline
Cloud & $l$ & $b$ & Off pos.  & Map & rms$_{\rm{1-0}}$ &rms$_{\rm{2-1}}$ &$\Delta v$ & $A_{\mathrm{rms}}^{\rm{1-0}}$ & $A_{\mathrm{rms}}^{\rm{2-1}}$ & $d_{k}$ & $d_{\rm{GC}}$ & $\frac{^{12}{\rm C}}{^{13}{\rm C}}$ \\
               & ($^{\circ}$) & ($^{\circ}$) & ($^{\rm{\prime\prime}}$, $^{\prime\prime}$) &($^{\prime} \times ^{\prime}$) & (K) & (K) & (km s$^{-1}$) & (K km s$^{-1}$) & (K km s$^{-1}$) & (kpc) & (kpc)\\
        \hline
        O &24.93 &-0.15 & -757,87  &5$\times$4     &0.6 &0.20 &40-60 &2.5 &0.6 &3.0 &5.5 &53\\
        V &23.48 &-0.54 & 735,-9   &3.5$\times$2.5 &0.6 &0.20 &58-68 &1.3 &0.5 &3.8 &4.9 &49\\
        X &24.48 &-0.70 & 1050,41  &3.5$\times$3   &0.6 &0.14 &42-54 &1.7 &0.4 &3.2 &5.4 &52\\
        Y &25.16 &-0.29 & 2550,5   &4.5$\times$3.5 &0.6 &0.20 &58-68 &2.0 &0.6 &3.8 &4.9 &50\\
    \hline
    \end{tabular}
    \caption{For each of the four IRDCs we report: Cloud Name, Galactic coordinate, relative coordinate of off position, map size, rms per channel per pixel in the $^{13}$CO(1-0) and $^{13}$CO(2-1) maps in unit of main beam temperature, velocity range of the corresponding $^{13}$CO emission as identified in Section~\ref{results}, integrated noise toward these velocity ranges, kinematic and Galactocentric distances, and assumed isotopic ratio of $^{12}$C/$^{13}$C.}
    \label{tab:tab1}
\end{table*}

The $^{13}$CO(2-1) IRAM30m data were complemented using the $^{13}$CO(1-0) cubes from the FUGIN (FOREST Unbiased Galactic plane Imaging survey with the Nobeyama 45-m telescope) survey \citep{umemoto2017}. This survey uses the 45m antenna at the Nobeyama Radio Observatory (Japan) to map the $^{12}$CO, $^{13}$CO and C$^{18}$O $J=1\rightarrow$0 transitions toward part of the first (10$^{\circ}\leq l \leq$ 50$^{\circ}$, $|b| \leq$ 1°) and third (198$^{\circ}\leq l \leq$ 236$^{\circ}$, $|b|\leq$ 1$^{\circ}$) quadrants of the Galaxy. Observations were performed in OTF mode with scanning speed 100$^{\prime\prime}$/sec. The publicly available cubes have $B_{\rm eff}$=0.43, angular resolution of 20$^{\prime\prime}$, pixel size of 8.5$^{\prime\prime}$, velocity resolution of 0.65 km s$^{-1}$ and rms of 0.65 K per channel per pixel. For a more detailed description of the survey we report to \cite{umemoto2017}.

\section{Results and Discussion}\label{results}
\subsection{Identification of $^{13}$CO emission velocity}

\begin{figure*}
    \centering
    \includegraphics[width=0.31\textwidth]{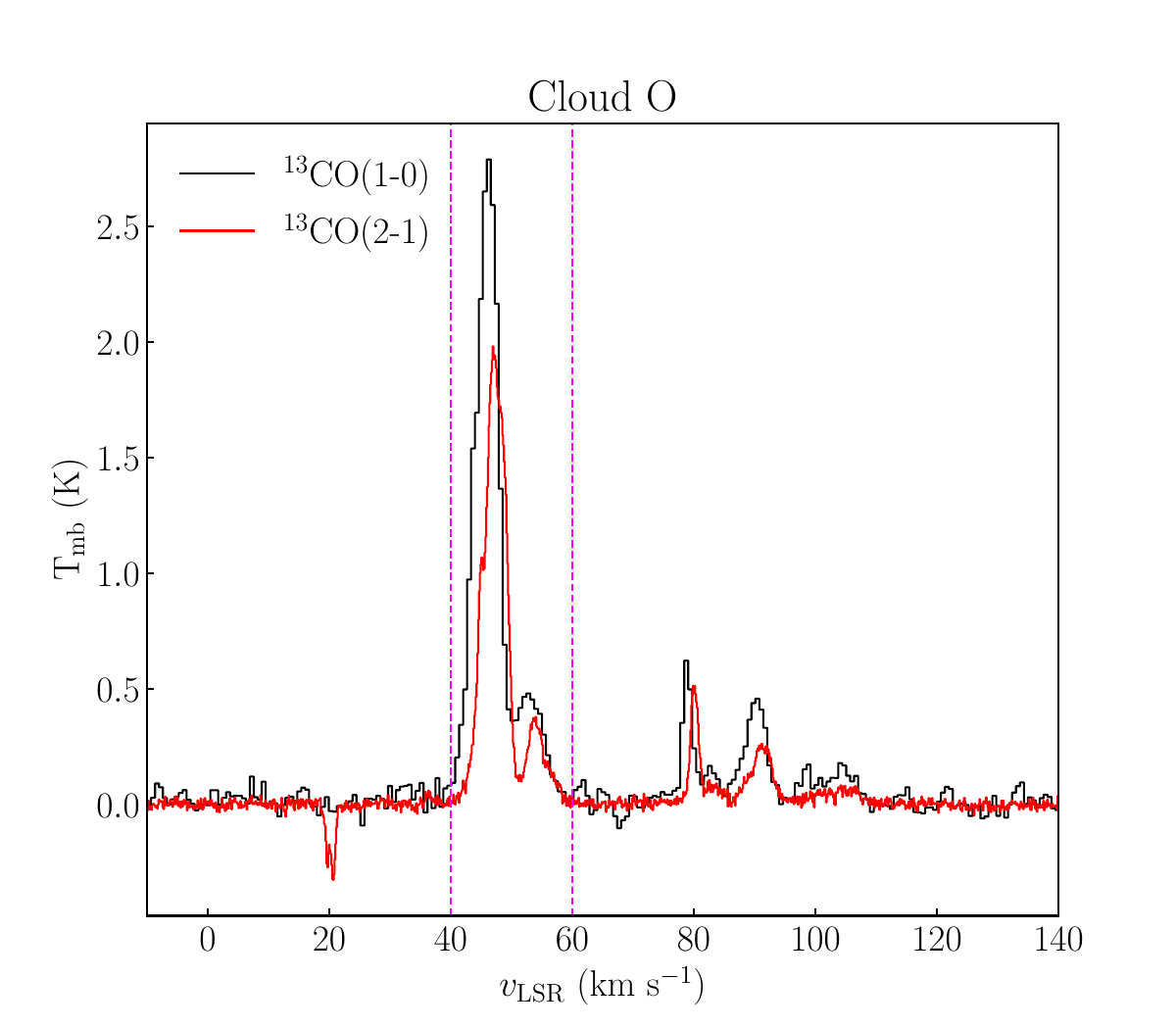}\includegraphics[width=0.35\textwidth]{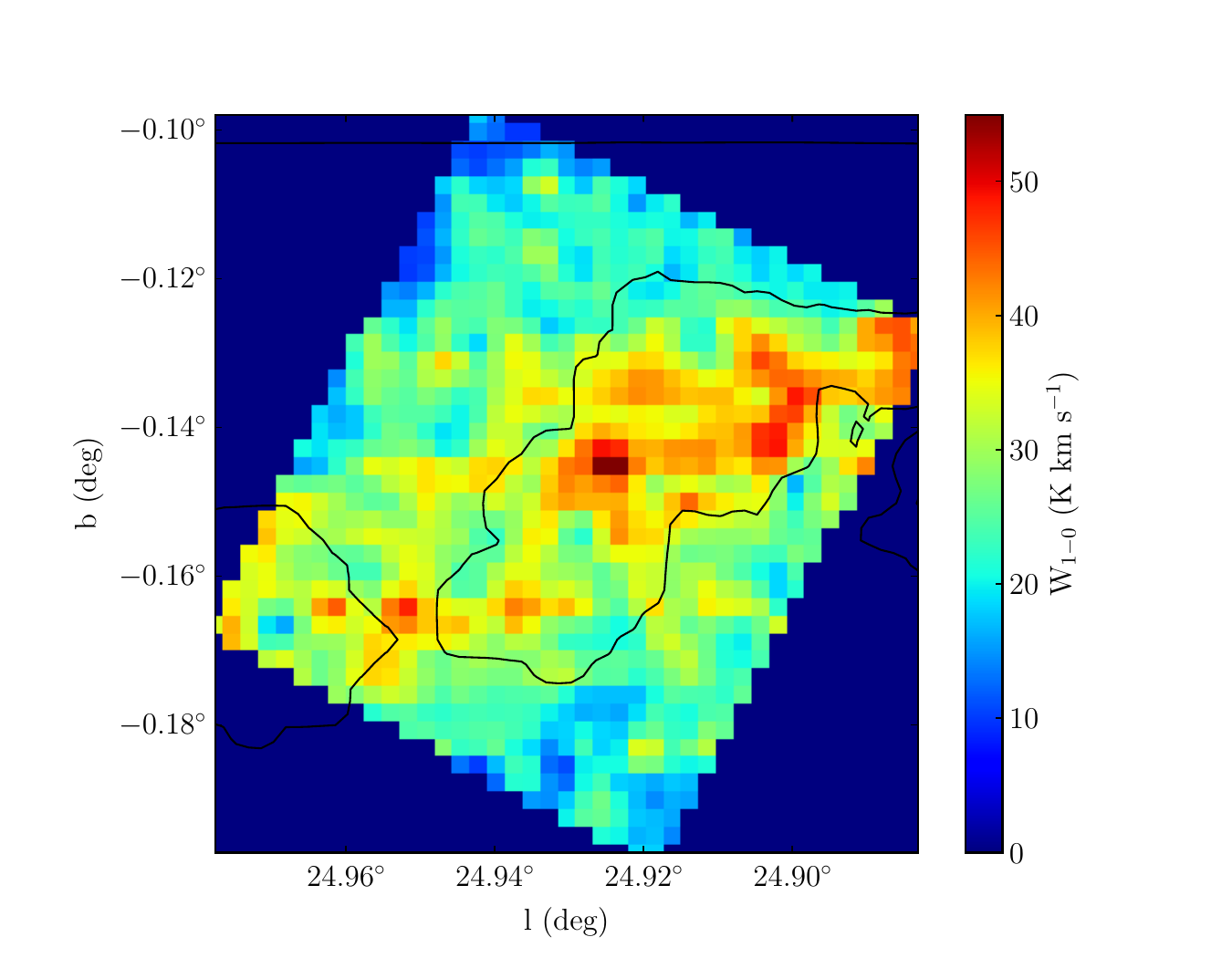}\includegraphics[width=0.35\textwidth]{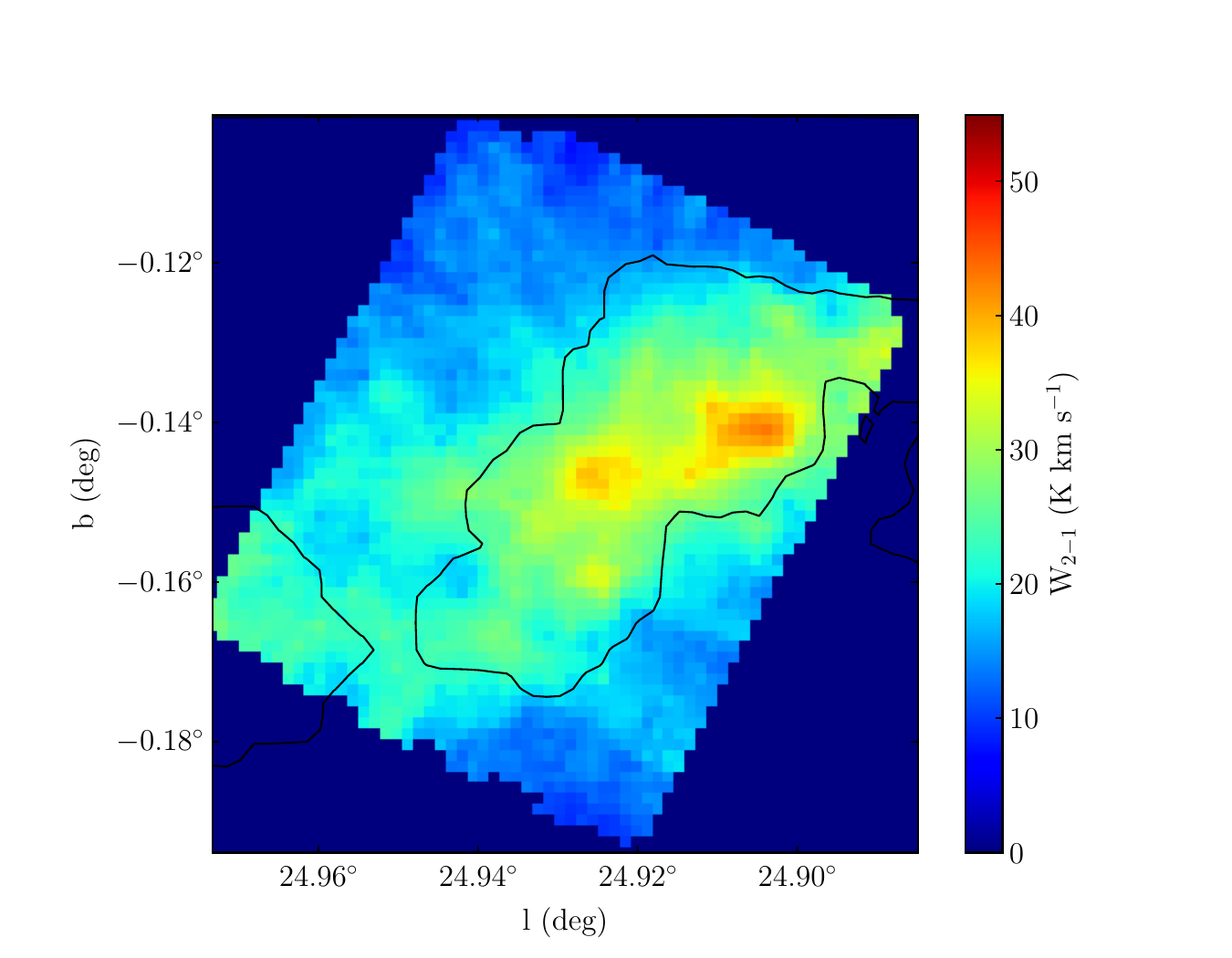}\\
    \includegraphics[width=0.31\textwidth]{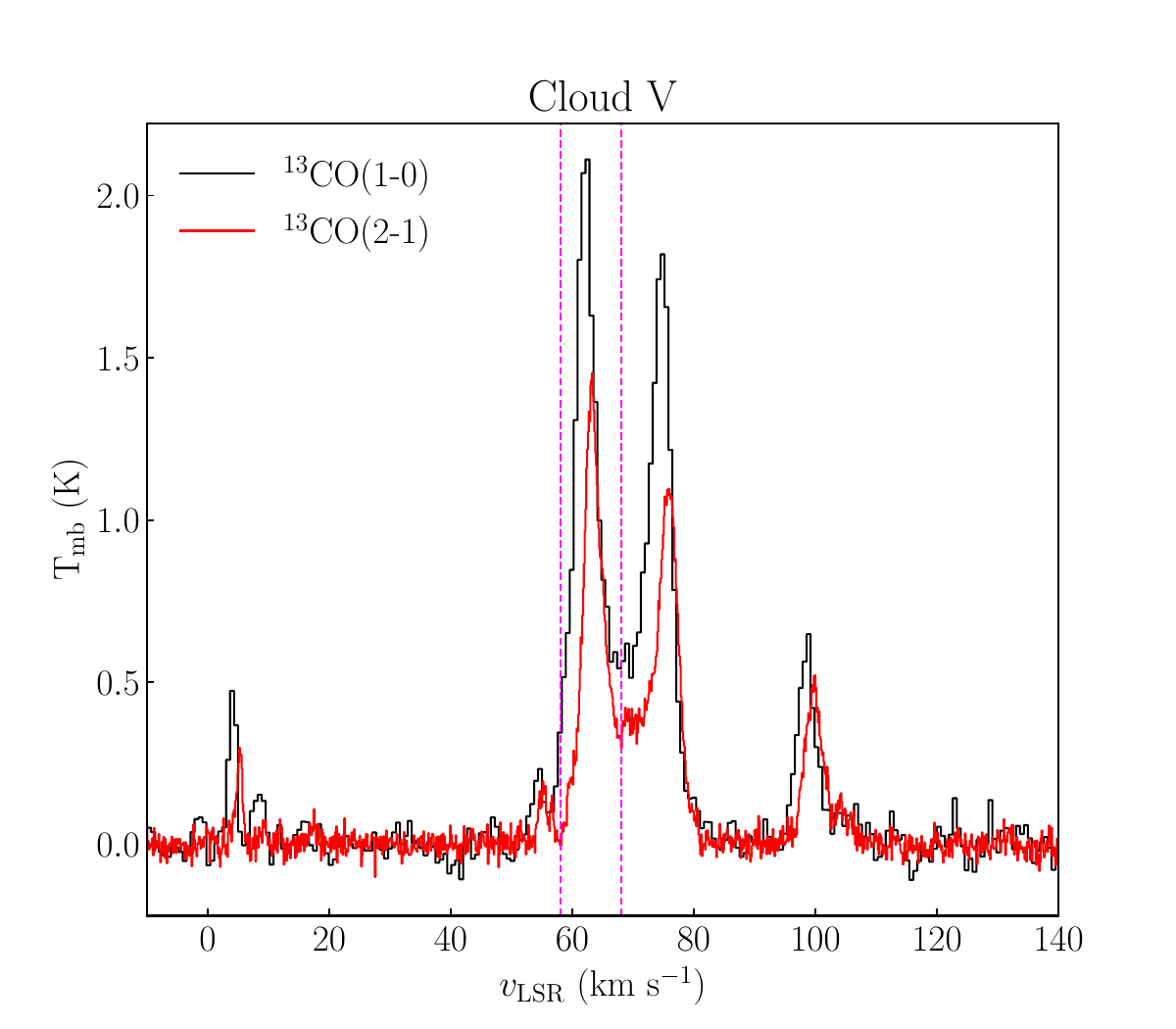}\includegraphics[width=0.35\textwidth]{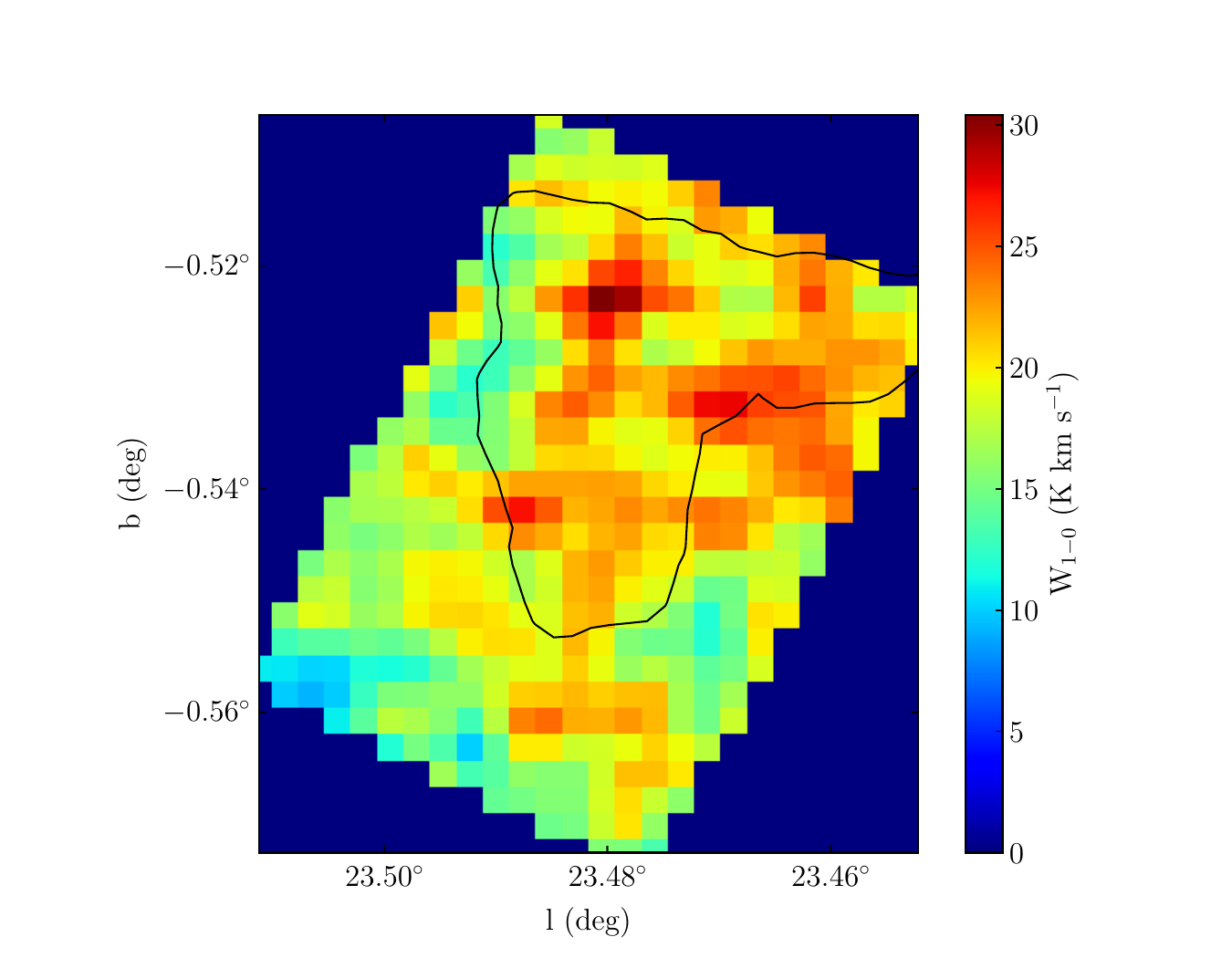}\includegraphics[width=0.35\textwidth]{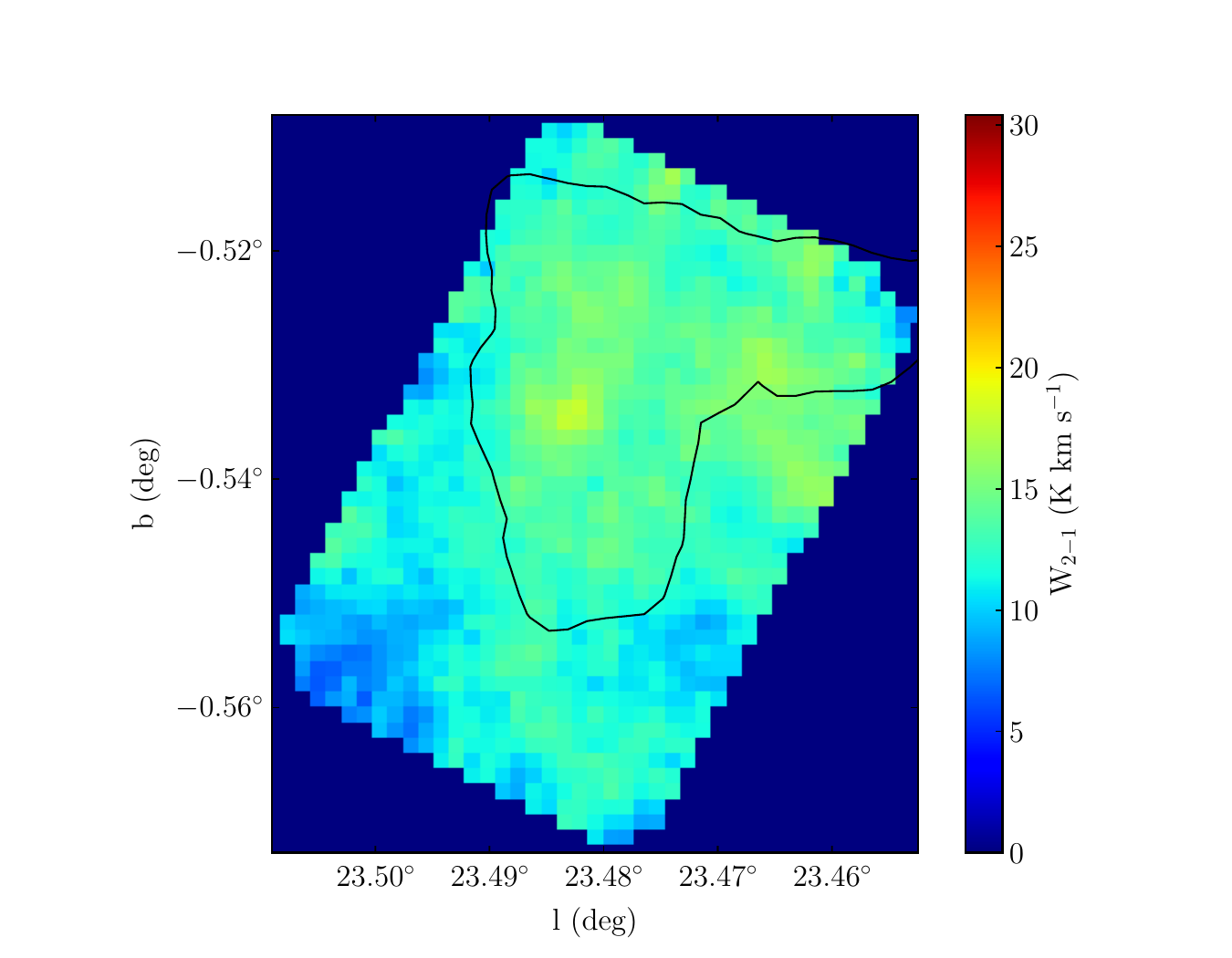}\\
    \includegraphics[width=0.31\textwidth]{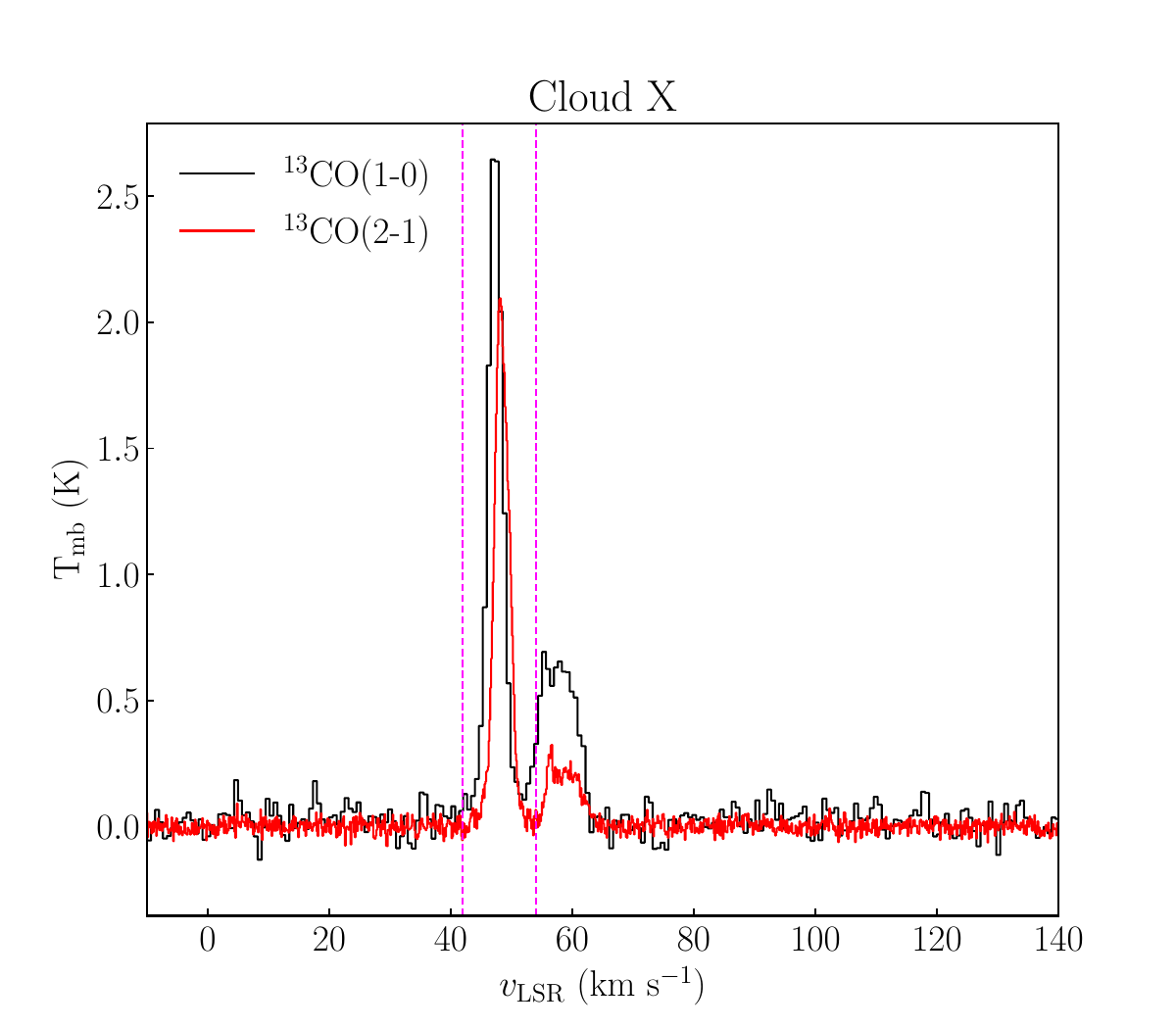}\includegraphics[width=0.35\textwidth]{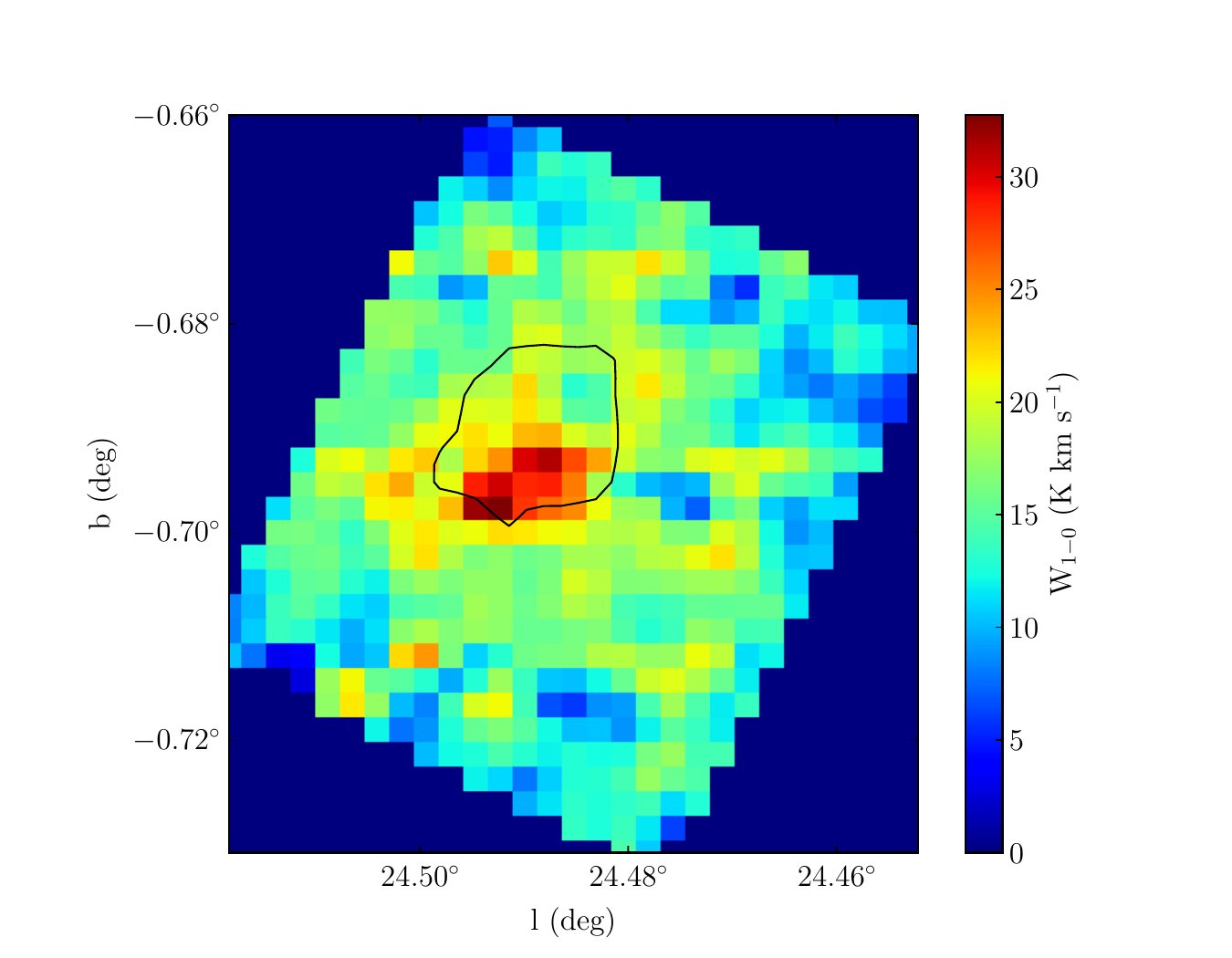}\includegraphics[width=0.35\textwidth]{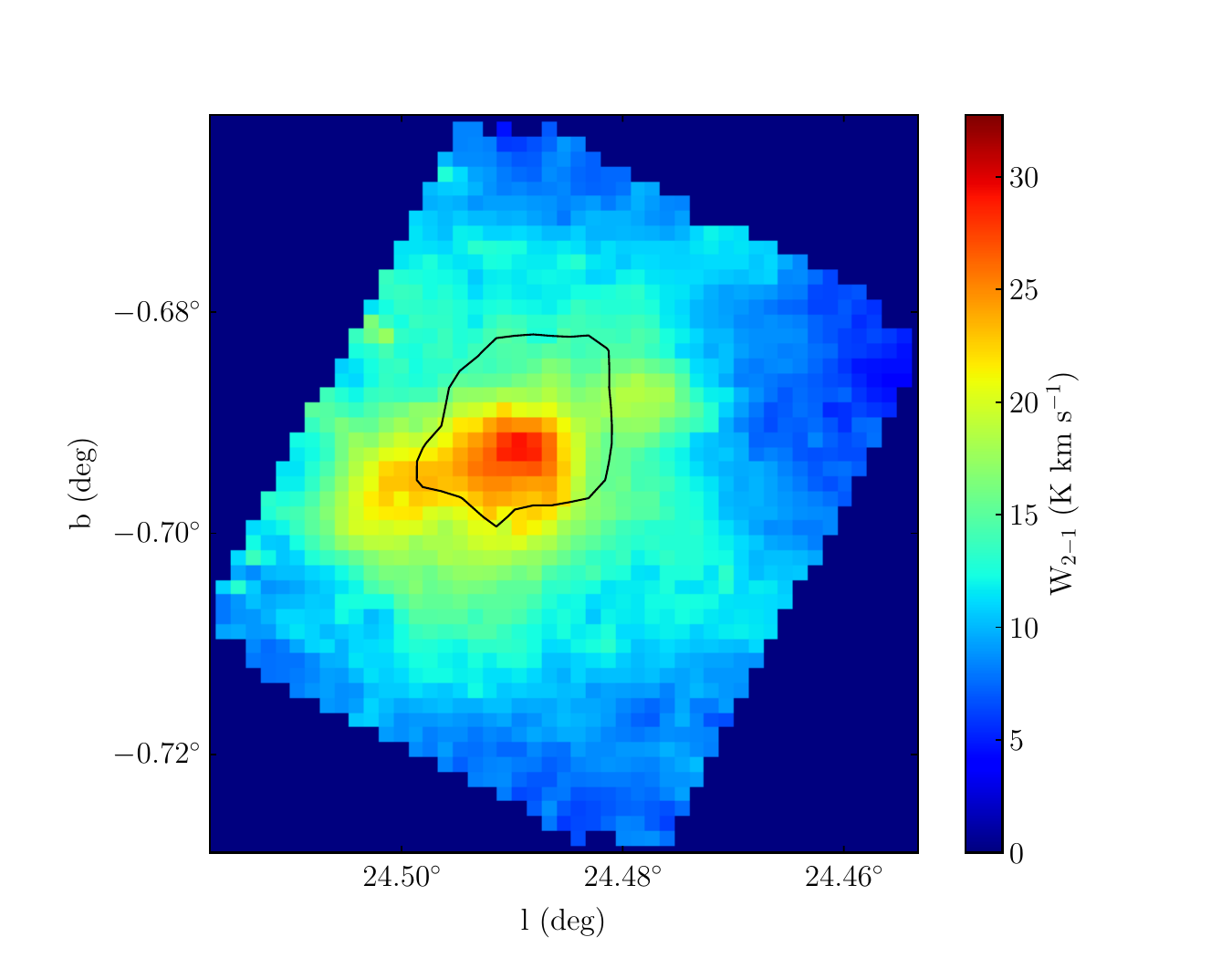}\\
    \includegraphics[width=0.31\textwidth]{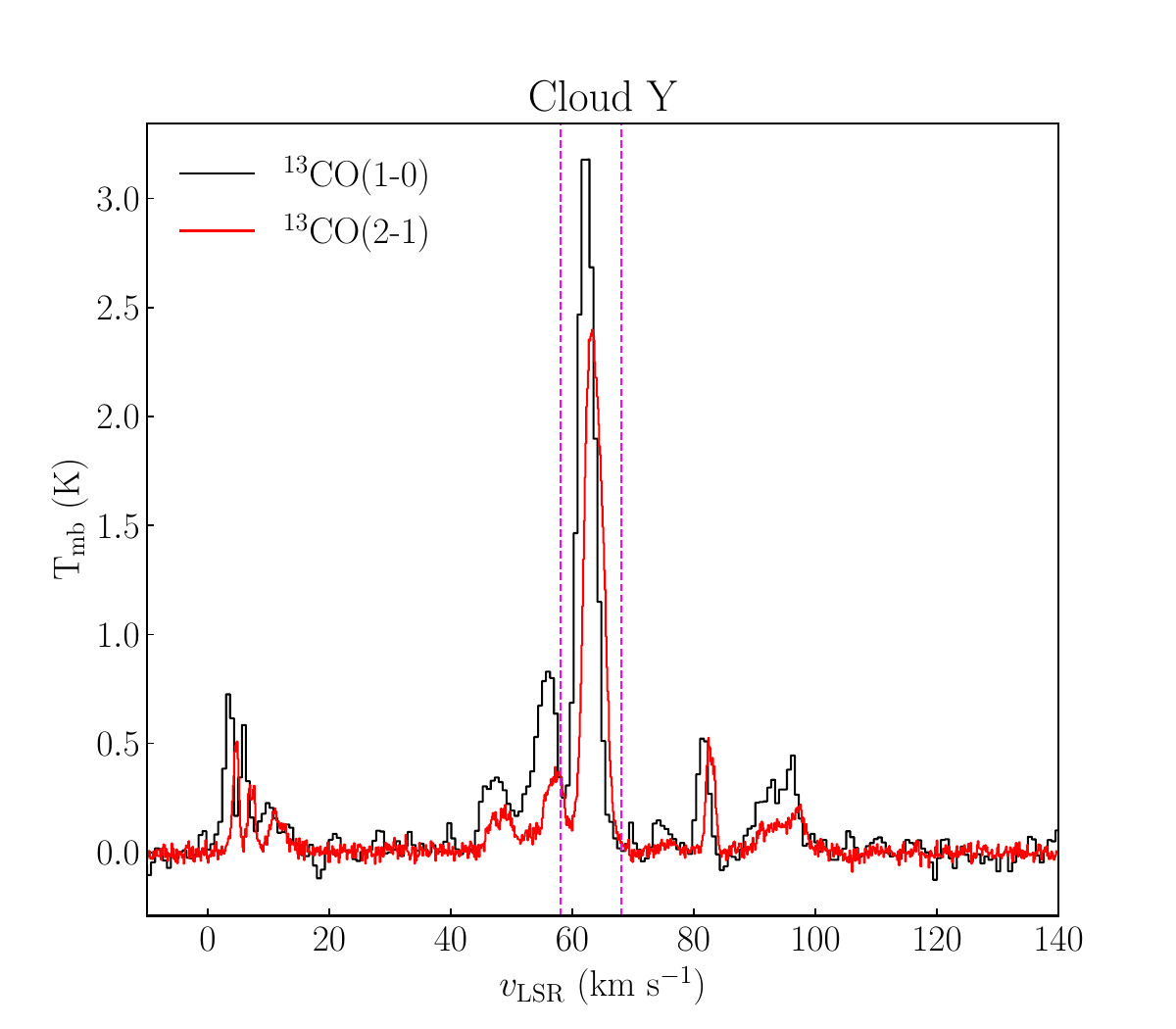}\includegraphics[width=0.35\textwidth]{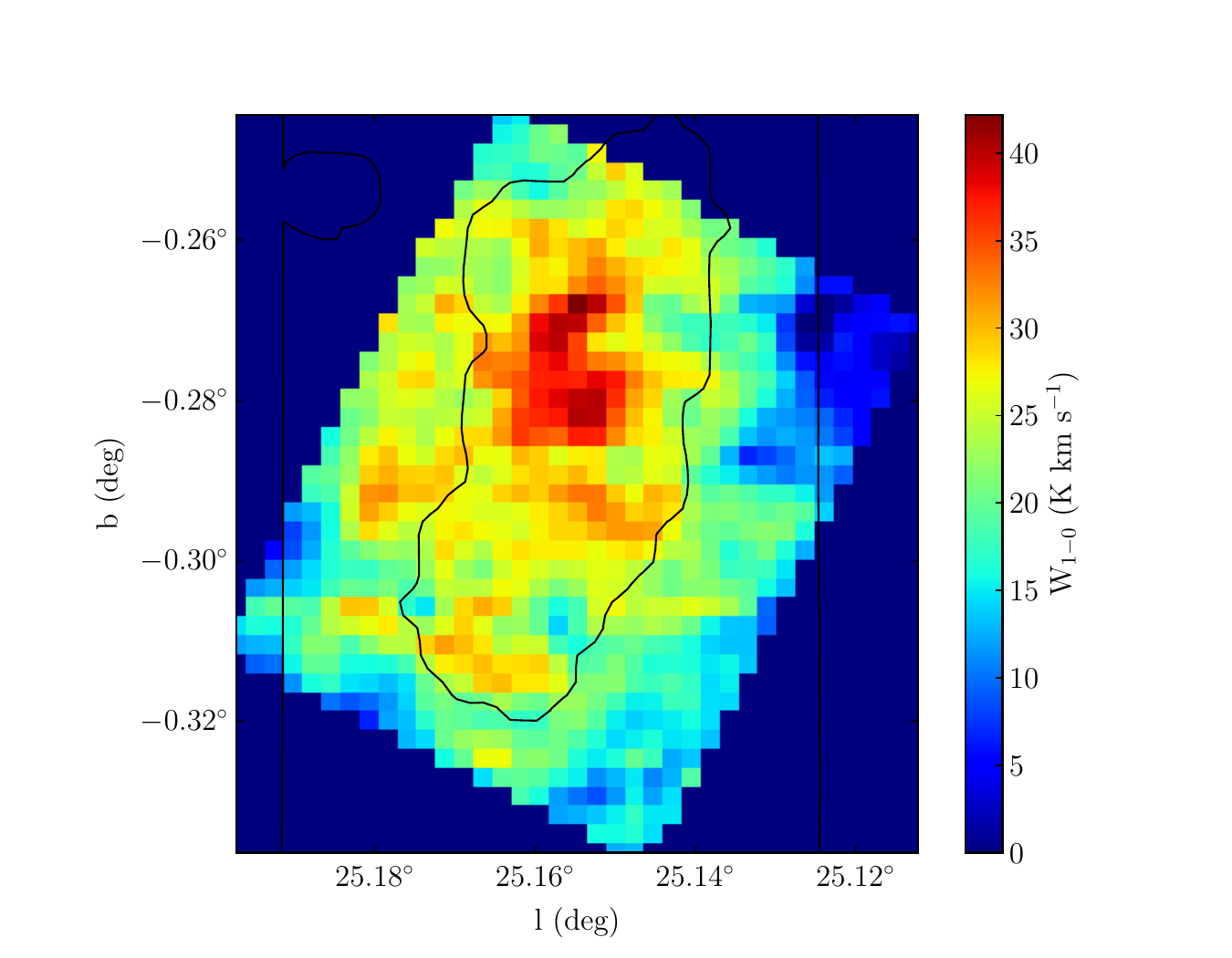}\includegraphics[width=0.35\textwidth]{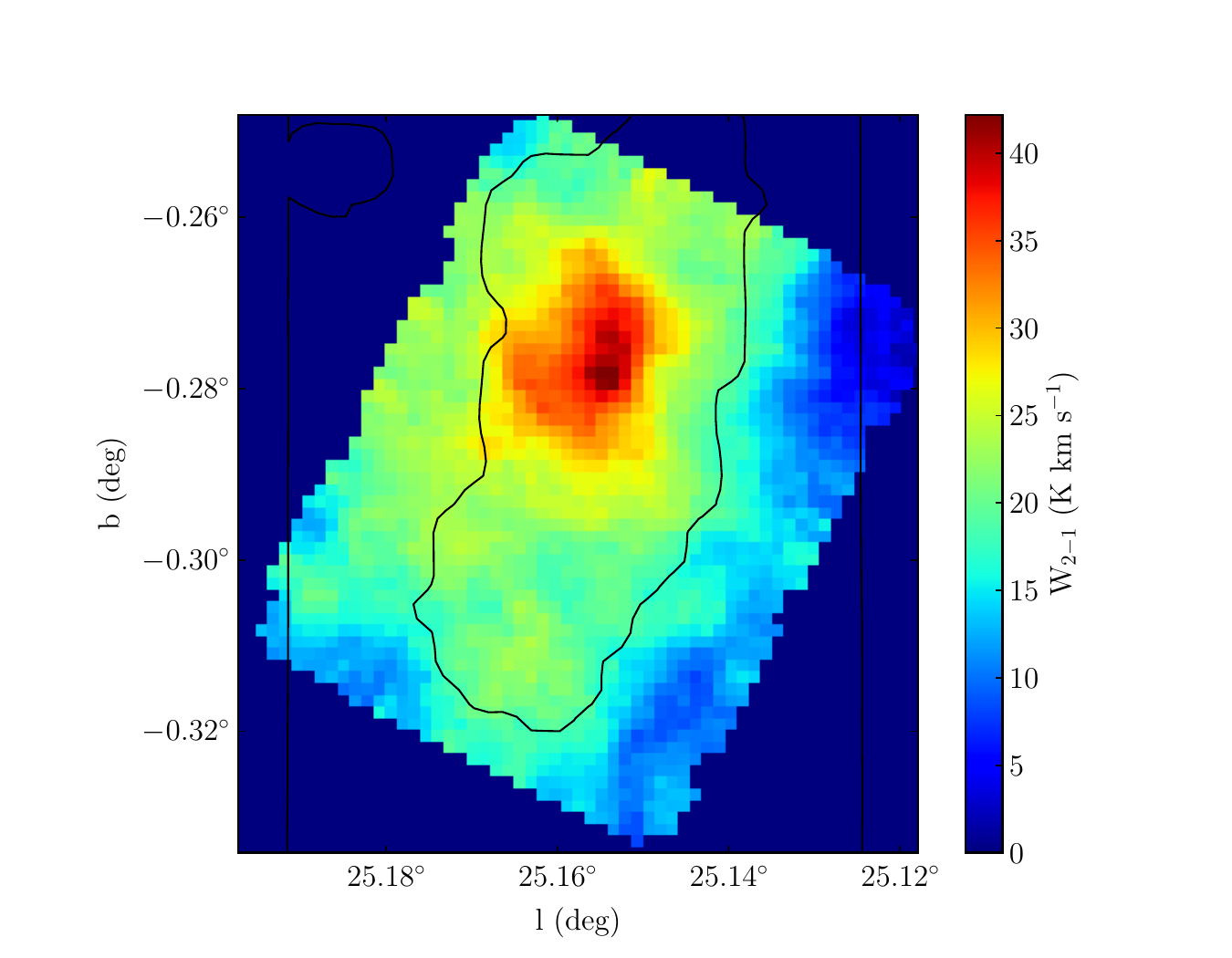}
    \caption{\textit{Left column:} $^{13}$CO(1-0) (black) and $^{13}$CO(2-1) (red) spectra averaged toward the full mapped region of the IRDCs. The velocity ranges considered for each cloud are indicated with magenta vertical dotted lines and reported in Table~\ref{tab:tab1}. \textit{Middle column:} Integrated intensity maps of the $^{13}$CO(1-0) obtained over the defined velocity range for each IRDC. \textit{Right column:} Integrated intensity maps of the $^{13}$CO(2-1) obtained over the defined velocity range for each IRDC.}
    \label{fig:fig2}
\end{figure*}

In Figure~\ref{fig:fig2} (left column), we show the $^{13}$CO(1-0) (black) and (2-1) (red) spectra averaged across the full map regions of each IRDC. Most spectra show multiple velocity components in both transitions. Indeed, $^{13}$CO is a very abundant species that probes relatively low-density molecular material in the ISM. We now want to discern which of these components are associated to the clouds and which are simply arising from emission along the line of sight. Hence, we have visually investigated the spatial correspondence and overlap between the cloud structure seen in the {\it Herschel}-derived maps (Figure~\ref{fig:fig1}) and each velocity component. From this correspondence, we have identified the velocity ranges of the $^{13}$CO emission associated with each cloud to be in the ranges 40-60~km~s$^{-1}$ for Cloud O, 58-68 km s$^{-1}$ for Cloud V, 42-54 km s$^{-1}$ for Cloud X, and 58-68 km s$^{-1}$ for Cloud Y.  The $^{13}$CO(1-0) and (2-1) integrated intensity maps obtained over these velocity ranges are also shown in Figure~\ref{fig:fig2}. These velocity ranges are reported in Table~\ref{tab:tab1}, along with the corresponding integrated noise for both transitions. We note that a more detailed analysis of the gas kinematics toward these sources will be presented in a forthcoming paper.

\subsection{$\mathrm{\Sigma_{13CO}}$ and $T_{\rm{ex}}$ maps}
\begin{figure*}
    \centering
    \includegraphics[width=0.32\textwidth,trim=1cm 1cm 1cm 1cm , clip=True]{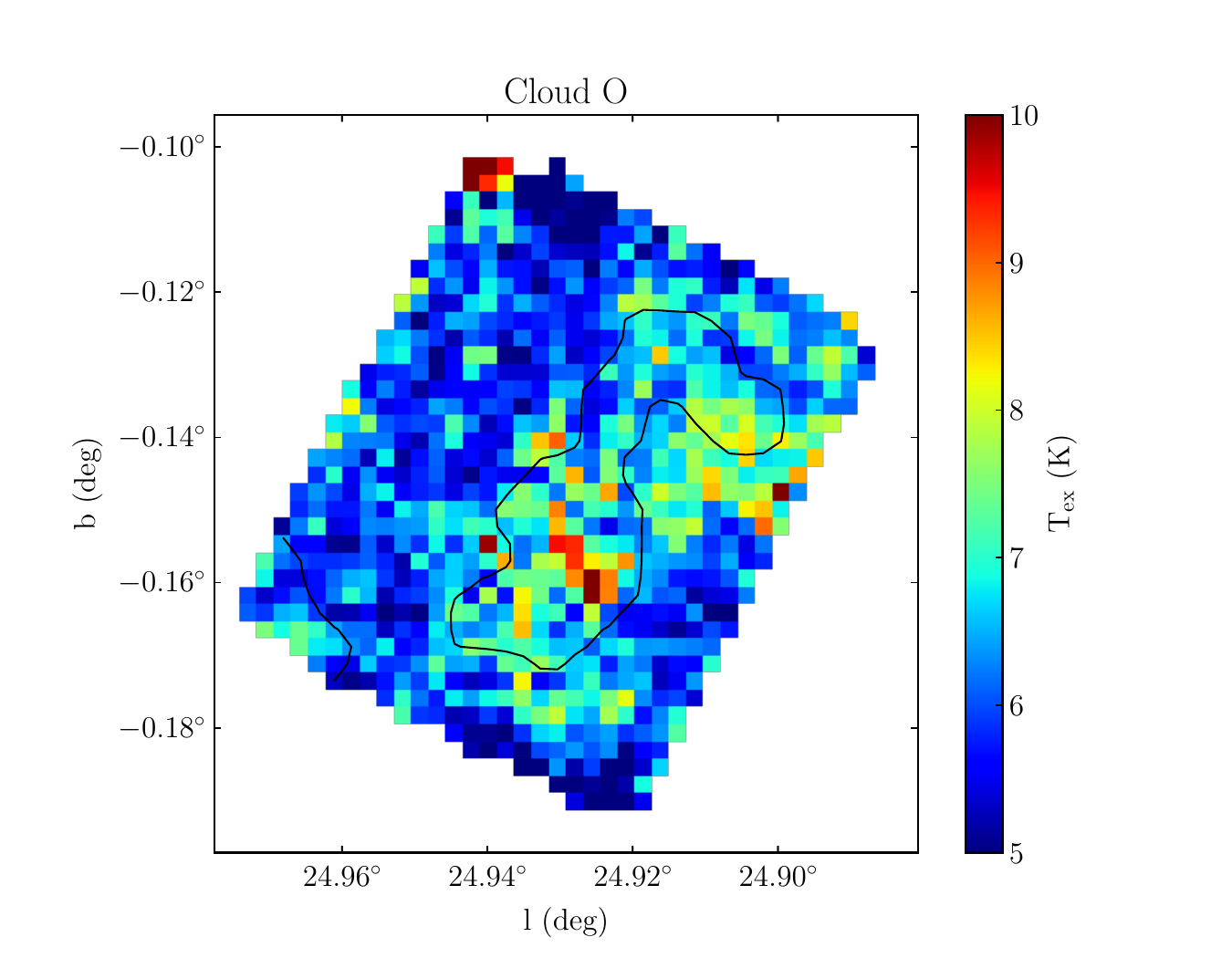}\includegraphics[width=0.32\textwidth,trim=1cm 1cm 1cm 1cm , clip=True]{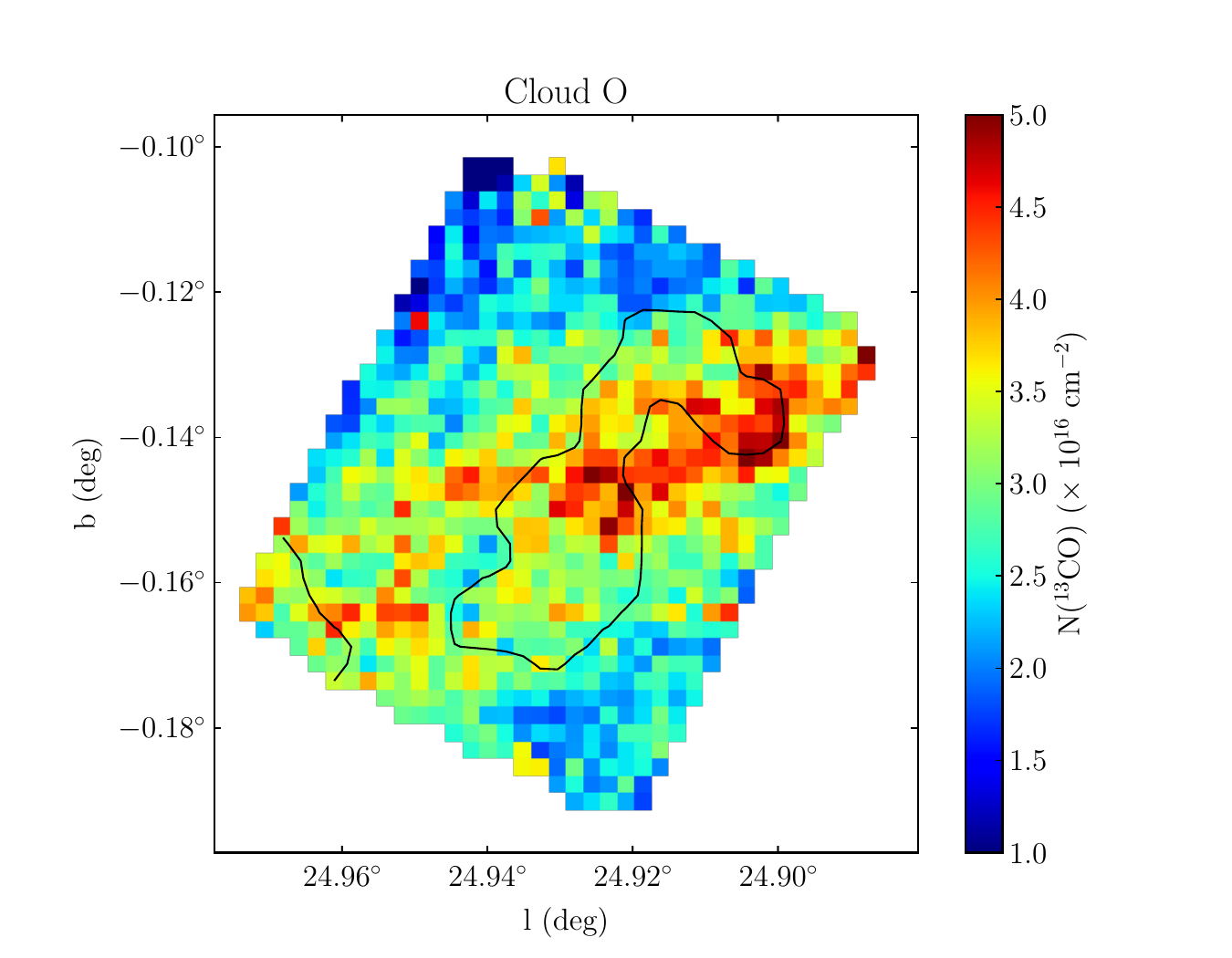}\includegraphics[width=0.32\textwidth,trim=1cm 1cm 1cm 1cm , clip=True]{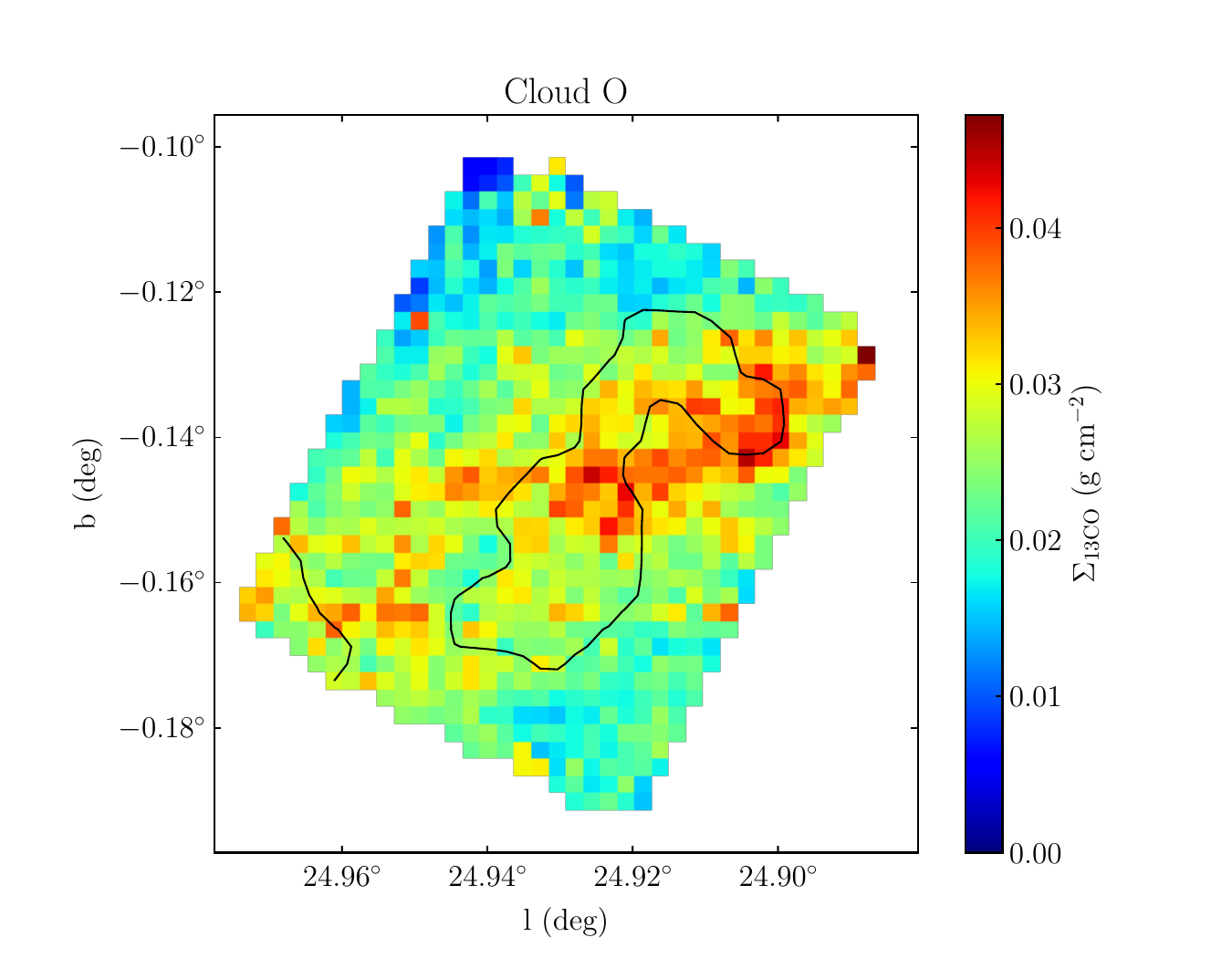}

    \includegraphics[width=0.32\textwidth,trim=1.7cm 1cm 1cm 1cm , clip=True]{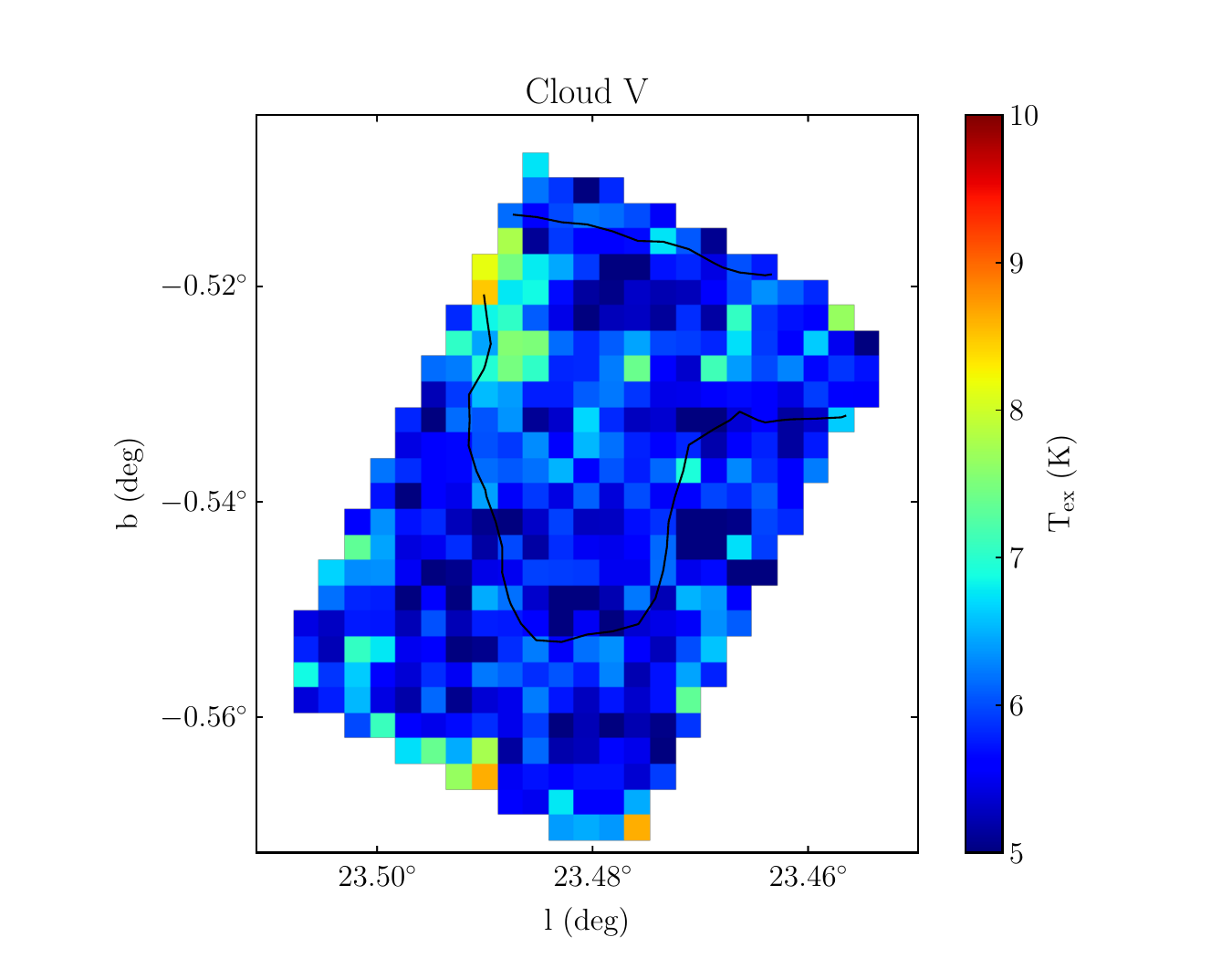}\includegraphics[width=0.32\textwidth,trim=1.7cm 1cm 1cm 1cm , clip=True]{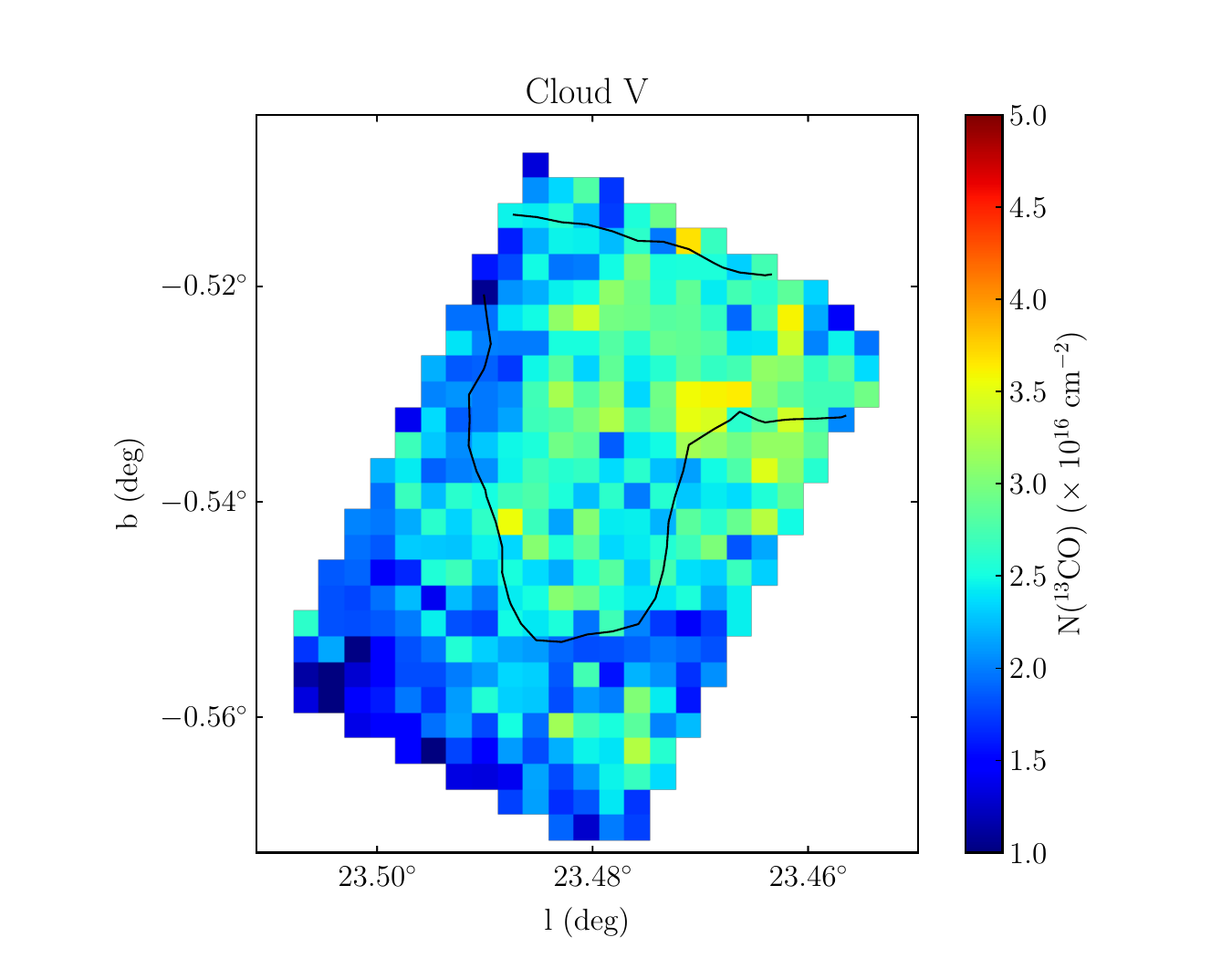}\includegraphics[width=0.32\textwidth,trim=1.7cm 1cm 1cm 1cm , clip=True]{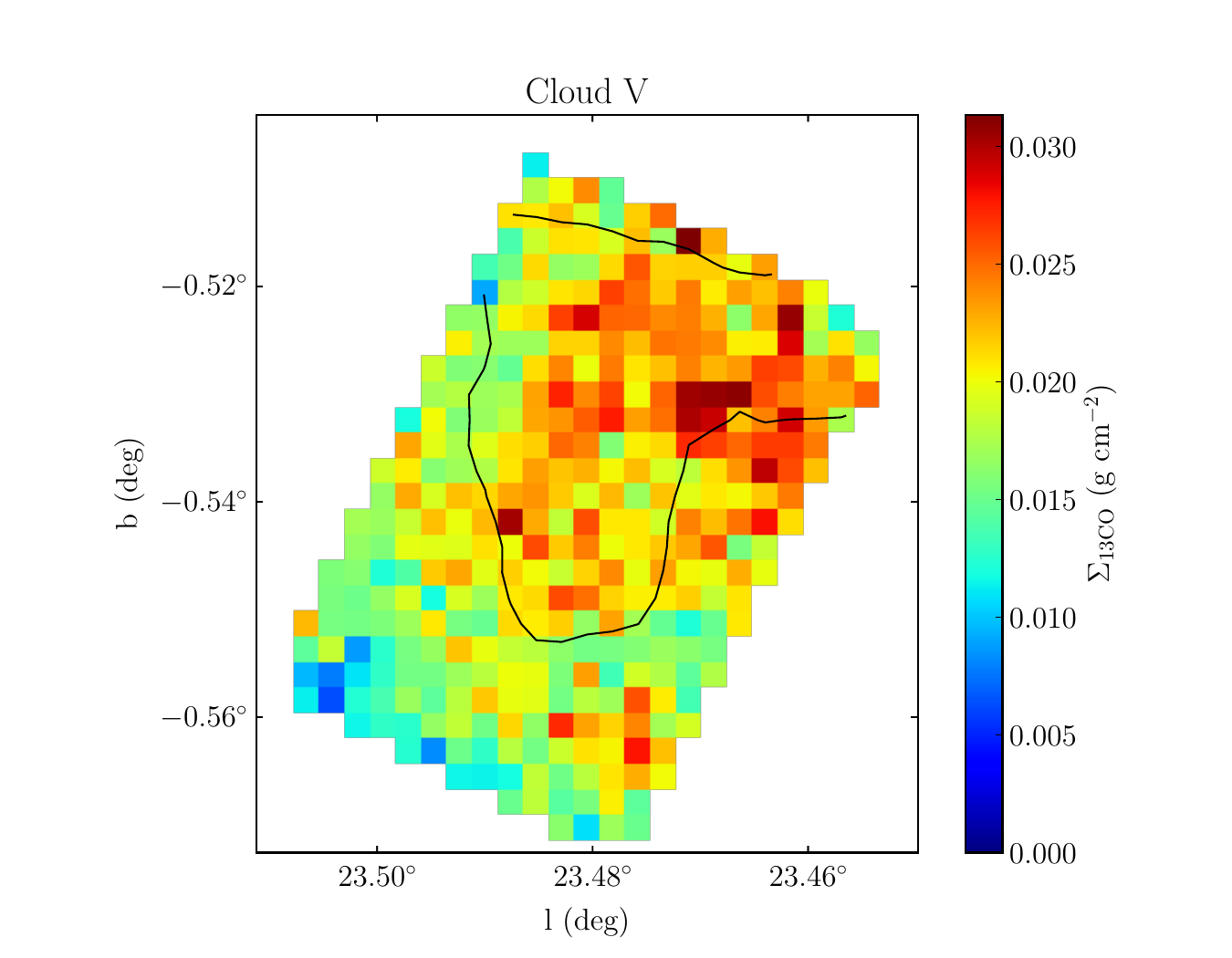}\\

    \includegraphics[width=0.32\textwidth,trim=1cm 1cm 1cm 1cm , clip=True]{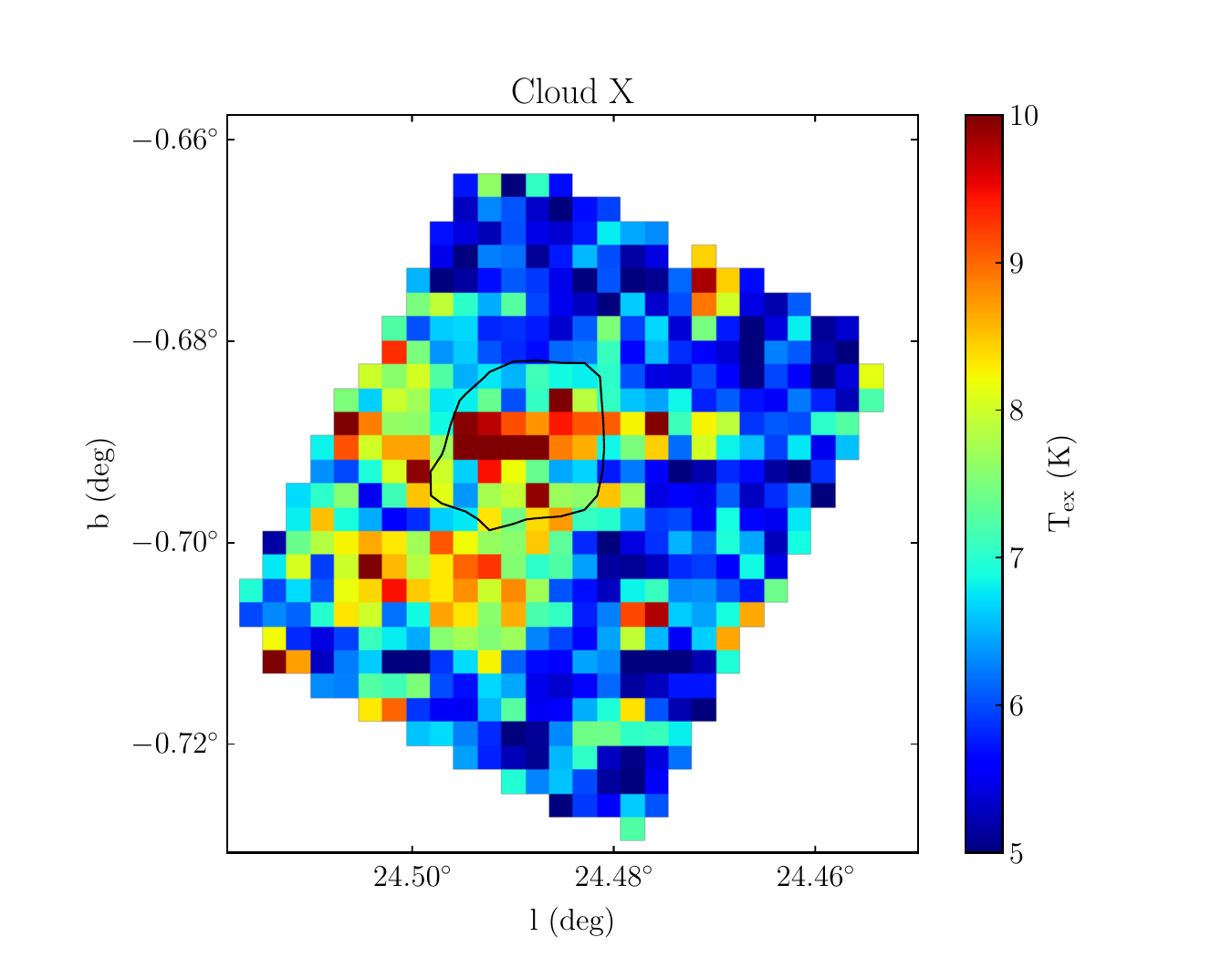}\includegraphics[width=0.32\textwidth,trim=1cm 1cm 1cm 1cm , clip=True]{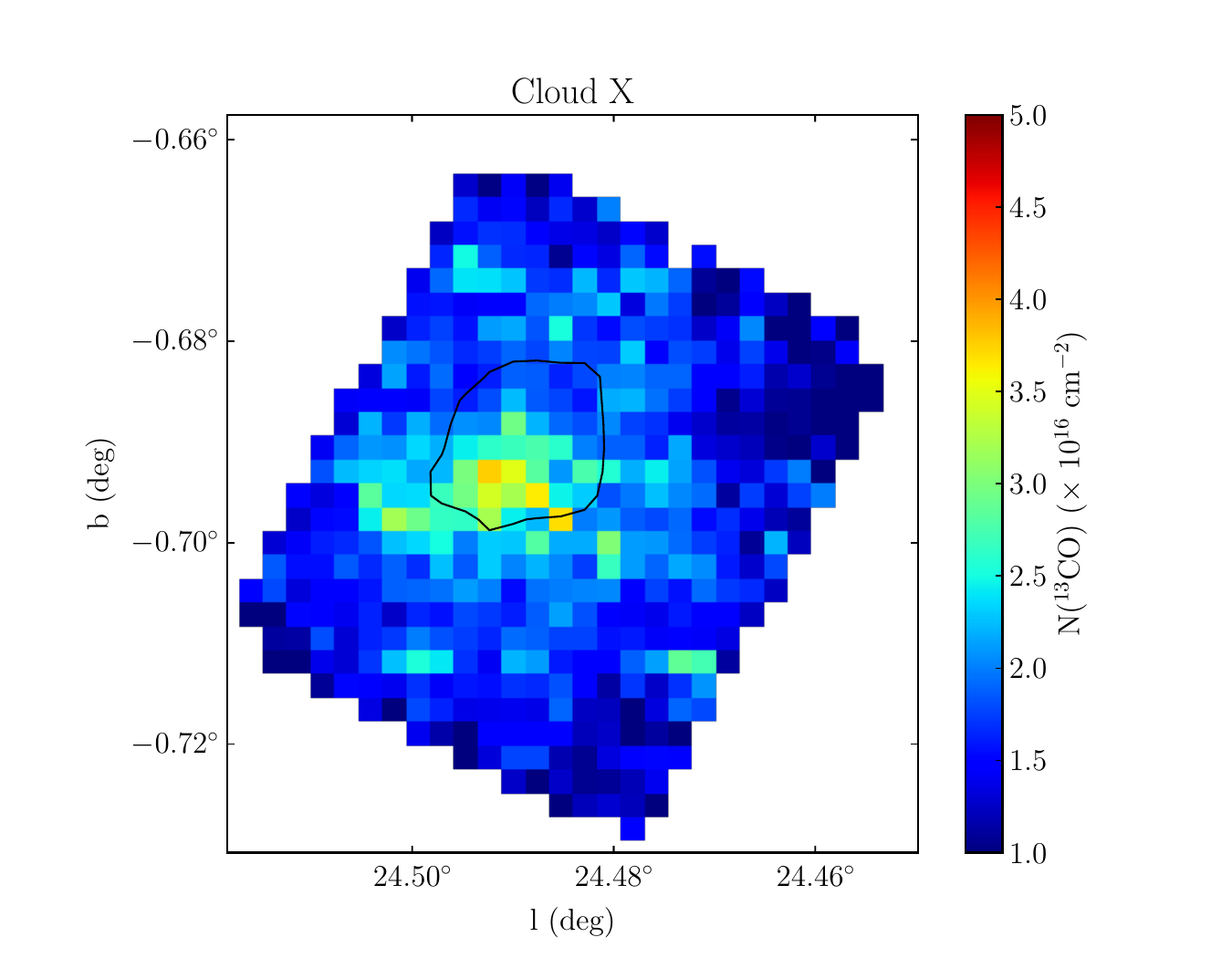}\includegraphics[width=0.32\textwidth,trim=1cm 1cm 1cm 1cm , clip=True]{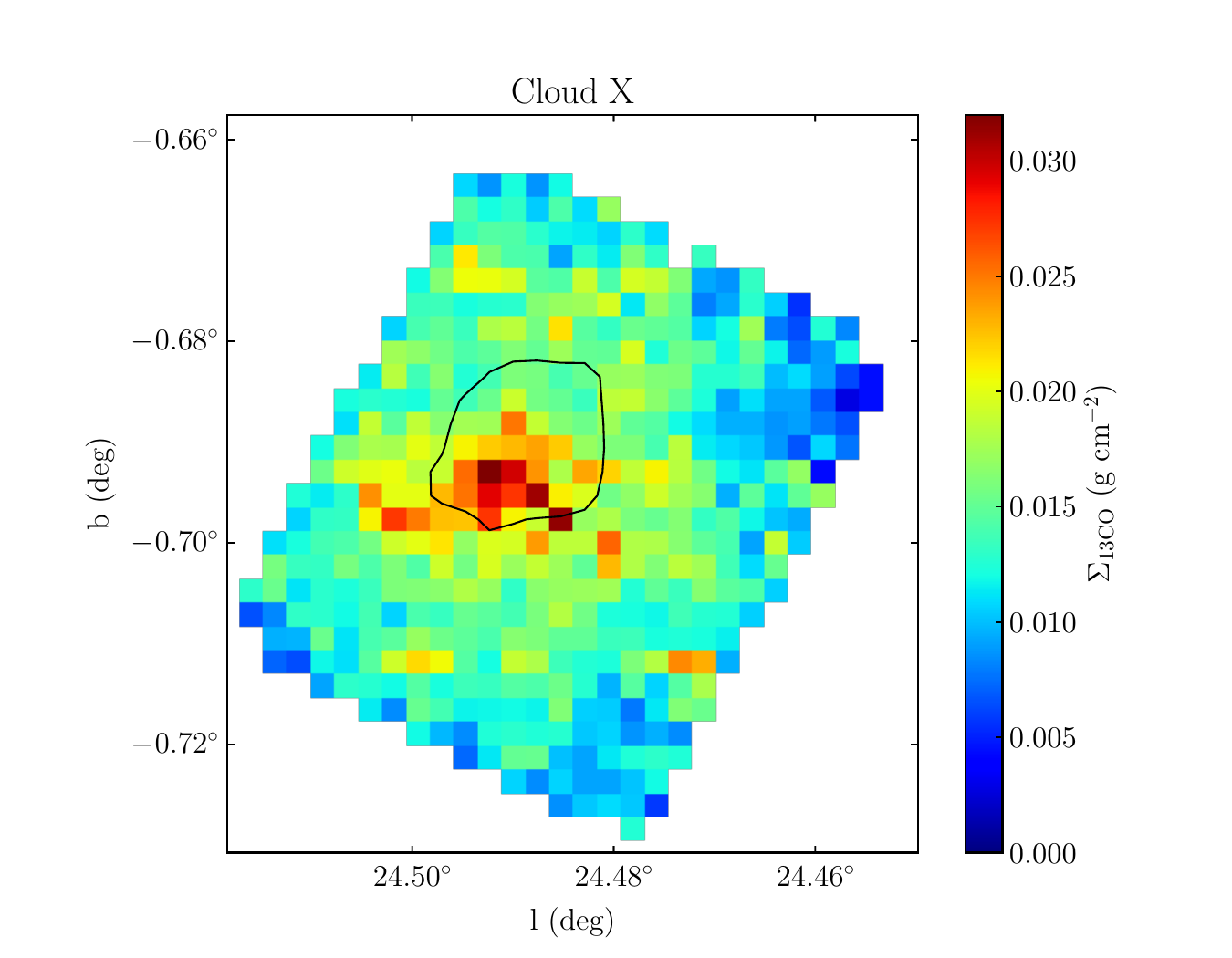}\\

    \includegraphics[width=0.32\textwidth,trim=1cm 0cm 1cm 1cm , clip=True]{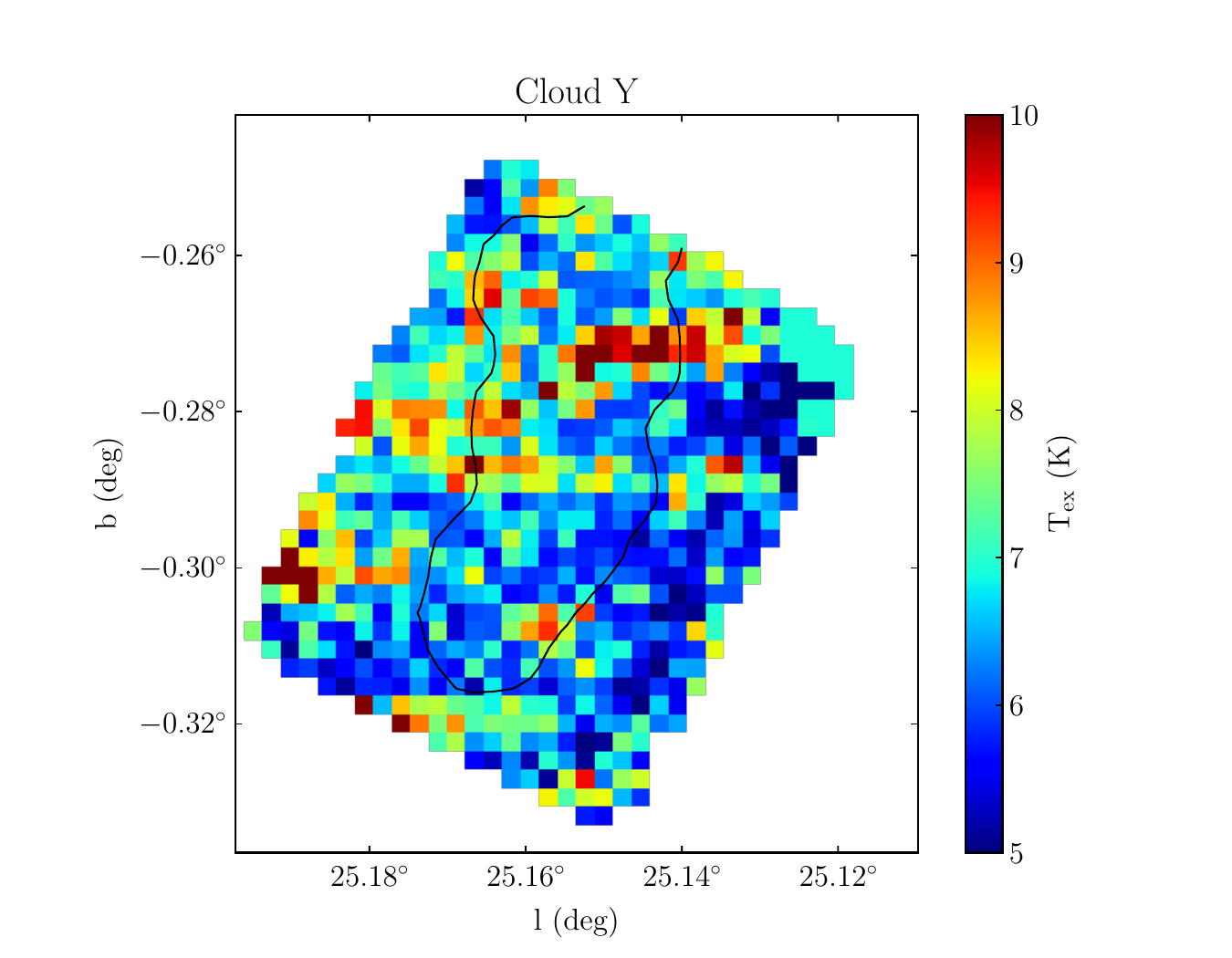}\includegraphics[width=0.32\textwidth,trim=1cm 0cm 1cm 1cm , clip=True]{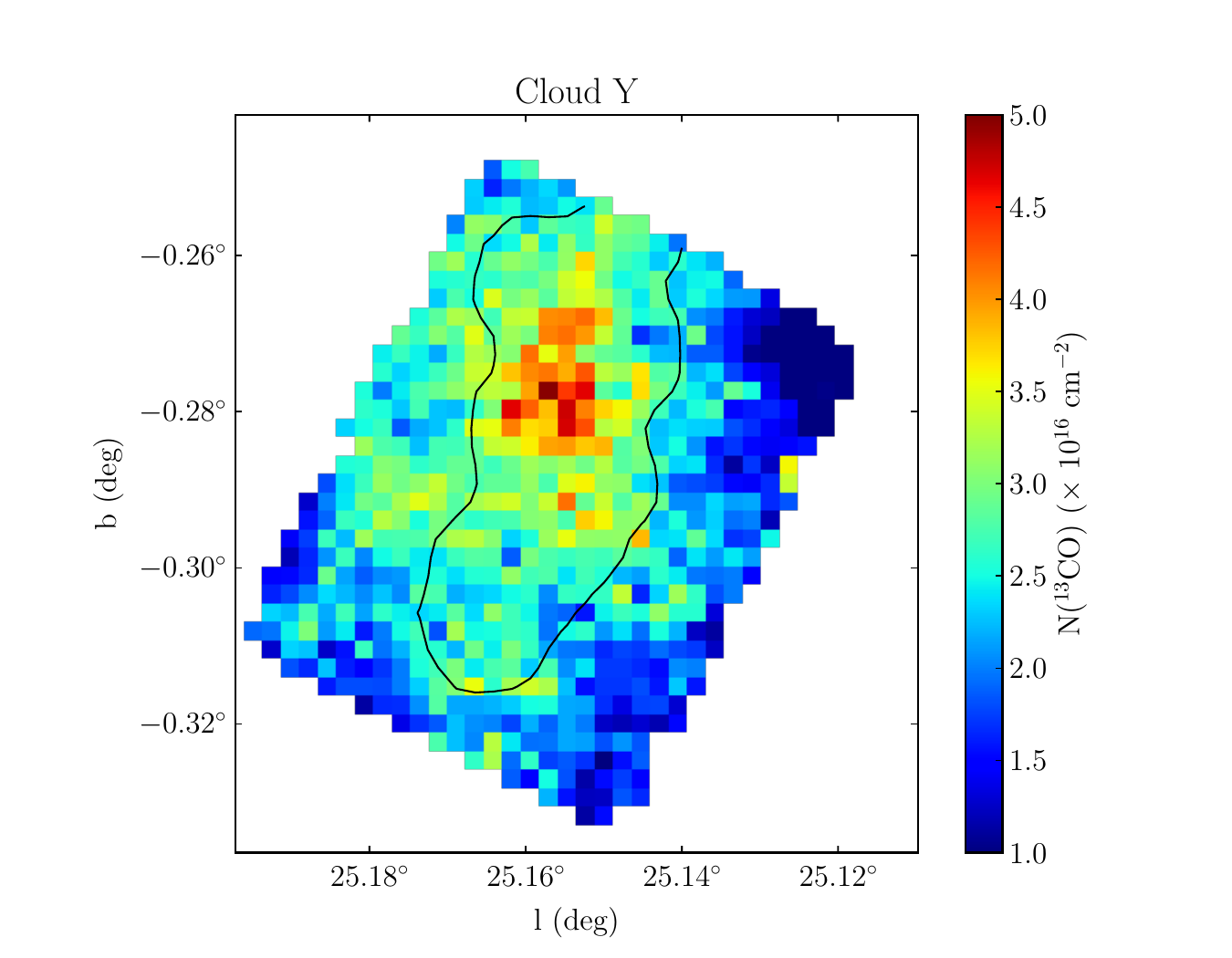}\includegraphics[width=0.32\textwidth,trim=1cm 0cm 1cm 1cm , clip=True]{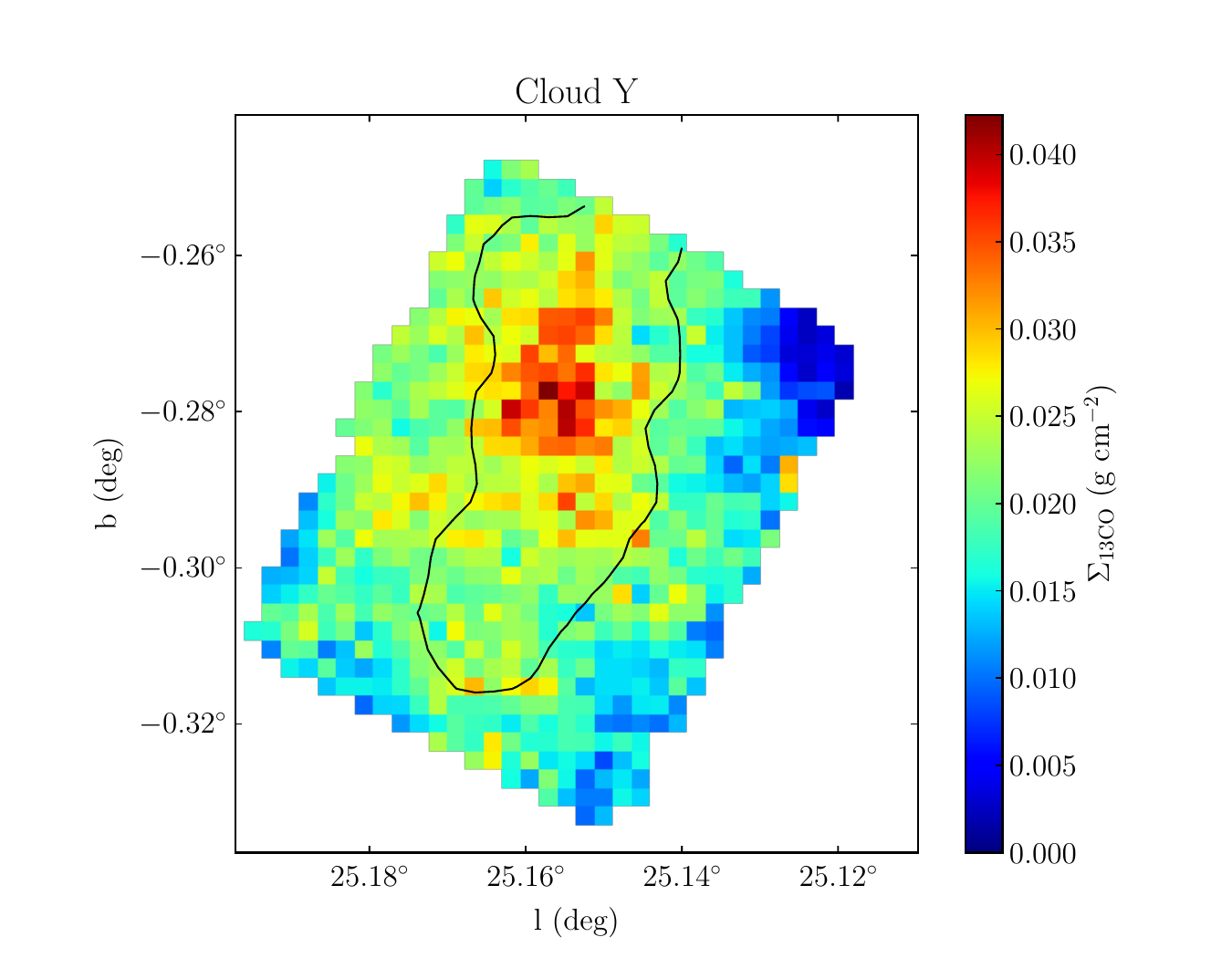}
 
    \caption{Maps of column density weighted excitation temperature ({\it left column}), $^{13}$CO column density ({\it middle column}), and $^{13}$CO-derived mass surface density ({\it right column}) obtained for the four IRDCs O, V, X, Y ({\it top to bottom rows}). In each panel, the black contour corresponds to the FIR-derived mass surface density of 0.1 g cm$^{-2}$.}
    \label{fig:fig3a}
\end{figure*}

\begin{figure*}
    \centering
    \includegraphics[width=0.35\textwidth,trim=1cm 1cm 1cm 1cm , clip=True]{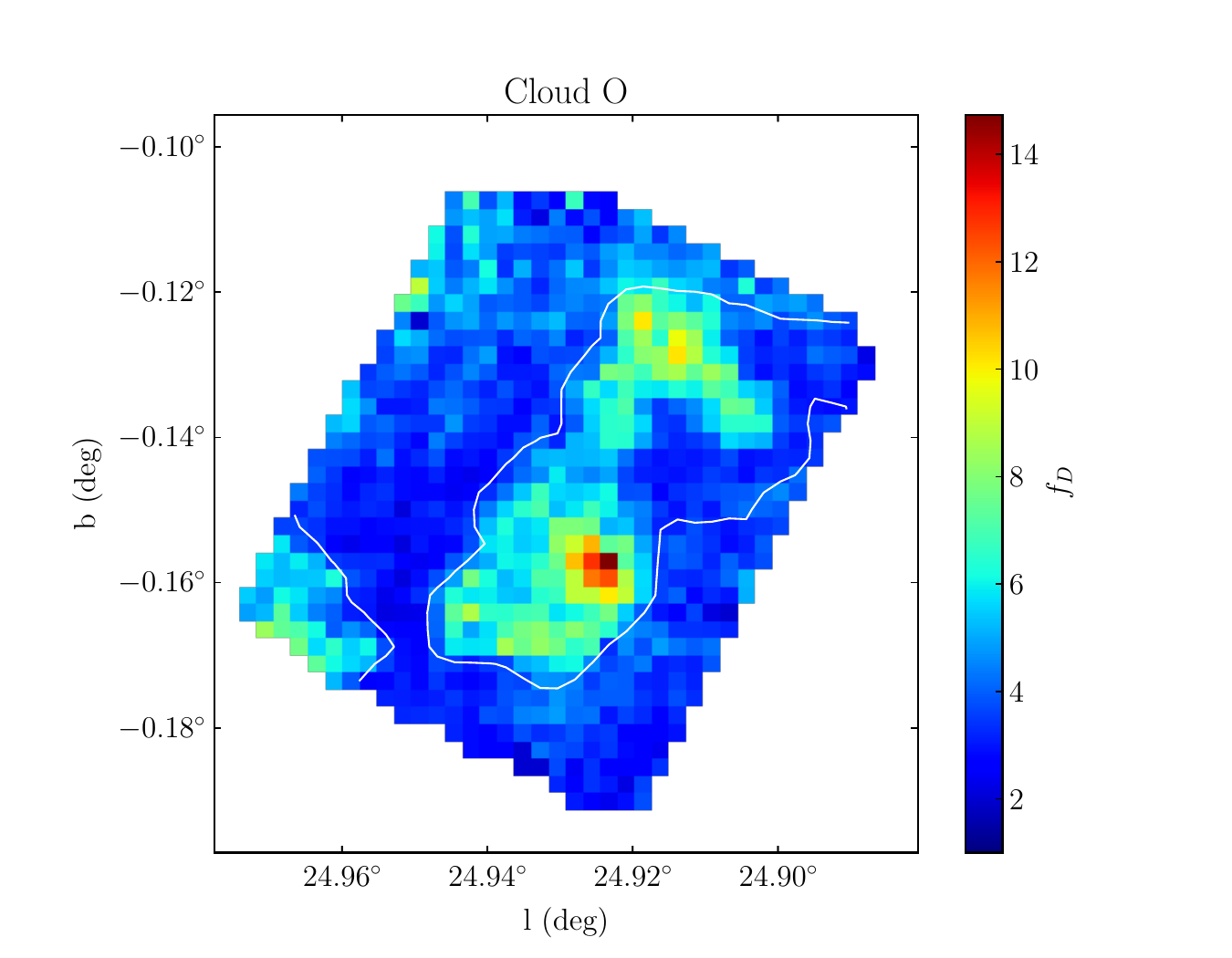}\includegraphics[width=0.35\textwidth,trim=1cm 1cm 1cm 1cm , clip=True]{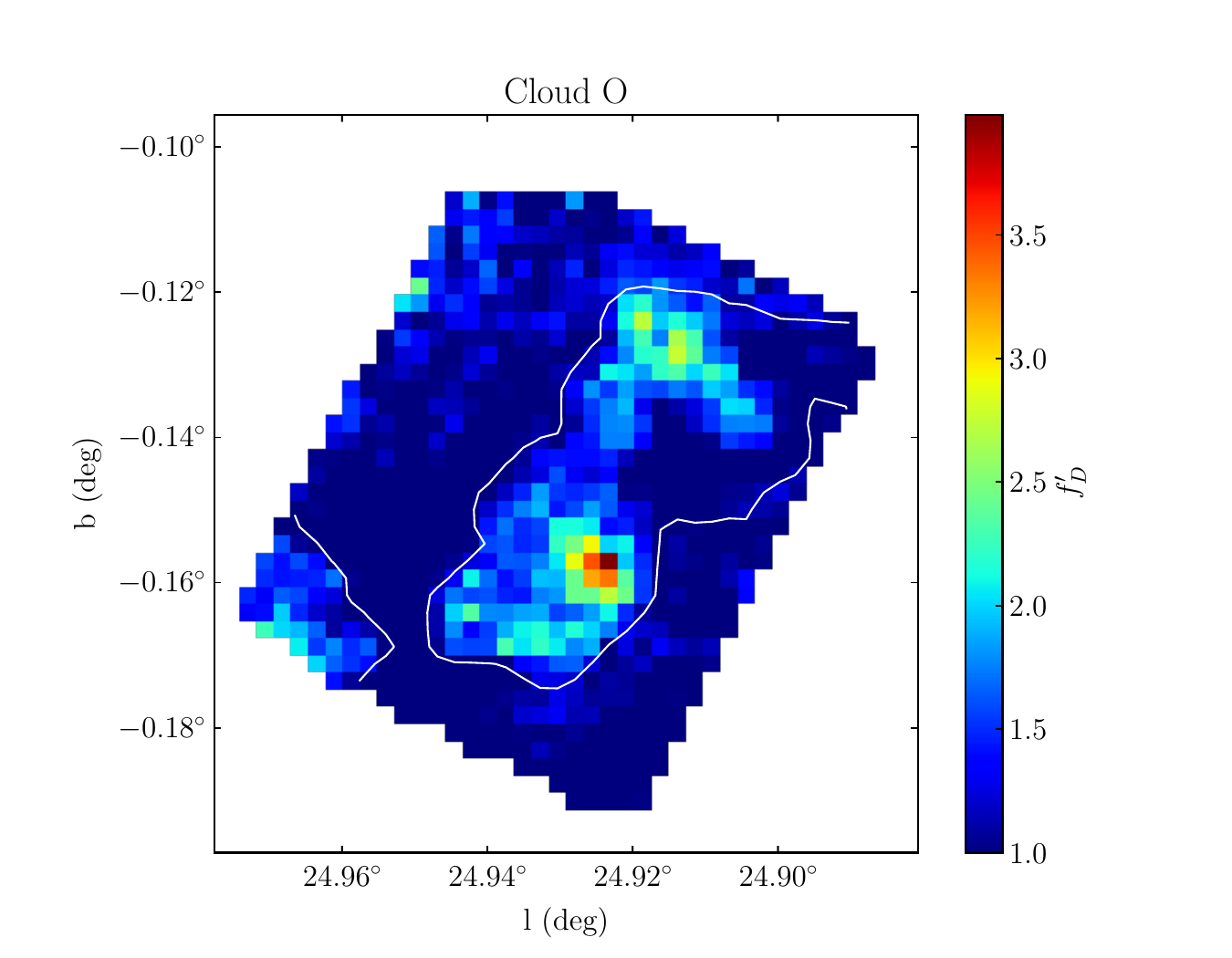}

    \includegraphics[width=0.35\textwidth,trim=1.7cm 1cm 1cm 1cm , clip=True]{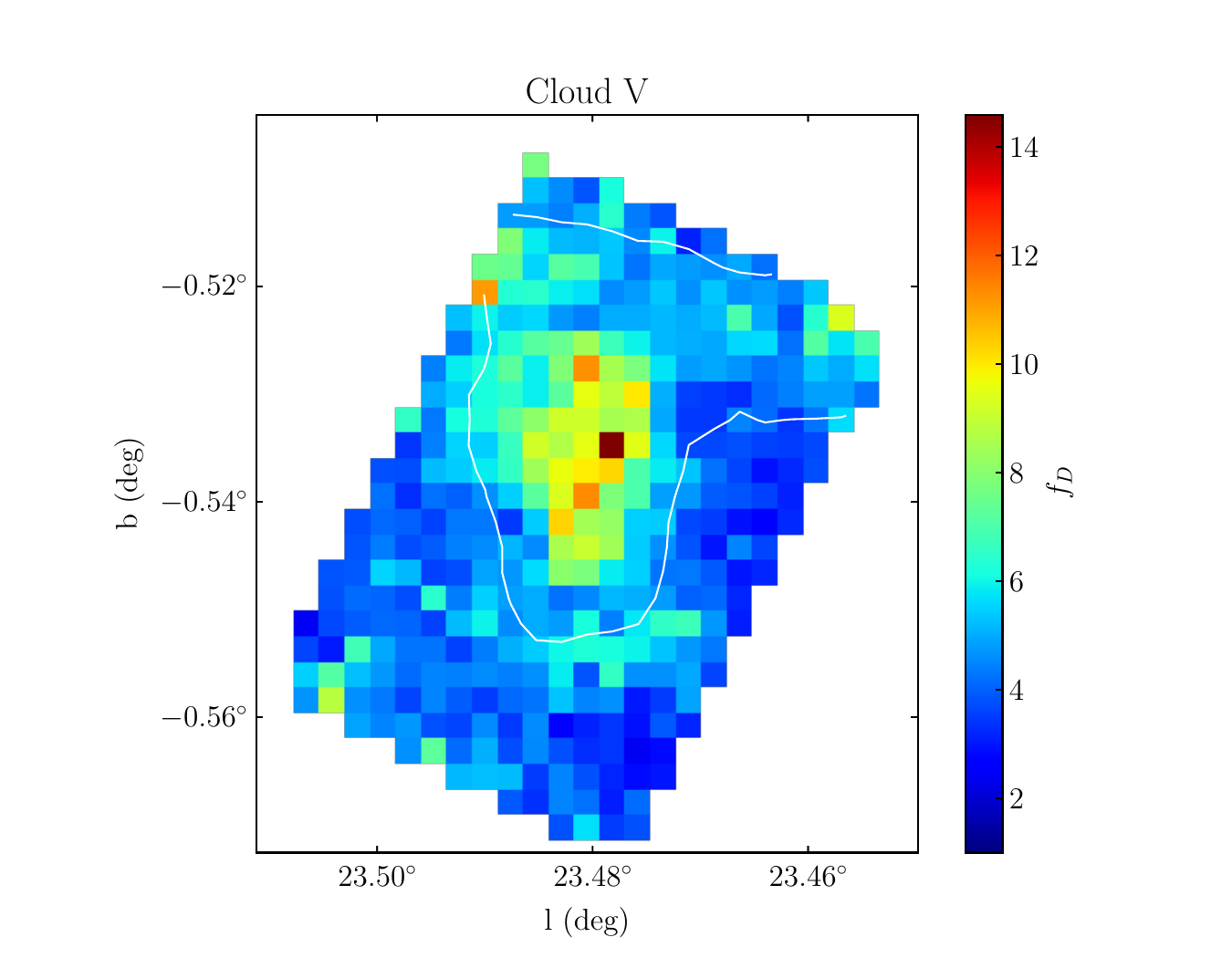}\includegraphics[width=0.35\textwidth,trim=1.7cm 1cm 1cm 1cm , clip=True]{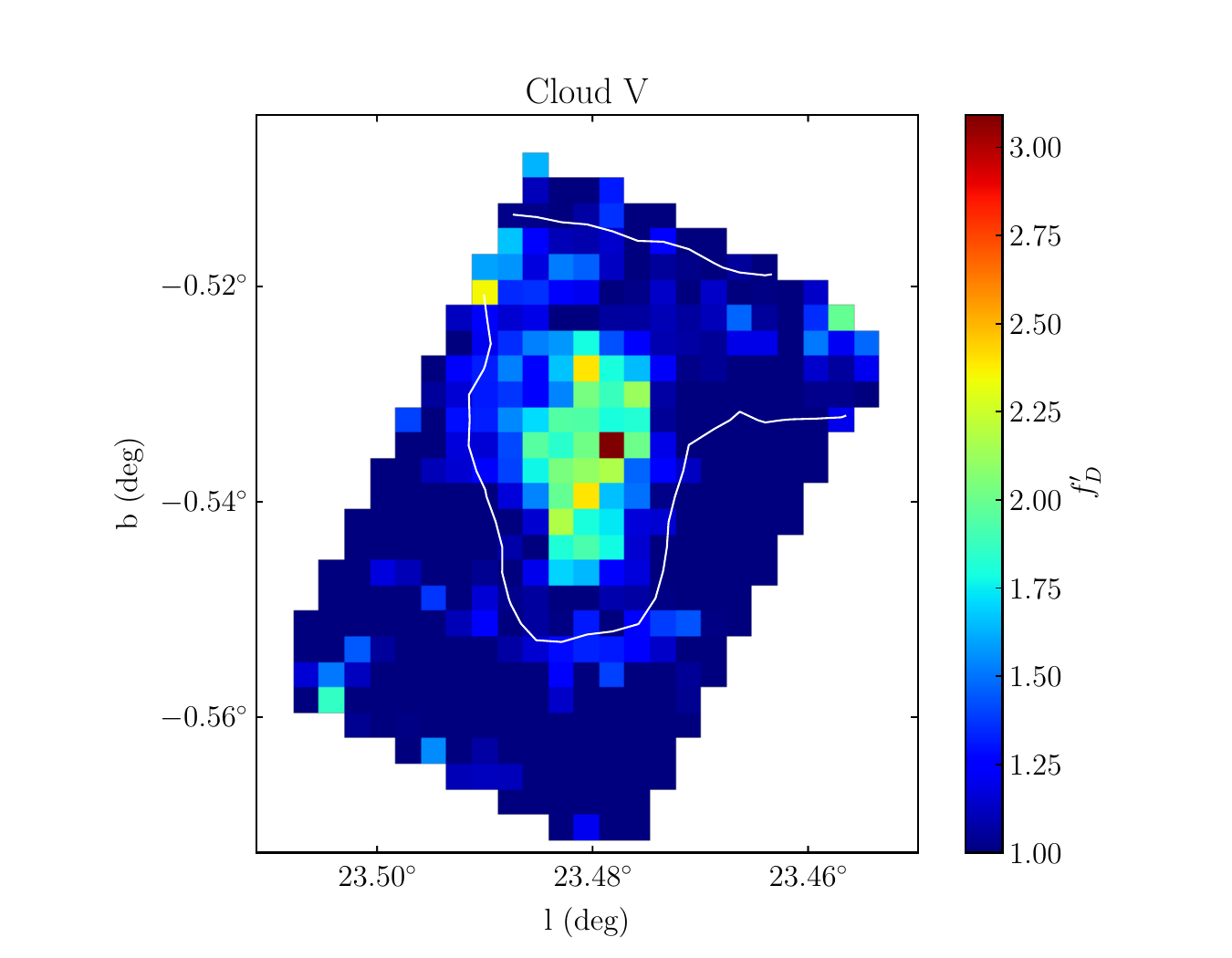}\\

    \includegraphics[width=0.35\textwidth,trim=1cm 1cm 1cm 1cm , clip=True]{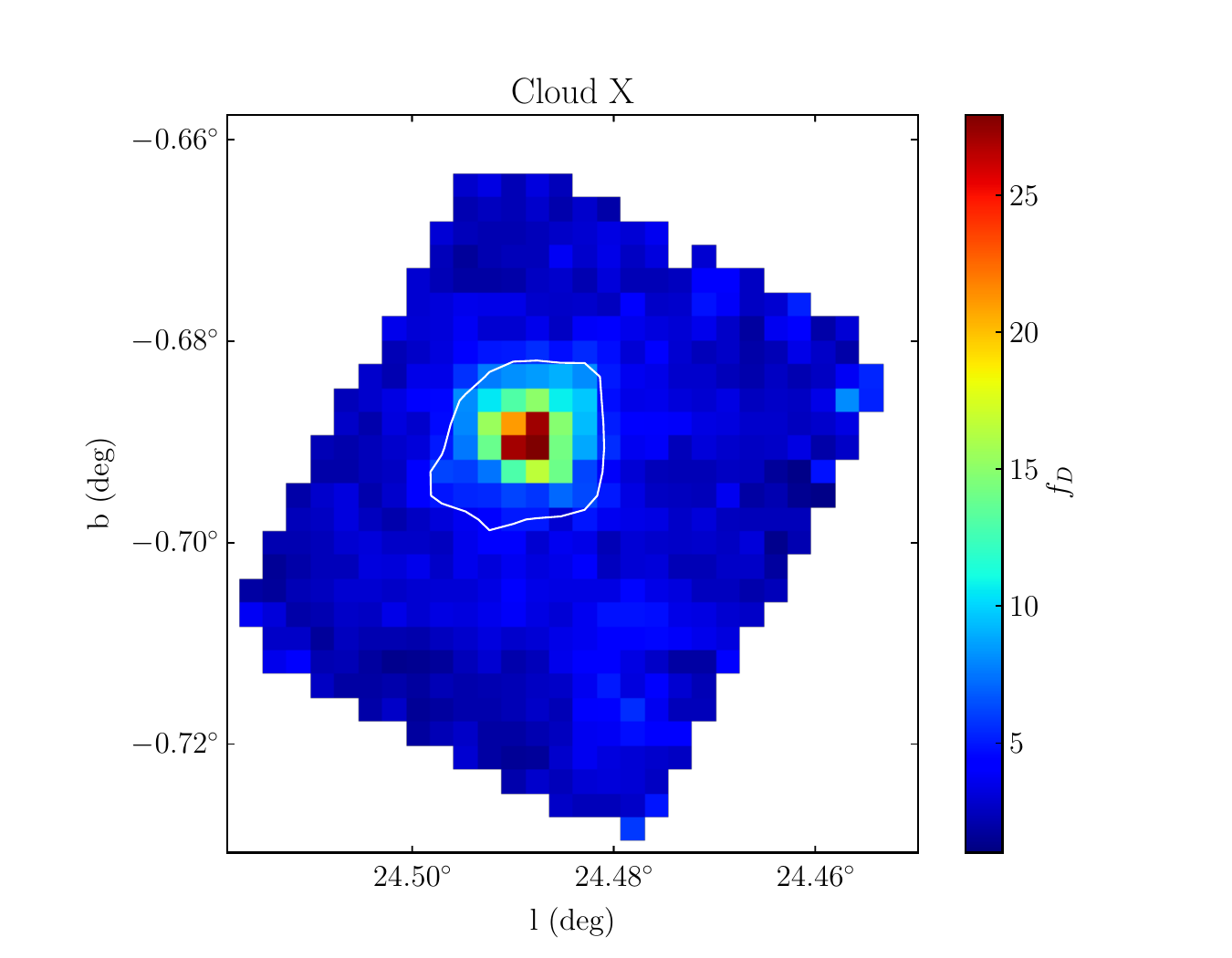}\includegraphics[width=0.35\textwidth,trim=1cm 1cm 1cm 1cm , clip=True]{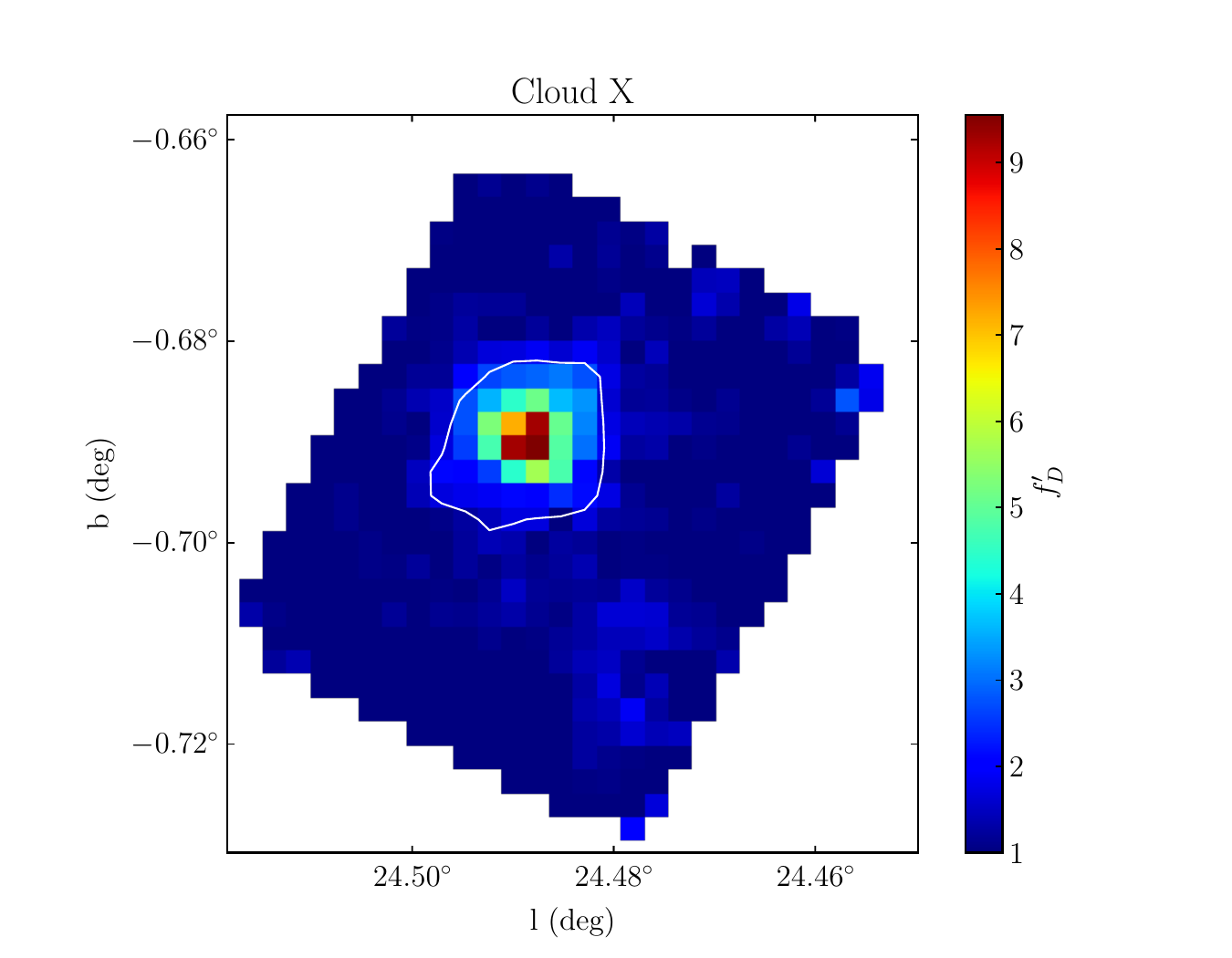}\\

    \includegraphics[width=0.35\textwidth,trim=1cm 1cm 1cm 1cm , clip=True]{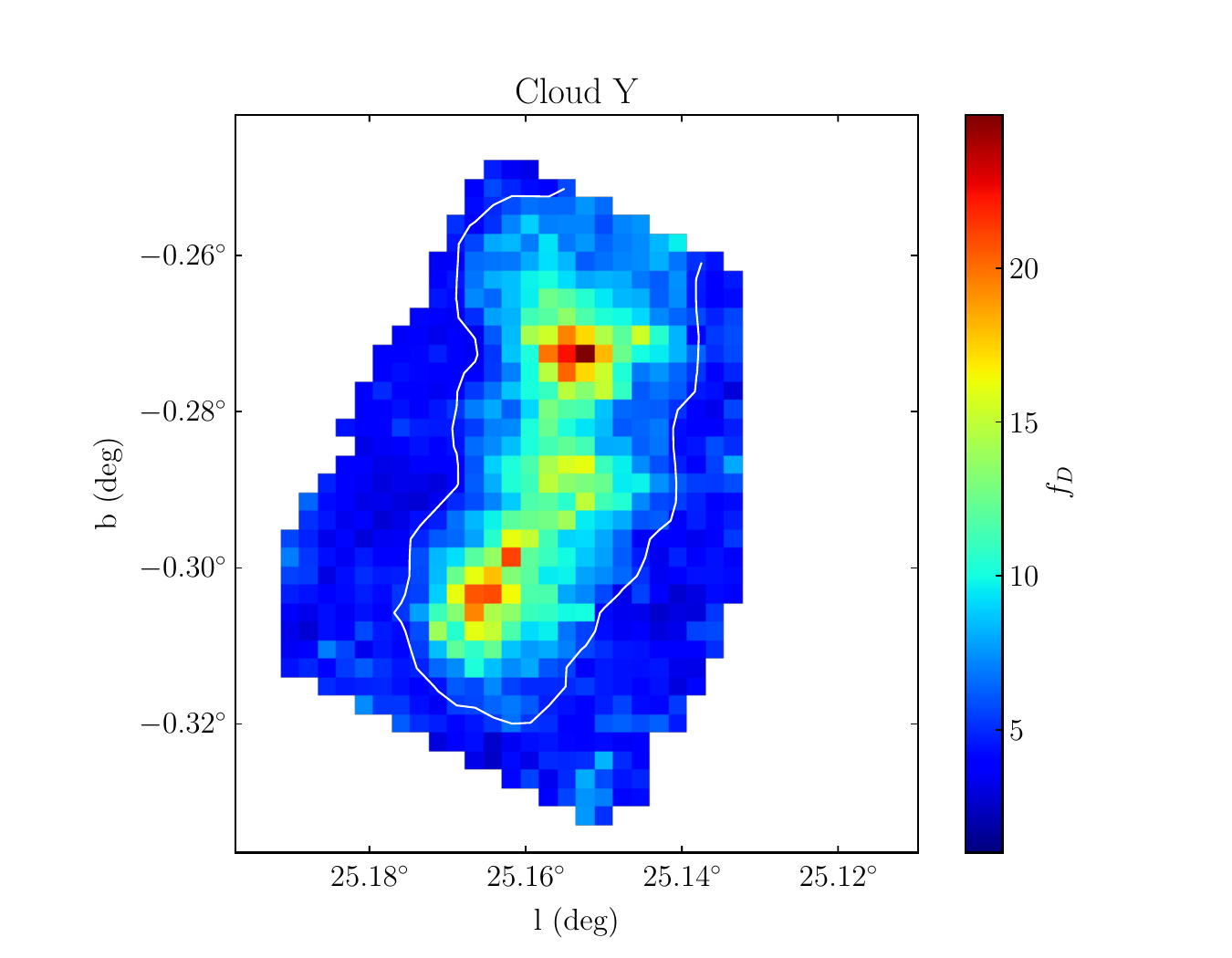}\includegraphics[width=0.35\textwidth,trim=1cm 1cm 1cm 1cm , clip=True]{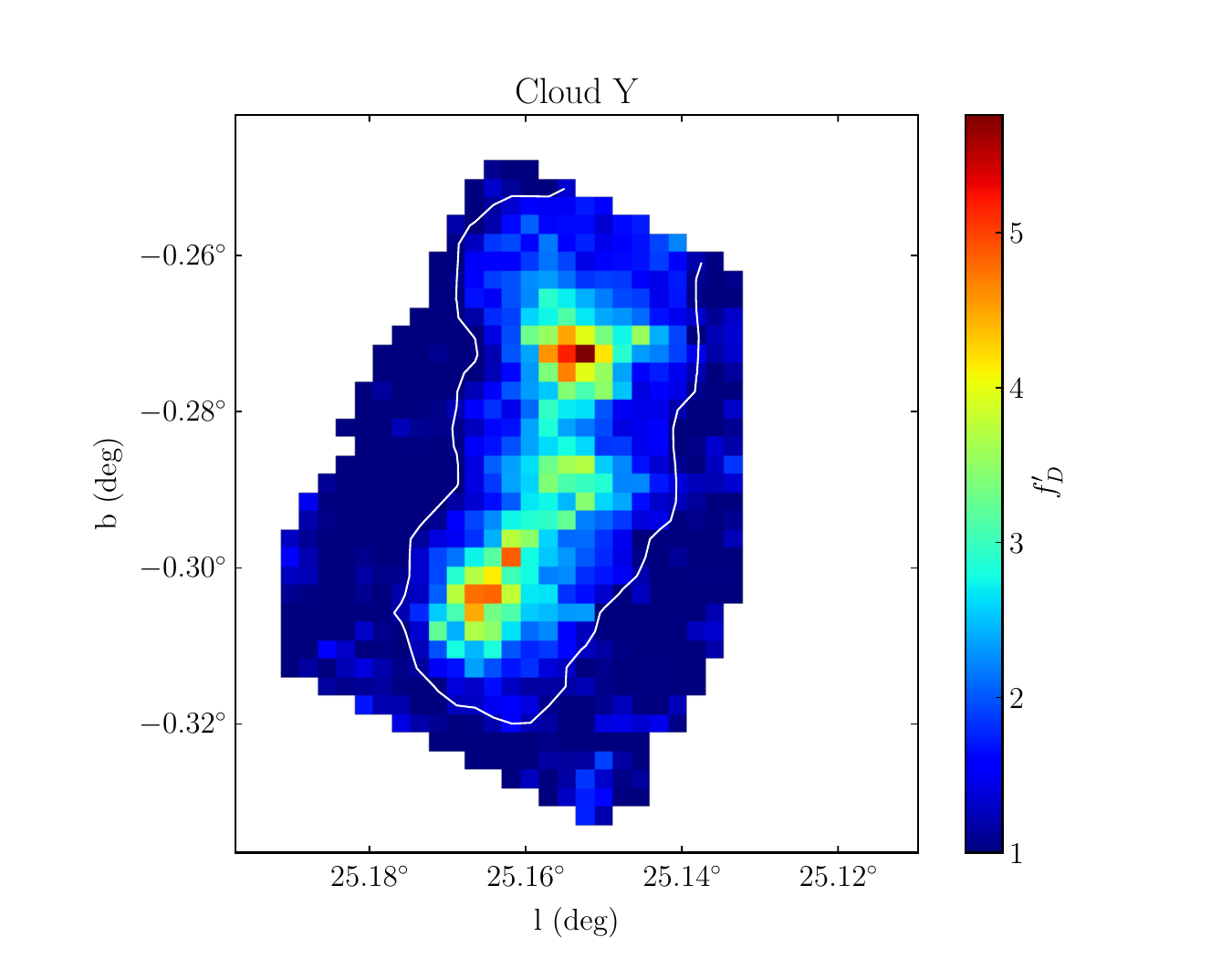}
 
    \caption{CO depletion factor (left column) and corrected CO depletion factor (right column). In each panel, white contours mark $\Sigma_{\rm FIR}= 0.1\:$g cm$^{-2}$. }
    \label{fig:fig3b}
\end{figure*}

Following the method described in \cite{hernandez2011}, we use the $^{13}$CO(1-0) and (2-1) emission to map the CO depletion factor across the four clouds. In particular, we regrid the $^{13}$CO(2-1) IRAM maps to match the poorer pixel size, angular and velocity resolution of the corresponding $^{13}$CO(1-0) FUGIN maps. We then assume local thermodynamic equilibrium (LTE) conditions, and use the following Equation \citep[see eq. A4 of][]{caselli2002} to estimate the species column density:
\begin{equation}
    \frac{dN_{\rm ^{13}CO}(v)}{dv} = \frac{8\pi \nu^3}{c^3A_{ul}} \frac{g_l}{g_u} \frac{Q_{\rm rot}(T_{\rm ex})}{1-{\rm exp}(-h\nu/[kT_{\rm ex}])} \frac{\tau_{\nu}}{g_l {\rm exp}(-E_{l}/[kT_{\rm ex}])},
\label{ColDens}
\end{equation}
where $\nu$ is the frequency of the transition, i.e., 110.2 GHz and 220.4 GHz for $^{13}$CO(1-0) and (2-1) respectively, $A_{ul}$ is the Einstein coefficient for spontaneous emission (6.3324$\times$10$^{-8}$ s$^{-1}$ for $J=1\rightarrow$0 and 6.0745$\times$10$^{-7}$ s$^{-1}$ for $J=2\rightarrow$1), g$_l$ and g$_u$ are the statistical weights of the lower and upper levels, respectively, $Q_{\rm{rot}}$ is the rotational partition function, $E_{l}$ is the energy of the lower state in the transition and $\tau$ is the optical depth. All the used spectroscopic parameters have been obtained from the Cologne Database for Molecular Spectroscopy catalogue$\footnote{https://cdms.astro.uni-koeln.de}$ and are listed in Table~\ref{tab:tab2} (except for g$_l$ that cancels out in Eq.~\ref{ColDens}). We use the following expression for $Q_{\rm{rot}}$:
\begin{equation}
    Q_{\rm{rot}} = \sum^{\infty}_{J=0} (2J+1)exp(-E_J/[kT_{\rm ex}])
\label{Qrot}
\end{equation}
with $E_J=J(J+1)hB$, where $B$ is the $^{13}$CO rotational constant 55101.011~MHz\footnote{https://spec.jpl.nasa.gov}. 

The optical depth, $\tau_\nu$, in eq.~\ref{ColDens} is estimated via:
\begin{equation}
    T_{B,\nu} = \frac{h\nu}{k}\left[f(T_{\rm ex}) -f(T_{\rm bg})\right](1-e^{-\tau_{\nu}}).
\label{tau}
\end{equation}
Here, $T_{B,\nu}$ is the main beam brightness temperature at each velocity (or frequency), $f(T) \equiv \left[ {\rm exp}(h\nu/[kT]) -1\right]^{-1}$ and $T_{\rm bg}$=2.73 K is the background temperature assumed to be the cosmic microwave background temperature.

\begin{table}
    \centering
    \begin{tabular}{ccccc}
    \hline
    \hline
         Transition &Frequency & $A_{ul}$ & $g_u$ &$E_{l}$\\
                    & (GHz)    & (10$^{-8}$ s$^{-1}$) & & (K)\\
    \hline
        $^{13}$CO(1-0)&110.201 &6.3324&3 &0.00\\
        $^{13}$CO(2-1)&220.399 &60.745&5 &5.29\\
    \hline
    \end{tabular}
    \caption{Spectroscopic information for the $^{13}$CO(1-0) and (2-1) transitions.}
    \label{tab:tab2}
\end{table}

In these relations (Equations~\ref{ColDens} to ~\ref{tau}), the $^{13}$C) column densities estimated from the two transitions independently need to be the same. Furthermore, it is necessary to know the excitation temperature, $T_{\rm{ex}}$, which is the same for both transitions in the adopted LTE conditions. Hence, we estimate $T_{\rm{ex}}$ at each $l, b, v$ element of the data cube as the value for which the ratio between the two $^{13}$CO column densities converge to unity, i.e., $R_{2,1}$ = $dN_{\rm ^{13}CO,21}/dN_{\rm ^{13}CO,10}$=1.
We first consider all the $l, b, v$ elements for which both transitions have emission above 3$\times$rms and calculate their $R_{2,1}$ assuming $T_{\rm ex}$=30~K. Next, we iteratively decrease $T_{\rm{ex}}$ until $R_{2,1}$ converges to unity. We calculate the two column densities including accounting for optical depth. 
For all those $l, b, v$ elements in which one or both transitions are below the 3$\times$rms threshold, we estimate $T_{\rm ex}$ to be the same as the average of that of the local $l, b$ position. Finally, for all the remaining $l, b, v$ elements without $T_{\rm{ex}}$ estimates, we assume this to be equal to the cloud-averaged $T_{\rm{ex}}$ across the full map. We thus obtain a cube of $T_{\rm{ex}}$ that we then use to estimate the $^{13}$CO column density toward each $l, b, v$ element.

We then converted these 3D column density cubes into 2D maps of $N_{\rm 13CO}$ by summing the contributions at each $v$ channel. Finally, we use these $N_{\rm 13CO}$ maps to estimate the total mass surface density, $\Sigma_{\rm 13CO}$ using the following equation:
\begin{equation}
    \Sigma_{\rm 13CO} = \frac{\mu_{\rm H}}{\chi_{\rm CO}}  \frac{\rm ^{12}C}{\rm ^{13}C} N_{\rm 13CO} 
\label{sigma}
\end{equation}
where $\chi_{\rm CO}=1.4\times10^{-4}$ is the fiducial CO abundance with respect to H nuclei, $\mu_{\rm H} = 2.34\times 10^{-24}\:$g is the mass per H nucleus, and $^{12}$C/$^{13}$C=51 is the ratio of these C isotopes. For this ratio we adopt the following relation from \cite{milam2005}, which assumes $d_{GC,0}$ = 8.0 kpc:
\begin{equation}
    \frac{^{12}\mathrm{C}}{^{13}\mathrm{C}} = 6.2 (\pm 1.0) (d_{\rm GC}/{\rm kpc}) + 18.7(\pm 7.4),
    \label{IsoRatio}
\end{equation}
where $d_{\rm GC}$ is the galactocentric distance, estimated from the cloud kinematic distances \citep{simon2006}. As reported in Table~\ref{tab:tab1}, we find $d_{\rm GC}$ in the range 4.9-5.5 kpc and $^{12}$C/$^{13}$C isotopic ratios between 49-52. Given the similarity of these values, for simplicity, we assume all the clouds to have a $^{12}$C/$^{13}$C isotopic ratio 51, i.e., corresponding to the mean of the derived values. We note that, from Equation~\ref{IsoRatio}, the uncertainty on the isotopic ratio is $\sim$25\%.

The $^{13}$CO column density map, $T_{\rm{ex}}$ column-density-weighted map and $^{13}$CO-derived mass surface density map for each IRDC are shown in Figure~\ref{fig:fig3a}. Toward all sources, we find $T_{\rm{ex}}$ in the range 5-10 K, with average values between 6-7 K (see Table~\ref{tab:tab3}), $N_{\rm 13CO}$ up to $6\times10^{16}\:$cm$^{-2}$ and $\Sigma_{\rm 13CO}$ up to 0.04 g cm$^{-2}$. These values are similar to those previously obtained toward Cloud H by \cite{hernandez2011} using the same methodology.

The quantities derived above have uncertainties that are both statistical due to measurement noise and systematic. For $T_{\rm ex}$, we assume its uncertainty to be 1 K at $T_{\rm ex}$=7 K, which was the level adopted by \cite{hernandez2011}. This corresponds to a $\sim$25\% uncertainty in the $^{13}$CO column densities. To this we sum in quadrature an additional $\sim$3\% uncertainty due to the rms, which thus ends up being a very minor contribution so that the final uncertainty on the column density is $\sim$25\%. Next we sum in quadrature the uncertainty arising from the adopted isotopic ratio and other quantities needed to derive $\Sigma_{\rm 13CO}$, for which we then estimate a total overall uncertainty of $\sim 35\%$. The {\it Herschel} FIR-derived $\Sigma$ also has an uncertainty of $\sim$30\%, as reported by \cite{lim2016}.  

\subsection{CO depletion}

\begin{figure*}
    \centering
    \includegraphics[width=0.5\textwidth, trim = 0.5cm 0.5cm 1cm 1cm, clip=True]{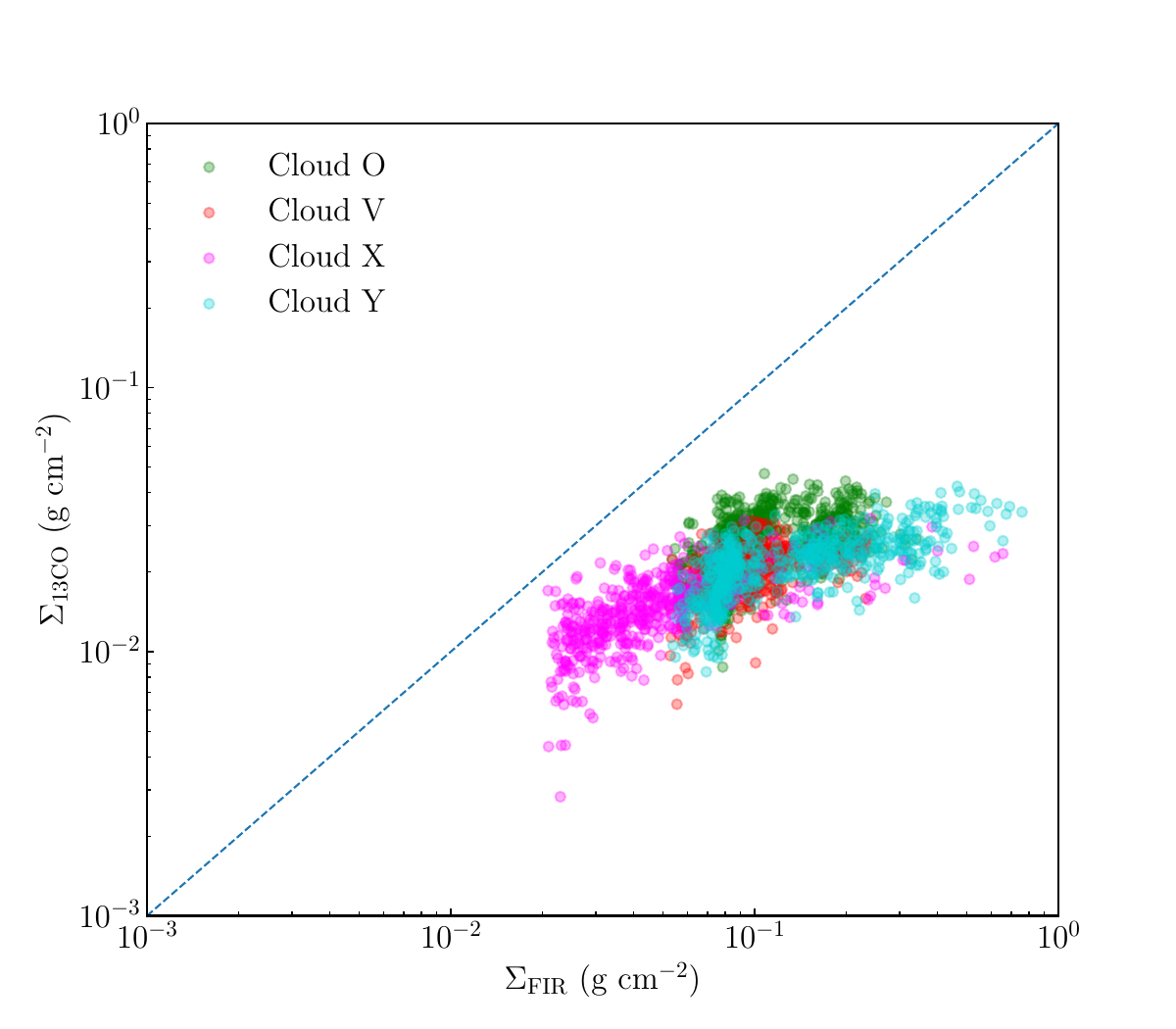}\includegraphics[width=0.5\textwidth, trim = 0.5cm 0.5cm 1cm 1cm, clip=True]{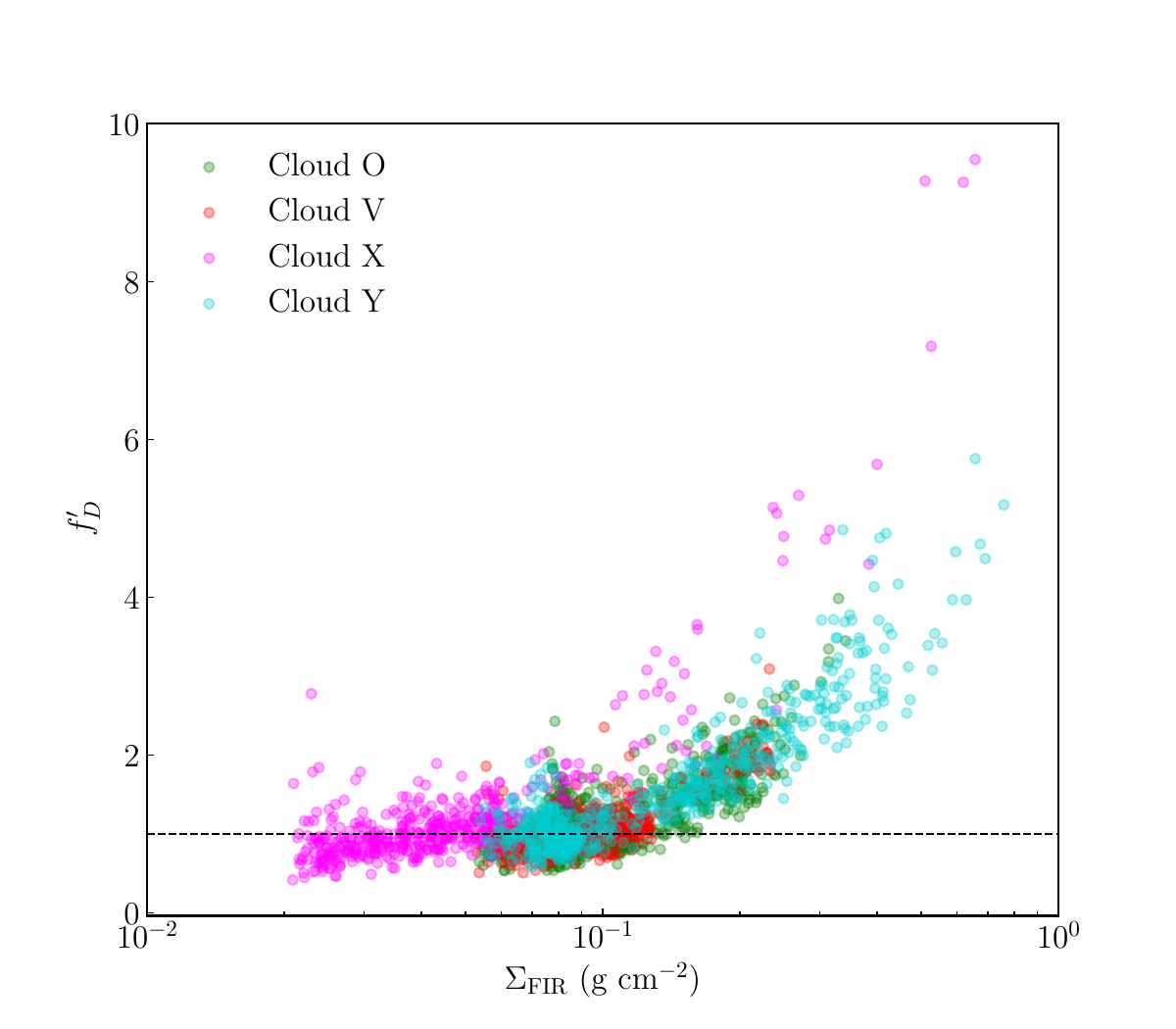}
    \caption{\textit{Left:} $^{13}$CO-derived mass surface density, $\Sigma_{\rm{13CO}}$, as a function of the Herschel-derived mass surface density, $\Sigma_{\rm{FIR}}$, for all clouds in our sample (different colors). The dotted line corresponds to the one-to-one line. \textit{Right:} Normalised CO depletion factor,$f^{'}_{D,{\rm CO}}$, as a function of the Herschel derived mass surface density, $\Sigma_{\rm{FIR}}$, for all clouds in our sample (different colors). The dotted horizontal corresponds to $f^{'}_{D,{\rm CO}}$=1.}
    \label{fig:fig4}
\end{figure*}

In Figure~\ref{fig:fig4}, we show a scatter plot of $\Sigma_{\rm 13CO}$ versus $\Sigma_{\rm FIR}$ for all the pixels in the maps of IRDCs O, V, X, and Y. We see that the $^{13}$CO-derived mass surface densities are in the range 0.003-0.05 g cm$^{-2}$. While $\Sigma_{\rm 13CO}$ increases with $\Sigma_{\rm FIR}$, it remains systematically lower by factors $\sim$3-25, with this factor growing towards the high-$\Sigma$ regime. This discrepancy between $\Sigma_{\rm 13CO}$ and $\Sigma_{\rm FIR}$ could be evidence of CO depletion. Alternatively, there could be other factors that are causing a systematic underestimation of $\Sigma$ via $^{13}$CO emission or an overestimation via FIR dust continuum emission.

One potential explanation, also discussed by \cite{hernandez2012}, is that a strong negative excitation temperature gradient from the more diffuse to the denser regions of the clouds could lead to the $\Sigma_{\rm{13CO}}$ being underestimated. However, as shown in Figure~\ref{fig:fig3a}, we do not observe such a trend in any of the IRDCs. On the contrary, we see hints of enhancement of $T_{\rm{ex}}$ toward the denser regions.

Local fractionation of the $^{13}$CO isotope could also explain the low values of $\Sigma_{\rm 13CO}$. Two possible mechanisms could account for this: isotope-selective photo-dissociation and chemical fractionation. However, in the case of isotope-selective photo-dissociation, the effect is known to be negligible for H volume densities $>$10$^2$ cm$^{-3}$ \citep{szucs2014}, with this limit well below the average IRDC densities in our clouds. Chemical fractionation, on the other hand, is predicted to become effective for $^{12}$CO column densities in the range 10$^{15}$-10$^{17}$ cm$^{-2}$. At the considered $^{12}$C/$^{13}$C ratio of 51, this corresponds to $^{13}$CO column densities in the range 2$\times$10$^{13}$-2$\times$10$^{15}$ cm$^{-2}$. Toward the four IRDCs, we estimate $N_{\rm 13CO}\geq10^{16}\:$cm$^{-3}$, hence, while chemical fractionation may be occurring to some extent, it is most likely not the main cause of the low $\Sigma_{\rm 13CO}$ that we obtain. 

Finally, the dust opacity assumptions adopted by \cite{lim2016} may affect the $\Sigma_{\rm FIR}$ estimates. The effects of these assumptions are already considered in the 30\% uncertainty reported by the authors and can only account for a small fraction of the $\Sigma_{\rm FIR}$ variations. In light of all this, we conclude that the relatively low values of $\Sigma_{\rm 13CO}$ are primarily due to the depletion of CO from the gas phase onto dust grains. 

To quantify CO depletion, following previous studies \citep[e.g.,][]{caselli1999,fontani2006,hernandez2011,jimenezserra2014}, we define the CO depletion factor as:
\begin{equation}
    f_{D} \equiv \frac{\Sigma_{\rm CO,expected}}{\Sigma_{\rm CO,observed}} = \frac{\Sigma_{\rm FIR}}{\Sigma_{\rm 13CO}}.
    \label{COdep}
\end{equation}
The corresponding CO depletion maps are shown in Figure~\ref{fig:fig3b}. In order to take into account the systematic uncertainties due to the assumed CO abundance and the Herschel-derived $\Sigma_{\rm FIR}$, we re-normalise $f_D$, so that it is $\sim$1 for $T_{\rm{dust}}\geq$20~K, i.e., the CO freeze-out temperature \citep{caselli1999}. We note that this temperature roughly corresponds to a value of $\Sigma\sim$0.1 g cm $^{-2}$ for all the clouds. We indicate this normalised CO depletion factor as $f^{\prime}_{D,{\rm CO}}=\beta f_D$ and show the obtained maps in Figure~\ref{fig:fig3b}. For each cloud, the normalisation factors, $\beta$, are reported in Table~\ref{tab:tab3}, along with the column density weighted mean excitation temperatures.

\begin{figure*}
    \centering
    \includegraphics[width=0.5\textwidth, trim = 0.5cm 0.5cm 1cm 1cm, clip=True]{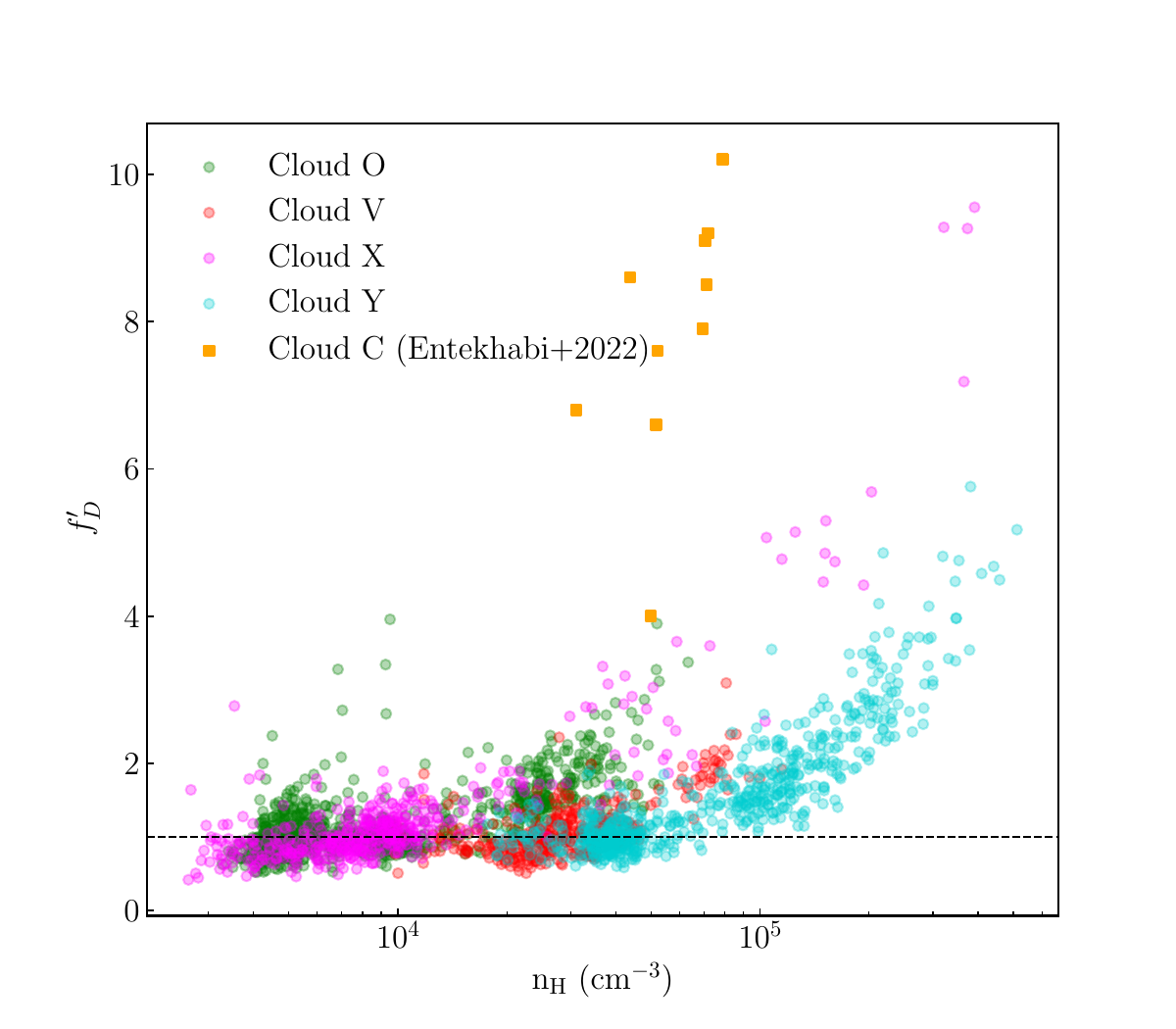}\includegraphics[width=0.5\textwidth, trim = 0.5cm 0.5cm 1cm 1cm, clip=True]{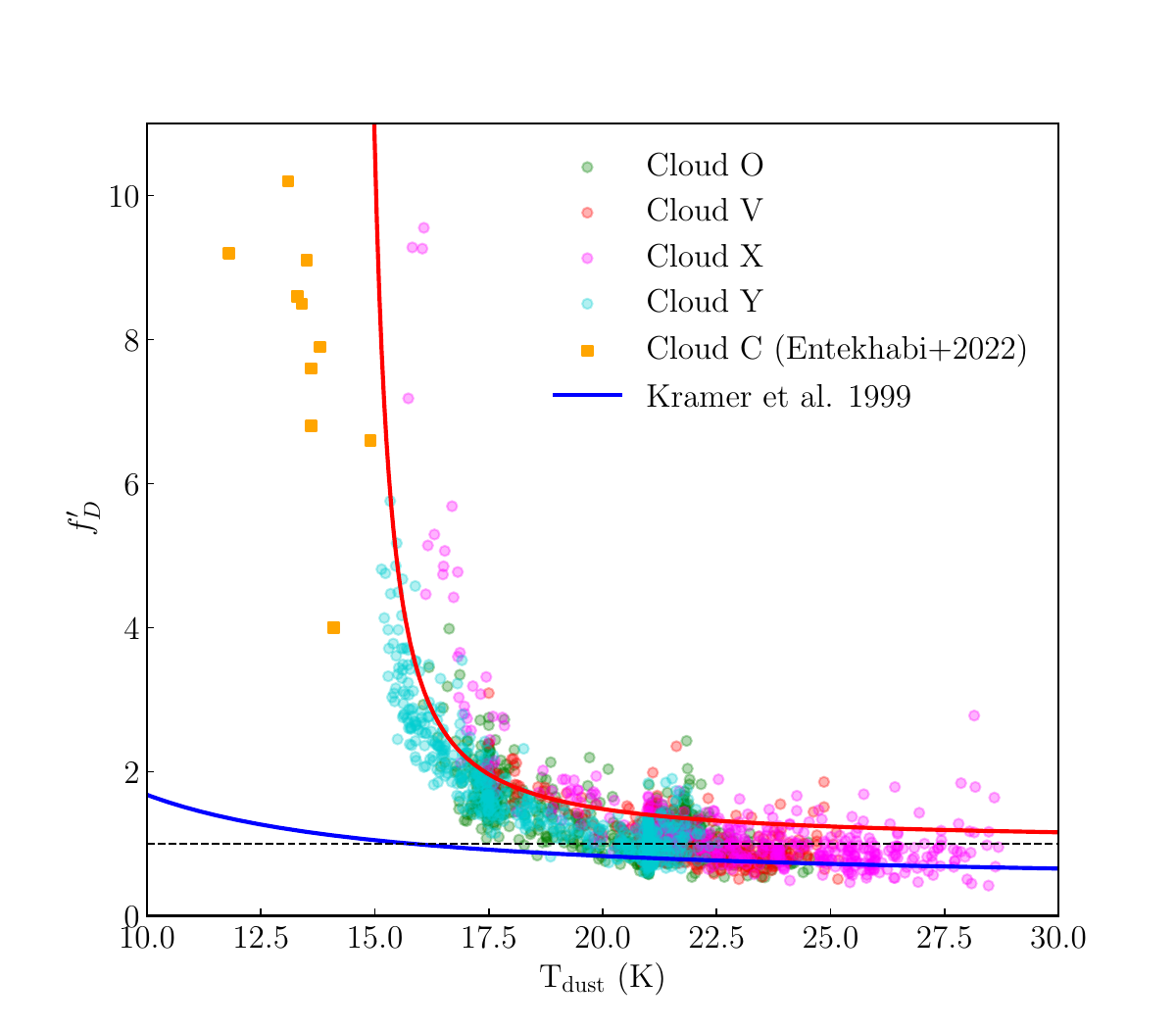}
    \caption{Corrected CO depletion factors, $f_{D}^{\prime}$, as a function of gas density (i.e., number density of H nuclei, $n_{\rm H}$) (left panel) and dust temperature (right panel). We also report the CO depletion values measured by \cite{entekhabi2022} in Cloud C as orange squares. In both panels the horizontal black dotted line corresponds to $f_{D}^{\prime}=1$. In the right panel, we also show the empirical relation reported by \cite{kramer1999} that is derived in the low-mass core IC~5146 (blue curve) (see eq.~\ref{eq:kramer}). We also show our derived ``IRDC depletion relation'' (see eq.~\ref{eq:irdc}) (red curve).}
    \label{fig:fig5}
\end{figure*}

From Figure~\ref{fig:fig3b}, we see that the CO depletion factor takes values $f_D\sim 3-25$ or $f_D^{\prime}\sim 1-9$ after normalisation. For an average cloud kinematic distance of 3~kpc (Table~\ref{tab:tab1}), the 8.5$^{\prime\prime}$ pixel scale in our maps corresponds to $\sim$0.125 pc and contains $7.4\times (\Sigma_{\rm 13CO}/0.1 {\rm g \:cm}^{-2})\:M_{\odot}$. Hence, several tens of solar masses per pixel are missed when using CO observations if depletion is not taken into account.

The CO depletion values shown in Figure~\ref{fig:fig3b} are consistent with those reported by previous studies on IRDCs within the \cite{butlerTan2012} sample. \cite{hernandez2011} report normalised CO depletion factors up to 5 toward the IRDC G035.39-00.33 \cite[cloud H in][]{butlerTan2012}, using C$^{18}$O emission. Toward the same source, \cite{jimenezserra2014} obtained $f_{D}$ up to 12, from a non-LTE analysis of the $^{13}$CO emission. \cite{entekhabi2022} used astrochemical models to infer the CO depletion factors toward ten massive clumps in the IRDC G028.37 \cite[cloud C in][]{butlerTan2012}. For these regions, \cite{entekhabi2022} reported $f_{D}$ up to 10, slightly larger than the values reported here. However, the 10 clumps studied in cloud C have colder dust temperatures, as discussed further below. \cite{sabatini2019} reported $f_{D}\leq6$, toward the IRDC G351.77-0.51, similar to our $f_{D}^{\prime}$ values. Our normalised CO depletion estimates are also consistent with those measured toward high-mass star forming clumps by \cite{feng2020} (up to 15) and \cite{fontani2012}, i.e., $\leq$10 for a sub-sample of clumps located at distances similar to our sources.

\begin{table}[]
    \centering
    \begin{tabular}{ccccc}
    \hline
    \hline
         Cloud &$\langle T_{\rm{ex}}\rangle$& $\beta$ & $f_{D}$ & $f_D^{\prime}$\\
         & (K) & \\
    \hline
         O&6.4 &3.7 &6 &1.6\\
         V&5.8 &4.7 &7 &1.5\\
         X&6.8 &2.9 &15 &5\\
         Y&7.0 &4.3 &11 &2.5\\
    \hline
    \end{tabular}
    \caption{Column density weighted excitation temperature, CO depletion normalisation factors, Herschel mass surface density weighted CO depletion factor and normalised CO depletion factor. The average CO depletion factors (both normalised and non-normalised) have been obtained by considering pixels with $\Sigma_{\rm{FIR}}\geq$0.1 g cm$^{-2}$.}
    \label{tab:tab3}
\end{table}

We now investigate how the CO depletion factor varies as a function of the cloud properties of H number density and dust temperature, as shown in Figure~\ref{fig:fig5}. Here, we also include the $f_{D}$ values measured by \cite{entekhabi2022} in IRDC G28.37 (Cloud C). The H number densities shown in Figure~\ref{fig:fig5} have been obtained by applying the machine-learning denoising diffusion probabilistic model (DDPM) described in \cite{xu2023} to the Herschel derived mass surface density maps. For a detailed description of the method, we refer to \cite{xu2023}.
As shown in Figure~\ref{fig:fig5}, the CO depletion factor exhibits clear trends as a function of both $n_{\rm H}$ and $T_{\rm{dust}}$. In particular, denser and colder regions show higher levels of CO depletion. However, there is relatively high scatter in the relation of $f_D$ with density, especially if the Cloud C data are also considered. On the other hand, the relation of CO depletion with temperature follows a monotonic relation more tightly.
 
The correlation between the large-scale CO depletion in IRDCs and cloud properties has been considered in some previous works \citep[e.g.,][]{kramer1999,fontani2011,sabatini2019}. These studies generally reported evidence of correlation between $n_{\rm H}$ and $f_{D}$ that are broadly consistent with our results. Only \citet{fontani2011} observed a slight anti-correlation between the two quantities, but also suggested this may be due to beam dilution effects. 
\citet{kramer1999} and \citet{sabatini2019} also investigated the relation between $f_{D}$ and $T_{\rm{dust}}$ for low-mass star-forming regions and IRDCs, respectively, finding qualitatively similar trends to our results.

In Figure~\ref{fig:fig5} right panel, we show the functional form of $f_D(T_{\rm dust})$ derived by \cite{kramer1999} (see Eq. 1). We see that it does not track the rapid rise of $f_D$, which we see occurring at $T_{\rm dust}\lesssim 18\:$K. This may reflect real differences in the molecular cloud environments between the low-mass core IC~5146 and our IRDC sample: especially the range of $\Sigma$ in the low-mass core is much smaller, i.e., $\Sigma\lesssim0.1\:{\rm g\:cm}^{-2}$, that our IRDCs, where $0.1\:{\rm g\:cm}^{-2}\lesssim\Sigma\lesssim 0.7\:{\rm g\:cm}^{-2}$. Alternatively, as mentioned in \S\ref{intro}, systematic uncertainties in measurement of $T_{\rm dust}$ may be at play.

In Figure~\ref{fig:fig5} we also present a new functional form for $f^\prime_D(T_{\rm dust})$ that is a better description of IRDC conditions:
\begin{equation}
    f_{D}^{\prime} = {\rm exp} \Bigg[\frac{T_0}{(T_{\rm dust}-T_1)}\Bigg].\label{eq:irdc}
\end{equation} 
The example curve shown in Fig.~~\ref{fig:fig5} has $T_0=4\:$K and $T_1=12\:$K.
The validity in the temperature range is $\sim 15\:{\rm K} \lesssim T_{\rm dust}\lesssim 30\:$K. The main feature of the relation is the rapid rise of depletion factor at temperatures $\lesssim 18\:$K. 

Overall, these results suggest that dust temperature, rather than density, is the most important variable in controlling CO depletion factor. In addition, the relatively small amount of scatter seen in the $f^\prime_D(T_{\rm dust})$ relation may indicate than CO depletion has reached a near equilibrium value, i.e., a balance in the rates of freeze-out and desorption, with the latter being dominated by thermal desorption rather than non-thermal (e.g., cosmic ray induced) desorption processes. We anticipate that these results will be important constraints for astrochemical models of molecular clouds  \citep[see also][]{entekhabi2022} as well as their chemodynamical history \citep[e.g.,][]{hsu2023b}.

\section{Conclusions}
We have used $^{13}$CO(1-0) and (2-1) observations to infer the levels of CO depletion toward a sample of four IRDCs. We find normalised CO depletion factors up to 10, that cannot be explained by systematic uncertainties in our analysis or chemical effects only. We find that CO depletion generally increases with increasing cloud density, although with significant scatter. There is a tighter correlation with dust temperature, with CO depletion rising rapidly for $T_{\rm dust}\lesssim18\:$K. To capture this behavior we have proposed a functional form for the normalised CO depletion factor of $f_D^\prime = A \:{\rm exp}(T_0/[T_{\rm dust}-T_1])$ with values of the coefficients $T_0\simeq 4\:$K and $T_1\simeq12\:$K. Overall these results indicate that dust temperature is the most important variable in controlling CO depletion factor. The relatively small amount of scatter may indicate that the level of gas phase CO has reached near equilibrium values with thermal desorption playing a dominant role in balancing freeze-out. These results provide important constraints for both astrochemical models and the chemodynamical history of gas during the early stages of star formation.

\begin{acknowledgements}
G.C. acknowledges support from the Swedish Research Council (VR Grant; Project: 2021-05589). J.C.T. acknowledges support from ERC project 788829 (MSTAR). I.J-.S acknowledges funding from grant No. PID2019-105552RB-C41 awarded by the Spanish Ministry of Science and Innovation/State Agency of Research MCIN/AEI/10.13039/501100011033. JDH gratefully acknowledges financial support from the Royal Society (University Research Fellowship; URF/R1/221620). P.G. acknowledges support from the Chalmers Cosmic Origins postdoctoral fellowship. R.F. acknowledges support from the grants Juan de la Cierva FJC2021-046802-I, PID2020-114461GB-I00 and CEX2021-001131-S funded by MCIN/AEI/ 10.13039/501100011033 and by "European Union NextGenerationEU/PRTR". S.V. acknowledges partial funding from the European Research Council (ERC) Advanced Grant MOPPEX 833460. This work is based on observations carried out under project number 013-20 with the IRAM 30m telescope. IRAM is supported by INSU/CNRS (France), MPG (Germany) and IGN (Spain). This publication makes use of data from FUGIN, FOREST Unbiased Galactic plane Imaging survey with the Nobeyama 45-m telescope, a legacy project in the Nobeyama 45-m radio telescope. 
\end{acknowledgements}

%
\bibliographystyle{aa} 
\bibliography{aa.bib}

\begin{thebibliography}{56}
\expandafter\ifx\csname natexlab\endcsname\relax\def\natexlab#1{#1}\fi

\bibitem[{{Butler} \& {Tan}(2009)}]{2009ApJ...696..484B}
{Butler}, M.~J. \& {Tan}, J.~C. 2009, \apj, 696, 484

\bibitem[{{Butler} \& {Tan}(2012)}]{butlerTan2012}
{Butler}, M.~J. \& {Tan}, J.~C. 2012, \apj, 754, 5

\bibitem[{{Caselli} {et~al.}(2002){Caselli}, {Benson}, {Myers}, \&
  {Tafalla}}]{caselli2002}
{Caselli}, P., {Benson}, P.~J., {Myers}, P.~C., \& {Tafalla}, M. 2002, \apj,
  572, 238

\bibitem[{{Caselli} {et~al.}(1999){Caselli}, {Walmsley}, {Tafalla}, {Dore}, \&
  {Myers}}]{caselli1999}
{Caselli}, P., {Walmsley}, C.~M., {Tafalla}, M., {Dore}, L., \& {Myers}, P.~C.
  1999, \apjl, 523, L165

\bibitem[{{Christie} {et~al.}(2012){Christie}, {Viti}, {Yates}, {Hatchell},
  {Fuller}, {Duarte-Cabral}, {Sadavoy}, {Buckle}, {Graves}, {Roberts},
  {Nutter}, {Davis}, {White}, {Hogerheijde}, {Ward-Thompson}, {Butner},
  {Richer}, \& {Di Francesco}}]{christie2012}
{Christie}, H., {Viti}, S., {Yates}, J., {et~al.} 2012, \mnras, 422, 968

\bibitem[{{Churchwell} {et~al.}(2009){Churchwell}, {Babler}, {Meade},
  {Whitney}, {Benjamin}, {Indebetouw}, {Cyganowski}, {Robitaille}, {Povich},
  {Watson}, \& {Bracker}}]{churchwell2009}
{Churchwell}, E., {Babler}, B.~L., {Meade}, M.~R., {et~al.} 2009, \pasp, 121,
  213

\bibitem[{{Crapsi} {et~al.}(2005){Crapsi}, {Caselli}, {Walmsley}, {Myers},
  {Tafalla}, {Lee}, \& {Bourke}}]{crapsi2005}
{Crapsi}, A., {Caselli}, P., {Walmsley}, C.~M., {et~al.} 2005, \apj, 619, 379

\bibitem[{{Dalgarno} \& {Lepp}(1984)}]{dalgarno1984}
{Dalgarno}, A. \& {Lepp}, S. 1984, \apjl, 287, L47

\bibitem[{{Egan} {et~al.}(1998){Egan}, {Shipman}, {Price}, {Carey}, {Clark}, \&
  {Cohen}}]{egan1998}
{Egan}, M.~P., {Shipman}, R.~F., {Price}, S.~D., {et~al.} 1998, \apjl, 494,
  L199

\bibitem[{{Entekhabi} {et~al.}(2022){Entekhabi}, {Tan}, {Cosentino}, {Hsu},
  {Caselli}, {Walsh}, {Lim}, {Henshaw}, {Barnes}, {Fontani}, \&
  {Jim{\'e}nez-Serra}}]{entekhabi2022}
{Entekhabi}, N., {Tan}, J.~C., {Cosentino}, G., {et~al.} 2022, \aap, 662, A39

\bibitem[{{Feng} {et~al.}(2020){Feng}, {Li}, {Caselli}, {Du}, {Lin},
  {Sipil{\"a}}, {Beuther}, {Sanhueza}, {Tatematsu}, {Liu}, {Zhang}, {Wang},
  {Hogge}, {Jimenez-Serra}, {Lu}, {Liu}, {Wang}, {Zhang}, {Zahorecz}, {Li},
  {Liu}, \& {Yuan}}]{feng2020}
{Feng}, S., {Li}, D., {Caselli}, P., {et~al.} 2020, \apj, 901, 145

\bibitem[{{Fontani} {et~al.}(2006){Fontani}, {Caselli}, {Crapsi}, {Cesaroni},
  {Molinari}, {Testi}, \& {Brand}}]{fontani2006}
{Fontani}, F., {Caselli}, P., {Crapsi}, A., {et~al.} 2006, \aap, 460, 709

\bibitem[{{Fontani} {et~al.}(2012){Fontani}, {Giannetti}, {Beltr{\'a}n},
  {Dodson}, {Rioja}, {Brand}, {Caselli}, \& {Cesaroni}}]{fontani2012}
{Fontani}, F., {Giannetti}, A., {Beltr{\'a}n}, M.~T., {et~al.} 2012, \mnras,
  423, 2342

\bibitem[{{Fontani} {et~al.}(2011){Fontani}, {Palau}, {Caselli},
  {S{\'a}nchez-Monge}, {Butler}, {Tan}, {Jim{\'e}nez-Serra}, {Busquet},
  {Leurini}, \& {Audard}}]{fontani2011}
{Fontani}, F., {Palau}, A., {Caselli}, P., {et~al.} 2011, \aap, 529, L7

\bibitem[{{Ford} \& {Shirley}(2011)}]{fordShirley2011}
{Ford}, A.~B. \& {Shirley}, Y.~L. 2011, \apj, 728, 144

\bibitem[{{Fortune-Bashee} {et~al.}(2024){Fortune-Bashee}, {Sun}, \&
  {Tan}}]{2024ApJ...977L...6F}
{Fortune-Bashee}, X., {Sun}, J., \& {Tan}, J.~C. 2024, \apjl, 977, L6

\bibitem[{{Foster} {et~al.}(2014){Foster}, {Arce}, {Kassis}, {Sanhueza},
  {Jackson}, {Finn}, {Offner}, {Sakai}, {Sakai}, {Yamamoto}, {Guzm{\'a}n}, \&
  {Rathborne}}]{foster2014}
{Foster}, J.~B., {Arce}, H.~G., {Kassis}, M., {et~al.} 2014, \apj, 791, 108

\bibitem[{{Herbst} \& {van Dishoeck}(2009)}]{HerbstDishoeck2009}
{Herbst}, E. \& {van Dishoeck}, E.~F. 2009, \araa, 47, 427

\bibitem[{{Hernandez} \& {Tan}(2015)}]{hernandez2015}
{Hernandez}, A.~K. \& {Tan}, J.~C. 2015, \apj, 809, 154

\bibitem[{{Hernandez} {et~al.}(2011){Hernandez}, {Tan}, {Caselli}, {Butler},
  {Jim{\'e}nez-Serra}, {Fontani}, \& {Barnes}}]{hernandez2011}
{Hernandez}, A.~K., {Tan}, J.~C., {Caselli}, P., {et~al.} 2011, \apj, 738, 11

\bibitem[{{Hernandez} {et~al.}(2012){Hernandez}, {Tan}, {Kainulainen},
  {Caselli}, {Butler}, {Jim{\'e}nez-Serra}, \& {Fontani}}]{hernandez2012}
{Hernandez}, A.~K., {Tan}, J.~C., {Kainulainen}, J., {et~al.} 2012, \apjl, 756,
  L13

\bibitem[{{Hsu} {et~al.}(2023){Hsu}, {Tan}, {Holdship}, {Duo}, {Xu}, {Viti},
  {Wu}, \& {Gaches}}]{hsu2023b}
{Hsu}, C.-J., {Tan}, J.~C., {Holdship}, J., {et~al.} 2023, arXiv e-prints,
  arXiv:2308.11803

\bibitem[{{Inutsuka} {et~al.}(2015){Inutsuka}, {Inoue}, {Iwasaki}, \&
  {Hosokawa}}]{inutsuka2015}
{Inutsuka}, S.-i., {Inoue}, T., {Iwasaki}, K., \& {Hosokawa}, T. 2015, \aap,
  580, A49

\bibitem[{{Jim{\'e}nez-Serra} {et~al.}(2014){Jim{\'e}nez-Serra}, {Caselli},
  {Fontani}, {Tan}, {Henshaw}, {Kainulainen}, \&
  {Hernandez}}]{jimenezserra2014}
{Jim{\'e}nez-Serra}, I., {Caselli}, P., {Fontani}, F., {et~al.} 2014, \mnras,
  439, 1996

\bibitem[{{Kainulainen} \& {Tan}(2013)}]{kainulainen2013}
{Kainulainen}, J. \& {Tan}, J.~C. 2013, \aap, 549, A53

\bibitem[{{Kong} {et~al.}(2015){Kong}, {Caselli}, {Tan}, {Wakelam}, \&
  {Sipil{\"a}}}]{kong2015}
{Kong}, S., {Caselli}, P., {Tan}, J.~C., {Wakelam}, V., \& {Sipil{\"a}}, O.
  2015, \apj, 804, 98

\bibitem[{{Kong} {et~al.}(2017){Kong}, {Tan}, {Caselli}, {Fontani}, {Liu}, \&
  {Butler}}]{2017ApJ...834..193K}
{Kong}, S., {Tan}, J.~C., {Caselli}, P., {et~al.} 2017, \apj, 834, 193

\bibitem[{{Kramer} {et~al.}(1999){Kramer}, {Alves}, {Lada}, {Lada}, {Sievers},
  {Ungerechts}, \& {Walmsley}}]{kramer1999}
{Kramer}, C., {Alves}, J., {Lada}, C.~J., {et~al.} 1999, \aap, 342, 257

\bibitem[{{Li} {et~al.}(2018){Li}, {Tan}, {Christie}, {Bisbas}, \&
  {Wu}}]{2018PASJ...70S..56L}
{Li}, Q., {Tan}, J.~C., {Christie}, D., {Bisbas}, T.~G., \& {Wu}, B. 2018,
  \pasj, 70, S56

\bibitem[{{Lim} {et~al.}(2016){Lim}, {Tan}, {Kainulainen}, {Ma}, \&
  {Butler}}]{lim2016}
{Lim}, W., {Tan}, J.~C., {Kainulainen}, J., {Ma}, B., \& {Butler}, M.~J. 2016,
  \apjl, 829, L19

\bibitem[{{Milam} {et~al.}(2005){Milam}, {Savage}, {Brewster}, {Ziurys}, \&
  {Wyckoff}}]{milam2005}
{Milam}, S.~N., {Savage}, C., {Brewster}, M.~A., {Ziurys}, L.~M., \& {Wyckoff},
  S. 2005, \apj, 634, 1126

\bibitem[{{Morii} {et~al.}(2021){Morii}, {Sanhueza}, {Nakamura}, {Jackson},
  {Li}, {Beuther}, {Zhang}, {Feng}, {Tafoya}, {Guzm{\'a}n}, {Izumi}, {Sakai},
  {Lu}, {Tatematsu}, {Ohashi}, {Silva}, {Olguin}, \& {Contreras}}]{morii2021}
{Morii}, K., {Sanhueza}, P., {Nakamura}, F., {et~al.} 2021, \apj, 923, 147

\bibitem[{{Moser} {et~al.}(2020){Moser}, {Liu}, {Tan}, {Lim}, {Zhang}, \&
  {Farias}}]{moser2020}
{Moser}, E., {Liu}, M., {Tan}, J.~C., {et~al.} 2020, \apj, 897, 136

\bibitem[{{Perault} {et~al.}(1996){Perault}, {Omont}, {Simon}, {Seguin},
  {Ojha}, {Blommaert}, {Felli}, {Gilmore}, {Guglielmo}, {Habing}, {Price},
  {Robin}, {de Batz}, {Cesarsky}, {Elbaz}, {Epchtein}, {Fouque}, {Guest},
  {Levine}, {Pollock}, {Prusti}, {Siebenmorgen}, {Testi}, \&
  {Tiphene}}]{perault1996}
{Perault}, M., {Omont}, A., {Simon}, G., {et~al.} 1996, \aap, 315, L165

\bibitem[{{Peretto} {et~al.}(2016){Peretto}, {Lenfestey}, {Fuller},
  {Traficante}, {Molinari}, {Thompson}, \& {Ward-Thompson}}]{peretto2016}
{Peretto}, N., {Lenfestey}, C., {Fuller}, G.~A., {et~al.} 2016, \aap, 590, A72

\bibitem[{{Pillai} {et~al.}(2019){Pillai}, {Kauffmann}, {Zhang}, {Sanhueza},
  {Leurini}, {Wang}, {Sridharan}, \& {K{\"o}nig}}]{pillai2019}
{Pillai}, T., {Kauffmann}, J., {Zhang}, Q., {et~al.} 2019, \aap, 622, A54

\bibitem[{{Pillai} {et~al.}(2006){Pillai}, {Wyrowski}, {Carey}, \&
  {Menten}}]{pillai2006}
{Pillai}, T., {Wyrowski}, F., {Carey}, S.~J., \& {Menten}, K.~M. 2006, \aap,
  450, 569

\bibitem[{{Rathborne} {et~al.}(2006){Rathborne}, {Jackson}, \&
  {Simon}}]{rathborne2006}
{Rathborne}, J.~M., {Jackson}, J.~M., \& {Simon}, R. 2006, \apj, 641, 389

\bibitem[{{Retes-Romero} {et~al.}(2020){Retes-Romero}, {Mayya}, {Luna}, \&
  {Carrasco}}]{RetesRomero2020}
{Retes-Romero}, R., {Mayya}, Y.~D., {Luna}, A., \& {Carrasco}, L. 2020, \apj,
  897, 53

\bibitem[{{Sabatini} {et~al.}(2019){Sabatini}, {Giannetti}, {Bovino}, {Brand},
  {Leurini}, {Schisano}, {Pillai}, \& {Menten}}]{sabatini2019}
{Sabatini}, G., {Giannetti}, A., {Bovino}, S., {et~al.} 2019, \mnras, 490, 4489

\bibitem[{{Simon} {et~al.}(2006){Simon}, {Rathborne}, {Shah}, {Jackson}, \&
  {Chambers}}]{simon2006}
{Simon}, R., {Rathborne}, J.~M., {Shah}, R.~Y., {Jackson}, J.~M., \&
  {Chambers}, E.~T. 2006, \apj, 653, 1325

\bibitem[{{Suwannajak} {et~al.}(2014){Suwannajak}, {Tan}, \&
  {Leroy}}]{2014ApJ...787...68S}
{Suwannajak}, C., {Tan}, J.~C., \& {Leroy}, A.~K. 2014, \apj, 787, 68

\bibitem[{{Sz{\H{u}}cs} {et~al.}(2014){Sz{\H{u}}cs}, {Glover}, \&
  {Klessen}}]{szucs2014}
{Sz{\H{u}}cs}, L., {Glover}, S. C.~O., \& {Klessen}, R.~S. 2014, \mnras, 445,
  4055

\bibitem[{{Tan}(2000)}]{tan2000}
{Tan}, J.~C. 2000, \apj, 536, 173

\bibitem[{{Tan}(2010)}]{2010ApJ...710L..88T}
{Tan}, J.~C. 2010, \apjl, 710, L88

\bibitem[{{Tan} {et~al.}(2014){Tan}, {Beltr{\'a}n}, {Caselli}, {Fontani},
  {Fuente}, {Krumholz}, {McKee}, \& {Stolte}}]{tan2014}
{Tan}, J.~C., {Beltr{\'a}n}, M.~T., {Caselli}, P., {et~al.} 2014, in Protostars
  and Planets VI, ed. H.~{Beuther}, R.~S. {Klessen}, C.~P. {Dullemond}, \&
  T.~{Henning}, 149

\bibitem[{{Tan} {et~al.}(2013){Tan}, {Kong}, {Butler}, {Caselli}, \&
  {Fontani}}]{tan2013}
{Tan}, J.~C., {Kong}, S., {Butler}, M.~J., {Caselli}, P., \& {Fontani}, F.
  2013, \apj, 779, 96

\bibitem[{{Tasker} \& {Tan}(2009)}]{tasker2009}
{Tasker}, E.~J. \& {Tan}, J.~C. 2009, \apj, 700, 358

\bibitem[{{Umemoto} {et~al.}(2017){Umemoto}, {Minamidani}, {Kuno}, {Fujita},
  {Matsuo}, {Nishimura}, {Torii}, {Tosaki}, {Kohno}, {Kuriki}, {Tsuda},
  {Hirota}, {Ohashi}, {Yamagishi}, {Handa}, {Nakanishi}, {Omodaka}, {Koide},
  {Matsumoto}, {Onishi}, {Tokuda}, {Seta}, {Kobayashi}, {Tachihara}, {Sano},
  {Hattori}, {Onodera}, {Oasa}, {Kamegai}, {Tsuboi}, {Sofue}, {Higuchi},
  {Chibueze}, {Mizuno}, {Honma}, {Muller}, {Inoue}, {Morokuma-Matsui},
  {Shinnaga}, {Ozawa}, {Takahashi}, {Yoshiike}, {Costes}, \&
  {Kuwahara}}]{umemoto2017}
{Umemoto}, T., {Minamidani}, T., {Kuno}, N., {et~al.} 2017, \pasj, 69, 78

\bibitem[{{V{\'a}zquez-Semadeni} {et~al.}(2019){V{\'a}zquez-Semadeni}, {Palau},
  {Ballesteros-Paredes}, {G{\'o}mez}, \& {Zamora-Avil{\'e}s}}]{vazquez2019}
{V{\'a}zquez-Semadeni}, E., {Palau}, A., {Ballesteros-Paredes}, J.,
  {G{\'o}mez}, G.~C., \& {Zamora-Avil{\'e}s}, M. 2019, \mnras, 490, 3061

\bibitem[{{Whittet} {et~al.}(2010){Whittet}, {Goldsmith}, \&
  {Pineda}}]{whittet2010}
{Whittet}, D.~C.~B., {Goldsmith}, P.~F., \& {Pineda}, J.~L. 2010, \apj, 720,
  259

\bibitem[{{Wu} {et~al.}(2017){Wu}, {Tan}, {Nakamura}, {Van Loo}, {Christie}, \&
  {Collins}}]{wu2017}
{Wu}, B., {Tan}, J.~C., {Nakamura}, F., {et~al.} 2017, \apj, 835, 137

\bibitem[{{Wu} {et~al.}(2015){Wu}, {Van Loo}, {Tan}, \& {Bruderer}}]{wu2015}
{Wu}, B., {Van Loo}, S., {Tan}, J.~C., \& {Bruderer}, S. 2015, \apj, 811, 56

\bibitem[{{Xu} {et~al.}(2023){Xu}, {Tan}, {Hsu}, \& {Zhu}}]{xu2023}
{Xu}, D., {Tan}, J.~C., {Hsu}, C.-J., \& {Zhu}, Y. 2023, \apj, 950, 146

\bibitem[{{Yu} {et~al.}(2020){Yu}, {Wang}, \& {Tan}}]{yu2020}
{Yu}, H., {Wang}, J., \& {Tan}, J.~C. 2020, \apj, 905, 78

\bibitem[{{Zhang} {et~al.}(2009){Zhang}, {Wang}, {Pillai}, \&
  {Rathborne}}]{zhang2009}
{Zhang}, Q., {Wang}, Y., {Pillai}, T., \& {Rathborne}, J. 2009, \apj, 696, 268

\end{thebibliography}
%

\end{document}